\documentclass[useAMS,usenatbib] {mn2e}

\usepackage[table]{xcolor}

\usepackage{float}

\usepackage{verbatim}
\usepackage{graphicx}
\usepackage{subcaption}

\DeclareGraphicsExtensions{.png, .pdf}

\title[A lower fragmentation mass scale for clumps]{A lower fragmentation mass scale in high redshift galaxies and its implications on giant clumps: 
a systematic numerical study}
\author[Tamburello et al.]{Valentina Tamburello$^{1,2}$\thanks{E-mail:
vtambure@physik.uzh.ch}, Lucio Mayer $^{1,2}$\thanks{E-mail: lmayer@physik.uzh.ch}, Sijing Shen$^{3}$ and James Wadsley$^4$\\
$^{1}$Center for Theoretical Astrophysics and Cosmology, Institute for Computational Science, University of Zurich, \\
Winterthurerstrasse 190, 8057 Zurich, Switzerland\\
$^{2}$Physik-Institut, University of Zurich, Winterthurerstrasse 190, 8057 Zurich, Switzerland\\
$^{3}$Institute of Astronomy, University of Cambridge, Madingley Road, Cambridge CB3 0HA, United Kingdom\\
$^{4}$Department of Physics and Astronomy, McMaster University, 1280 Main St. W Hamilton, ON, L8S 4M1, Canada}

\begin{document}

\pagerange{\pageref{firstpage}--\pageref{lastpage}} 

\maketitle

\label{firstpage}

\begin{abstract}
We study the effect of sub-grid physics, galaxy mass, structural parameters and resolution on the fragmentation of gas-rich galaxy discs into massive star forming clumps. The initial conditions are set up with the aid of the ARGO cosmological hydrodynamical simulation. Blast-wave feedback does not suppress fragmentation, but reduces both the number of clumps and the duration of the unstable phase. Once formed, bound clumps cannot be destroyed by our feedback model. Widespread fragmentation is promoted by high gas fractions and low halo concentrations. Yet giant clumps $M > 10^8 M_{\odot}$ lasting several hundred Myr are rare and mainly produced by clump-clump mergers. They occur in massive discs with maximum rotational velocities $V_{max} > 250$ km/s at $z \sim 2$, at the high mass end of the observed galaxy population at those redshifts.
The typical gaseous and stellar masses of clumps in all runs are in the range $\sim 10^7-10^8 M_{\odot}$ for galaxies with disc mass in the range $10^{10}-8\times 10^{10} M_{\odot}$.
Clumps sizes are usually in the range  $100-400$ pc, in agreement with recent clump observations in lensed high-z galaxies. \\
We argue that many of the giant clumps identified in observations are not due to in-situ fragmetation, or are the result of blending of smaller structures owing to insufficient resolution.
Using an analytical model describing local collapse inside spiral arms, we can predict the characteristic gaseous masses of clumps at the onset of fragmentation 
($\sim 3-5 \times 10^7 M_{\odot}$) quite accurately, while the conventional Toomre mass overestimates them.  
Due to their moderate masses, clumps which migrate to the centre have marginal effect on bulge growth.

\end{abstract}

\begin{keywords}
galaxy evolution -- clumps -- feedback -- simulations.
\end{keywords}

\section{Introduction}\label{Intro}
In the last years observations in the UV-rest frame have shown that many massive galaxies in the redshift range $z \sim 1-3$ (the most active period of galaxy formation) are large discs with irregular morphologies known as clumps \citep{C95, vdB96, E05}. The clumpy galaxies were called ``clump-cluster" galaxies and ``chain" galaxies, when viewed face on and edge on, respectively \citep{E04b}. Even if they could be associated to mergers or galaxy interactions, only a minority is in the process of merging ($\sim 30 \%$) and, moreover, their morphology and kinematics indicate that they are rotationally supported, extended discs \citep{G06, FS06, S08, FS09}. \\
High redshift galactic discs are very different from their counterpart in the local universe. Most of the discs at $z > 1$ are highly turbulent, with high velocity dispersion of $\sigma = 30-90$ km/s and rotation to dispersion ratio of $v_{rot}/ \sigma \sim 1-7$ \citep{C09, FS09, Wisnioski15} against the value $v_{rot}/\sigma \sim 10-20$ in local galaxies. This suggests they should be thick turbulent discs, but still rotation dominated. Another important property of these discs is that they are also gas rich. Estimates of the molecular gas fraction, based on CO measurements, give us $f_{gas} = M_{gas}/(M_{gas} + M_{star}) \sim 0.4 - 0.6$ \citep{T08, T13, Saintonge13, DZ14}, more than a factor of 2 higher than in local massive spiral galaxies. Finally they are strongly star-forming discs, with star formation rate $SFR > 10 M_{\odot}/yr$, that follows the well established redshift evolution of the main-sequence (e.g. \citealt{N07, Daddi07, Rod11, Sch14}), but is more than one order of magnitude higher than today's SFR. Clumps are identified in high resolution HST images and in nebular line emission from integral field spectroscopy on scales of about $1$ kpc. They are characterised by high density (around 8 times denser than discs) and high star formation rate, $\sim 0.5-100 M_{\odot} /yr$ \citep{E04a, E05, FS11, G11, DZ11, Guo12}.
Modern telescopes with adaptive optics can achieve resolution of $\sim 1-1.5$ kpc at $z \sim 2$, limiting the number of spatially resolved elements, since not only the galaxies at high redshift are intrinsically smaller, but their cosmological surface brightness decreases $ \propto (1+z)^4$, enabling us to observe only the bright central regions. New observations \citep{Livermore12, Livermore15}, instead, have used gravitational lensing by massive foreground clusters to enhance spatial resolution. Moreover, unlensed surveys have focused on the more extreme star-forming population (due to detection limits), gravitational lensing has allowed to focus on more `normal' galaxies with lower intrinsic luminosity (since lensing conserves surface brightness). Star-forming clumps have been observed also in these galaxies but at much higher resolution ($\sim 100$ pc), allowing direct comparisons with the HII regions in the local Universe \citep{Swinbank09, Jones10, Livermore12, Livermore15}. 
While overall the range of sizes and masses of clumps found by these last works are wide,  $\sim 100 - 1.5$ kpc 
and  $3.3 \times 10^6$ - $3.1 \times 10^9$ $M_{\odot}$, the typical mass found in the lensing sample, which
benefits from high resolution, is only $5 \times 10^7 M_{\odot}$, much smaller than the mass of giant
clumps identified in previous lower resolution observations, which were in the range $10^8-10^9 M_{\odot}$
\citep{Guo12, T13}.
It is conceivable that the lower characteristic masses found in these recent observations might be
more representative of the clump phenomenon in hi-z galaxies, calling for theoretical models and simulations
to explain that.\\
The dominant interpretation so far to explain the origin of clumps is that they are produced by gravitational instability, in particular by the fragmentation of spiral arms in a massive gas-rich disc. According to the standard Toomre instability analysis \citep{T64}, a rotating disc becomes unstable to local gravitational collapse if its surface density is higher than the forces induced by differential rotation and velocity dispersion. This is expressed in terms of the Toomre parameter that, in case of fluid discs, is
\begin{equation}\label{eq:Q_gas}
Q_g = \frac{c_s k}{\pi G \Sigma}
\end{equation}
where $c_s$ is the sound speed, $k$ is the epicyclic frequency, $G$ is the gravitational constant and $\Sigma$ is the surface density of the disc at radius $r$. In case of kinematic discs (stellar discs) the Toomre parameter becomes
\begin{equation}\label{eq:Q_star}
Q_s = \frac{\sigma_r k}{3.36 G \Sigma}
\end{equation}
where $\sigma_r$ is the radial velocity dispersion. In both cases if $Q > 1.5$, the disc is locally stable.
As we will explain better in Section \ref{Clumps}, the Toomre wavelength
\begin{equation}\label{eq:lambda}
\lambda_T = \frac{4 \pi^2 G \Sigma}{k^2}
\end{equation}
should determine the maximum clump fragmentation scale, with characteristic fragmentation scale given by the most unstable wavelength, situated at one half of the Toomre wavelength: 
\begin{equation}\label{eq:most_unstable}
\lambda$(most unstable) $\equiv p \lambda_T 
\end{equation}
where $p$ is a constant equal to 0.5 or 0.55 for zero-thickness fluid or stellar discs respectively \citep{BT08}. This most unstable wavelength should then be comparable to the sizes of clumps that arise, but a first cautionary remark is that in a disc with both gas and stars one should use the two-fluid definition of the Toomre parameter (see \citealt{JS84}, \citealt{R01} and \citealt{C10}) given by:
\begin{equation}\label{eq:Q_total}
Q^{-1} = 2 Q_s^{-1} \frac{q}{1+q^2} + 2 Q_g^{-1} \frac{\sigma_{gs} q}{1+\sigma_{gs}^2 q^2}
\end{equation}
where $Q_g$ and $Q_s$ are defined in Equations (\ref{eq:Q_gas}) and (\ref{eq:Q_star}), respectively, $\sigma_{gs} \equiv \sigma_g/ \sigma_s$ is the ratio of the velocity dispersion of the stellar component to that of the gas, and $q$ is the dimensionless wave number. The system is unstable once $Q < 1$ and the most unstable wavelength corresponds to the $q$ that minimizes $Q$ (between $q = 1$ for $\sigma_{gs} = 1$ and $q \simeq \sigma_{gs}^{-1}$ for $\sigma_{gs} \ll 1$).
Another caveat is that, for a gaseous disc, one could use either the thermal sound speed or the radial gas velocity dispersion, and often the practice in the galactic disc modelling community has been to choose the maximum of the two (e.g. \citealt{C10}). We will adopt the same method throughout our analysis.\\
More in general, when approaching the study of disc instability one should take into account that the Toomre analysis is born out of linear perturbation theory and assumes an initially thin, axisymmetric disc, which undergoes an isothermal axisymmetric perturbation, while in reality we are normally interested in global, non-axisymmetric instabilities in a finite-thickness 3D disc, which is not strictly isothermal. Furthermore, gravitationally bound massive clumps, if formed by gravitational instability, cannot be fully understood assuming linear perturbation theory, because they are by definition in the strongly nonlinear regime.
For this reason we have used a different approach in Section \ref{Clumps} in order to understand clump masses, using a model already employed to explain the properties of sub-stellar companions fragmenting out of protoplanetary discs.\\
Finally, the typical age of the stellar population in clumps ranges from tens Myr to one Gyr \citep{E09, FS11, G11, Guo12}. Since older clumps ($>600 $ Myr) are found closer to the center and a fraction of these high-z clumpy discs shows a central stellar bulge \citep{G08, E08}, it was proposed a clump-origin bulge scenario \citep{N99, G08, E08, C10}: if clumps survive enough (this should happen only for those more massive and denser), they fall into the galactic centre, due to dynamical friction (losing angular momentum), and could have an important role in building up the bulge.\\
Previous works \citep{C10, G12} have based their results on cosmological simulations which zoom-in on selected massive galaxies with stellar masses comparable to the bright galaxies typically observed in high-z surveys. In this paper, instead, we present the results from different isolated galaxy simulations with setup based on structure of galaxies in high resolution cosmological simulation (ARGO, see \citealt{FM15}). We consider a much wider range of galaxy masses, probing into the lower mass but more common galaxy population now accessible with lensing surveys such as those of \citet{Livermore15}, while at the same time including galaxies as massive as those studied in previous simulations.\\
While less realistic to some extent, isolated disc simulations allow to reach high resolution and to explore systematically and in a controlled way the dependence of fragmentation on structural parameters, including galaxy mass, as well as on sub-grid physics, which is not possible in cosmological simulations. A systematic approach is essential if we want to understand the origin of clump masses for example.\\
In Section \ref{Sim} we describe our code and the simulation setup. In Section \ref{Results} we present our results, divided by their dependence on used sub-grid physics, structural parameters and resolution. In Section \ref{Clumps} we analyse the clump size and mass in our simulations, using an estimate of fragmentation mass scale alternative to the Toomre mass, showing why fragmentation cannot lead to giant clumps and demonstrating that, even after a relatively long phase of mass growth, the typical mass of clumps remains below $2 \cdot 10^8 M_{\odot}$. In Section \ref{Bulge} we study the possibility that the bulge is not built up by clumps (even if it can grow a bit), but it is produced by gas inflow due to spiral arms torques before clumpy phase. Finally in Section \ref{Discussion} we summarize and discuss our results. 
\section[]{THE SIMULATIONS}\label{Sim}
\subsection{Code}\label{Code}
We perform a large set of simulations of isolated galaxies using GASOLINE2, a new version of the N-body + smoothed particle hydrodynamic (SPH) code GASOLINE \citep{S01, SWR02, WSQ04}. The initial conditions adopted for the galaxy models are described in subsection \ref{subsec:IC}.
GASOLINE computes gravitational forces between particles using a binary tree, and smoothing them with a spline kernel function (we use an opening angle parameter 
$\theta = 0.7$). The code solves the equations of hydrodynamics and includes radiative cooling for a mixture of hydrogen, helium and metals as well as Compton cooling. While for hydrogen and helium we compute directly the rates without assuming ionization equilibrium, metal-line cooling rates are calculated assuming collisional ionization equilibrium. We simply used the tabulated values calculated with the photoionization code CLOUDY \citep{F98}, as in \citet{SWS10}.
We use well tested sub-grid prescriptions for star formation, supernova feedback, metal yields from supernovae and stellar mass return to the gas phase via stellar winds. 
The adopted recipes are described in detail in \citet{S2006} and are summarised here.
In our feedback model, known as the blastwave model, we disable cooling within the radius of the blastwave produced by supernovae type II for a time corresponding to the end of the snowplough phase ($10-30$ Myr). 
We release an energy of $4 \times 10^{50}$ erg per supernovae explosion into the gas encompassed by the blastwave.
The cooling is not disabled for type Ia supernovae, the frequency of which is estimated from the binary fraction
of \citet{RVN96}.
 A gas particle must satisfy three criteria in order to have star formation. It must be:
\begin{enumerate}
	\item denser than a fixed threshold density $n_{min}$
	\item lower than a fixed temperature threshold $T_{max}$
	\item part of a converging flow 
\end{enumerate}
Which individual gas particles form stars is determined using a probabilistic approach, assuming a Kroupa IMF \citep{Kroupa01}, after imposing that the local star formation rate (SFR) follows a Schmidt law \citep{S59}, using a star formation efficiency $\epsilon_{SF} = 0.01$.
In our case we use $T_{max} = 3 \times 10^4 K$ and a density threshold of $n_{SF} = 10$ cm$^{-3}$. 
We stress that the adopted sub-grid models for star formation and feedback are motivated by our previous work on galaxy formation. Indeed we have
shown extensively that a high star formation density threshold, such as that used here, in combination with the blastwave
feedback, can produce realistic replicas of massive spiral galaxies such as those in the ERIS simulations \citep{GC11, Bird13, M12}, explain the main properties of dwarf galaxies, including turning cores into cusps \citep{GBM10, S14}, and, more recently, produce a variety or realistic Hubble Types, including massive central galaxies inside galaxy groups (in the ARGO simulations, see \citealt{FM15} and \citealt{F15}). Indeed with a high star formation density threshold,
close to that at which gas turns into stars in Giant Molecular Clouds (GMCs), supernovae-driven winds can be naturally
generated, regulating both the baryonic content and the star formation rate of galaxies \citep{M12}.
The ability to reproduce realistic galaxies across a range of mass scales implies that our sub-grid recipes, despite
they should be regarded as purely phenomenological, reasonably capture the energetics of galaxy formation. This is
important for the investigation on which this paper focuses, since the stability of gaseous discs is ultimately determined by the energetics at play in the gas phase. We note that previous work on the stability of high redshift discs did not employ recipes that had previously undergone such an extensive validation in the broader context of galaxy formation. Furthermore, compared to the ERIS and ARGO simulations, which were run releasing $8 \times 10^{50}$ erg per supernova and used $\epsilon_{SF} = 0.05$ and $\epsilon_{SF}=0.1$ respectively, we opted for a more conservative choice of the star formation and feedback efficiency, which should favour gravitational instability in the cold gas phase.
GASOLINE2 allows to solve the hydrodynamical equations using SPH formulations, resolving mixing and two-fluid
instabilities. Such formulations are adopted in some of the runs of this paper. 
They include thermal and metal diffusion terms in the momentum and energy equations based on the sub-grid turbulence prescription described by \citet{SWS10} and the new implementation of the hydrodynamical force 
equation described by \citet{K14}.
The new approach is based on the geometric density average (GDSPH) in the SPH force expression $(P_i+P_j)/(\rho_i\rho_j)$ in place of the usual $P_i/\rho_i^2+P_j/ \rho_j^2$, where $P_i$ and $\rho_i$ are particle pressures and densities respectively. This modification leads to smoother gradients and removes artificial surface tension \citep{K14}.
Detailed tests of the GDSPH implementation combined with diffusion as in run3 (see Section \ref{subsec:IC}, Table \ref{table_runs} ) will be presented in Wadsley et al. (in preparation), showing that it can successfully capture Kelvin-Helmoltz and Rayleigh-Taylor instabilities.
Note that other new SPH implementations, while they often track entropy rather than energy as we do, also use a geometric density mean for the forces (e.g. \citealt{H13, Saitoh13}).
\\
Finally, in order to guarantee that any fragmentation remains physical, we ensure that the gas particle's implied Jeans scale is resolved at a given temperature and density by imposing in all our simulations a pressure floor constraint, following \citet{A09}. The minimum pressure is set to $P_{min} = \varepsilon G h^2\rho^2$, where $\varepsilon = 3.0$ is a safety factor, $G$ is the gravitational constant, $h$ is the smoothing length and $\rho$ is the particle density (see also \citealt{R14}).\\
We started from low resolution simulations using $1.2 \times 10^6$ dark matter particles ($p_{DM}$), $10^5$ gas particles ($p_{gas}$) and $10^5$ star particles ($p_s$), but in some cases we also increase the resolution (here after hres) using $p_{DM} = 2 \times 10^6$, $p_{gas} = p_s = 10^6$. The mass resolution is different for each case, see Table \ref{table_resolution}, depending on the initial conditions.
For all particle species, the gravitational softening length, $\epsilon$, is $100$ pc. Only one simulation has $\epsilon =50$ pc as time steps become prohibitively small with decreasing softening, see Table \ref{table_runs} for more details. 
Finally the hydro resolution, given by the SPH smoothing length, $h$, is different for each case: in the Table \ref{table_resolution} we indicate the most likely value of SPH resolution in an annulus centered on the galaxy centre, between 1 and 4 kpc (since clumps in our simulations form in that region), but note that in a SPH simulation the hydro resolution is higher in high density regions, so at the sites of clump formation it reaches values $\sim 10-25$ pc.\\
Note that this is comparable with that of high resolution AMR simulations (e.g. \citealt{G12, M14}).
\begin{table*}
\vspace {1 mm}
\centering
\begin{tabular}{|c|c|c|c|c|c|c|c|c|c|c|}
\hline
 \bf{Model} &  \bf{Vel [km/s]}    & \bf{c}    &  $\mathbf{f_{gas}}$    &  $\mathbf{M_{vir}[10^{11} M_{\odot}]}$  &  $\mathbf{M_{gas}[10^{9} M_{\odot}]}$  &  $\mathbf{M_{\star}[10^{9} M_{\odot}]}$  &  \bf{Res}  &  $\mathbf{m_{gas} [M_{\odot}]}$  &  $\mathbf{h_{1kpc<r<4kpc} [pc]}$  &  $\mathbf{R_d [kpc]}$  \\
\hline
 ML1 &  100   &  6  &  0.5&  4.27  &  6.7 &  6.7  &  low  &  $6.7 \cdot 10^4$  &  100  &  1.26   \\
 ML2 &  100  &  10  &  0.3  &  3.87  &  4.0  &  9.4  &  low  &  $4 \cdot 10^4$  &  91  &  1.03   \\
 MH3 &  100  &  10  &  0.5  &  3.87  &  6.7  &  6.7  &  high  &  $6.7 \cdot 10^3$  &  50  &  1.03   \\
 ML4 &  100  &  15  &  0.3  &  3.66  &  4.0  &  9.4  &  low  &  $4 \cdot 10^4$  &  74  &  0.86   \\
 ML5 &  150  &  6  &  0.3  &  14.4  &  13.5  &  31.6  &  low  &  $1.4 \cdot 10^5$  &  150  &  1.89   \\
 ML6 &  150  &  6  &  0.5  &  14.4  &  22.6  &  22.6  &  low  &  $2.3 \cdot 10^5$  &  145  &  1.89   \\
 ML7 &  150  &  10  &  0.3  &  13.0  &  13.5  &  31.6  &  low  &  $1.35 \cdot 10^5$  &  128  &  1.55  \\
 MH8 &  150  &  10  &  0.3  &  13.0  &  13.5  &  31.6  &  high  &  $1.35 \cdot 10^4$  &  56  &  1.55   \\
 ML9 &  150  &  10  &  0.5  &  13.0  &  22.6  &  22.6  &  low  &  $2.26 \cdot 10^5$  &  124  &  1.55   \\ 
 MH10 &  150  &  10  &  0.5  &  13.0  &  22.6  &  22.6  &  high  &  $2.26 \cdot 10^4$  &  53  &  1.55   \\
 ML11 &  180  &  6  &  0.5  &  25.0  &  39.0  &  39.0  &  low  &  $3.9 \cdot 10^5$  &  160  &  2.27   \\
\hline
\end{tabular}
\caption{The table shows the main features of individual models, whose names are indicated in Column 1. Column 2: velocity at virial radius; Column 3: concentration; Column 4 gas 
fraction; Column 5: 
virial mass; Column 6: gas mass; Column 7: stellar mass; Column 8: resolution in number of particles (see the text for more details); Column 9: gas mass resolution; Column 10: SPH 
resolution (h is the smoothing length) - we chose to indicate a range of typical values for $h$ between 1 and 4 kpc (distance from the galactic centre of mass), 
since clumps form in this distance range, but note that the minimum SPH resolution (reached in the highest density regions) at the sites of clumps can go down until 10 - 25 pc; Column 11: disc scale lengths.}
\label{table_resolution}
\end{table*}
\subsection[]{Initial Conditions} \label{subsec:IC}
\begin{figure*}
 \includegraphics[width=0.32\textwidth]{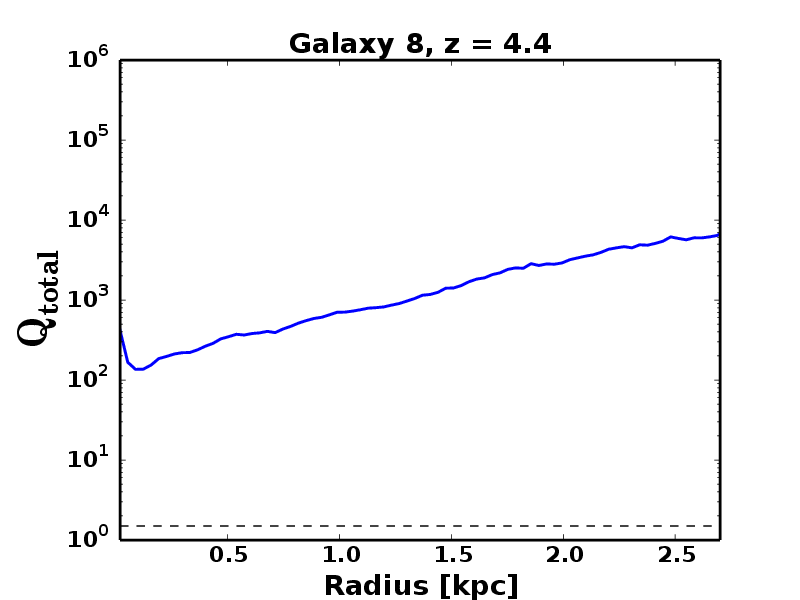}
 \includegraphics[width=0.32\textwidth]{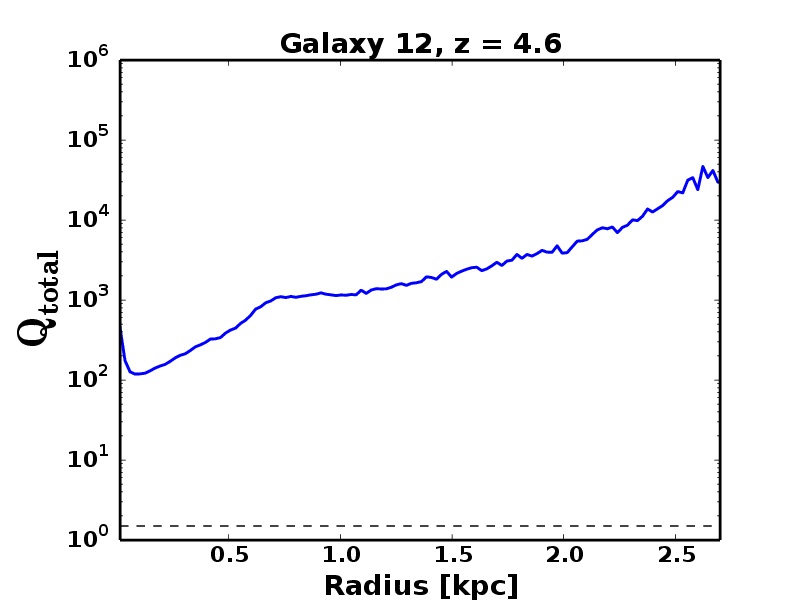}
 \includegraphics[width=0.32\textwidth]{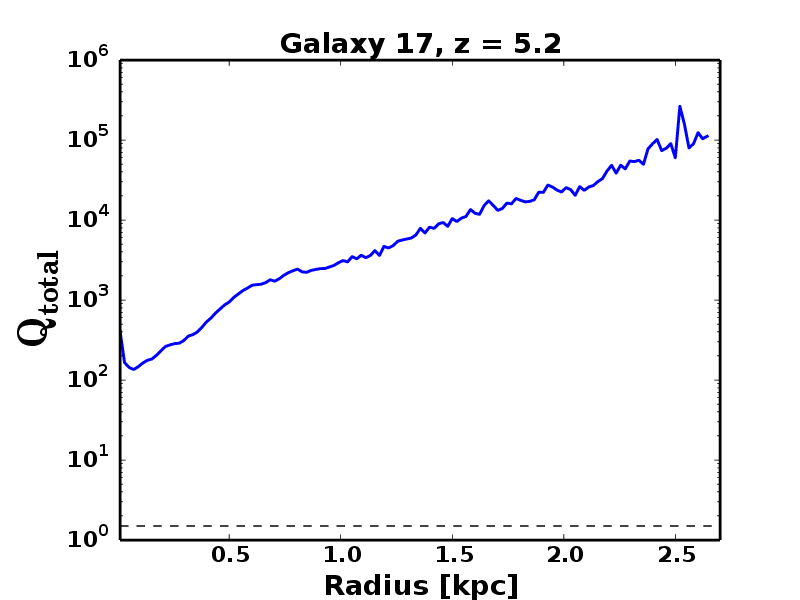}
\caption{Toomre parameter profiles ($Q_{total}$, as defined in the text) for three representative galaxies in the ARGO simulation. Note that $Q_{total}$ is well above of the critical value 1.5 (horizontal dashed line), meaning that selected galaxies are stable, as we explain in Section \ref{subsec:IC}.}
\label{fig:Q_toomre_ARGO}
\end{figure*}

\begin{figure*}
\includegraphics[width=.49\textwidth]{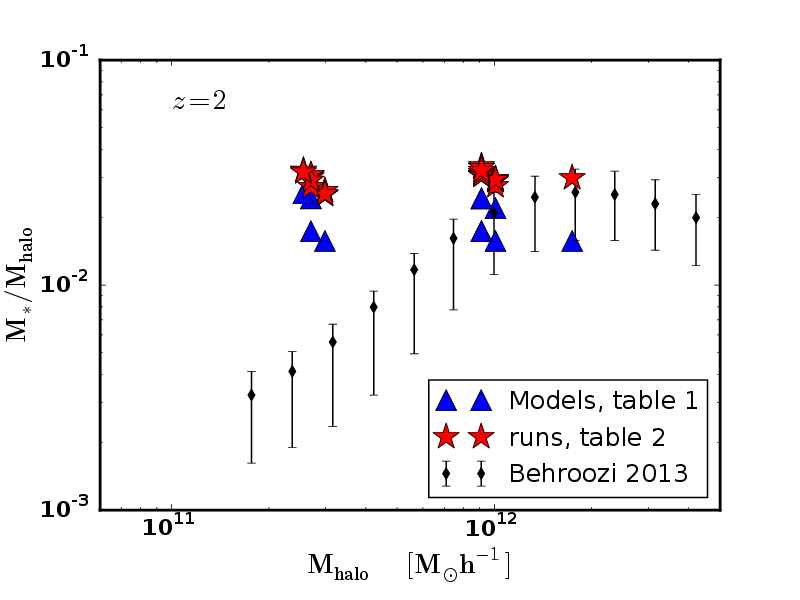}
\includegraphics[width=.49\textwidth]{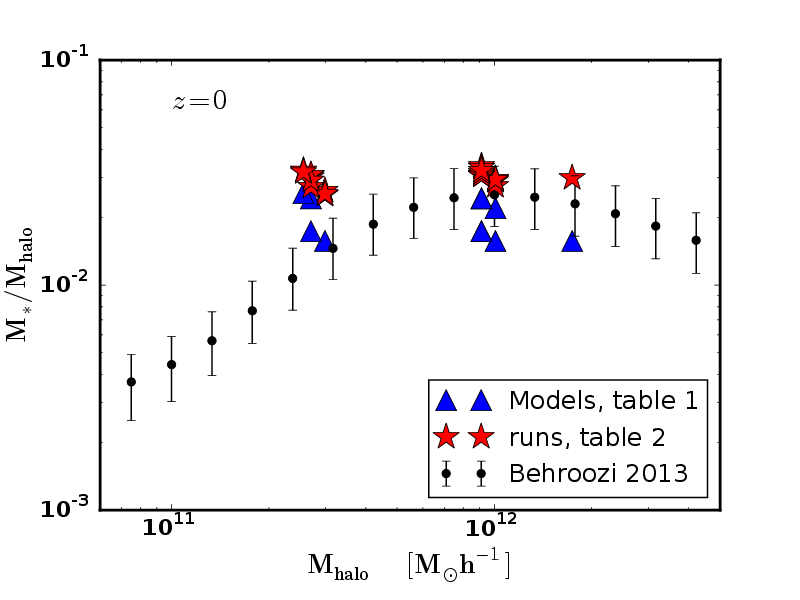}
   \caption{The stellar mass to halo mass ratio as a function of halo mass for the galaxy models: red stars are used for the initial conditions (described in detail in Table \ref{table_resolution}), blue triangles depict our galaxies at the end of the simulations (see Table \ref{table_runs}), and black dots show the abundance matching predictions from \citet{B13} at redshift $z = 2$ (left panel) and $z = 0$ (right panel). Note that we decide to show predictions at different redshift since, while our simulations are supposed to model the evolutionary phase at $z \sim 2$, they are not cosmological. Note that all the massive models (i.e. ML5, ML6, ML7, MH8, ML9, MH10, ML11) are in good agreement with the predictions and are those producing the most relevant results concerning clump formation. The most massive one, ML11, is still in agreement, even if it is on the border of the relation (see Table \ref{table_runs}).}
\label{fig:abundance_matching}
\end{figure*}

Galaxy models are built as in \citet{M01b, M02} using the technique originally developed by \citet{H93} (see also \citealt{SW99}). We use a system of units such that $G = 1$, $[M] =2.32 \times 10^5 M_{\odot}$ and $[R] = 1$ kpc. The models comprise a dark  matter halo and an embedded stellar and gaseous disc.
We recall that the aim of this paper is to study systematically the dependence of disc fragmentation by gravitational instability on both sub-grid physics and structural parameters suitable for high-z discs. While a systematic study requires controlled experiments with high resolution, non-cosmological galaxy simulations as those
presented in this paper,  we also want to ensure that such experiments are initialised with models that are as closely connected as possible with the results of cosmological simulations. 
Therefore, in order to set up the initial conditions of our simulations, we analysed 26 galaxies  and used the properties of the most massive (12), best resolved galaxies in the ARGO simulation, a recent state-of-the-art cosmological hydrodynamical simulation, as a template. 
The simulation, that is described extensively in \citet{FM15} and \citet{F15}, follows the formation of a $10^{13} M_{\odot}$ halo in a region of average local over-density. At $z \sim 4$ the halo mass of the central galaxy is $\sim 6 \times 10^{11} / M_{\odot}$, namely close to the knee of the mass function, suggesting that this is a very common type of  host halo environment at that epoch. 
We select the galaxies between $z=5.2$ and $z=3.8$, when they are still relatively isolated at the outskirts of the main host, hence they are not undergoing mergers, nor they are interacting strongly with any companion. It is remarkable that none of the galaxies in ARGO, including the ones selected, is observed to fragment into giant clumps by gravitational instability at any time, consistent with what was found also in the ERIS simulations \citep{GC11}.\\
It is well known that galaxies formed in cosmological simulations have a tendency to overproduce the stellar mass relative to the prediction of abundance matching \citep{B13}, especially at high redshift \citep{M14}, pointing to a missing aspect of the ISM thermodynamics. ARGO is no exception, although some of its galaxies, in particular the central one, are marginally consistent with such the abundance matching curve \citep{F15}.
At any rate, in most galaxies gas is converted into stars somewhat faster than it should, hence it is conceivable that the Toomre instability of the gas disc might be weakened relative to what we would see with more realistic, lower star formation efficiencies. In addition, star formation tends to be particularly high in the centre, leading to
an early build up of a central bulge, often via bar-driven gas inflow \citep{F15}, which also goes in the direction 
of stabilising the galaxy disc, both stellar and gaseous, by steepening the rotation curve and hence increasing the shear term in the Toomre parameter (i.e. $k$, the epicyclic frequency).\\
All these effects together could conspire to stabilise the galaxies in ARGO, explaining why no clear examples of fragmentation were found (see also Figure \ref{fig:Q_toomre_ARGO}). 
Therefore, in order to build our galaxy models, we follow a two-steps procedure. First, we construct  galaxies that have structural properties similar to those of the galaxies in ARGO, but make them consistent with the abundance matching technique, by proportionally raising the gas fraction and reducing their stellar mass, see Figure \ref{fig:abundance_matching}.
A second set of models comprises galaxies with higher virial mass, having rotational velocities more similar to those studied in previous simulation works (\citealt{BE10} and \citealt{C10}).
Other parameters, such as galaxy concentration, are also varied. The parameters of the initial models are reported in Table \ref{table_resolution} and Table \ref{table_runs}.
\begin{table*}
\vspace {1 mm}
\centering
\begin{tabular}{|c|c|c|c|c|c|c|c|c|c|c|c|c|}
\hline
 \bf{Model} &  \bf{run} &  \bf{Vel [km/s]}    & \bf{c}    &  $\mathbf{f_{gas}}$    &  \bf{FB}    &  \bf{MC}    &  \bf{MTD}    &  \bf{GDSPH}    &  \bf{Res}    &  $\mathbf{\epsilon [pc]}$    &  \bf{Clumps}    &  $\mathbf{T_c [Gyr]}$\\
\hline
 \rowcolor[cmyk]{1, 0, 0, 0} ML1 &  run1 &  100    & 6    &  0.5    &  yes    &  no    &  no    &  no    &  low    &  100    &  yes    &  1.0\\
 ML1 &  run2 &  100    & 6    &  0.5    &  yes    &  yes    &  no    &  no    &  low    &  100    &  no    &  -\\
 \rowcolor[cmyk]{1, 0, 0, 0} ML1 &  run3 &  100    & 6    &  0.5    &  yes    &  yes    &  yes    &  yes    &  low    &  100    &  yes    &  0.3\\
 ML2 &  run4 &  100    & 10    &  0.3    &  yes    &  no    &  no    &  no    &  low    &  100    &  no    &  -\\
 ML2 &  run5 &  100    & 10    &  0.3    &  no    &  no    &  no    &  no    &  low    &  100    &  no    &  -\\
 ML2 &  run6 &  100    & 10    &  0.3    &  yes    &  yes    &  no    &  no    &  low    &  100    &  no    &  -\\
 MH3 &  run7 &  100    & 10    &  0.5    &  yes    &  no    &  no    &  no    &  high    &  100    &  no    &  -\\
 MH3 &  run8 &  100    & 10    &  0.5    &  yes    &  yes    &  no    &  no    &  high    &  100    &  no    &  -\\ 
 ML4 &  run9 &  100    & 15    &  0.3    &  yes    &  no    &  no    &  no    &  low    &  100    &  no    &  -\\
 ML4 &  run10 &  100    & 15    &  0.3    &  no    &  no    &  no    &  no    &  low    &  100    &  no    &  -\\
 ML4 &  run11 &  100    & 15    &  0.3    &  yes    &  yes    &  no    &  no    &  low    &  100    &  no    &  -\\
 \rowcolor[cmyk]{1, 0, 0, 0} ML5 &  run12 &  150    & 6    &  0.3    &  yes    &  no    &  no    &  no    &  low    &  100    &  yes    &  1.0\\
 \rowcolor[cmyk]{1, 0, 0, 0} ML5 &  run13 &  150    & 6    &  0.3    &  yes    &  yes   &  no    &  no    &  low    &  100    &  yes    &  0.3\\
 \rowcolor[cmyk]{1, 0, 0, 0} ML6 &  run14 &  150    & 6    &  0.5    &  yes    &  no    &  no    &  no    &  low    &  100    &  yes    &  1.0\\
 \rowcolor[cmyk]{1, 0, 0, 0} ML6 &  run15 &  150    & 6    &  0.5    &  yes    &  yes   &  no    &  no    &  low    &  100    &  yes    &  1.0\\
 \rowcolor[cmyk]{1, 0, 0, 0} ML7 &  run16 &  150    & 10    &  0.3    &  yes    &  no    &  no    &  no    &  low    &  100    &  yes    &  0.5\\
 \rowcolor[cmyk]{1, 0, 0, 0} ML7 &  run17 &  150    & 10    &  0.3    &  no    &  no    &  no    &  no    &  low    &  100    &  yes    &  1.0\\
 ML7 &  run18 &  150    & 10    &  0.3    &  yes    &  yes    &  no    &  no    &  low    &  100    &  no    &  -\\
 ML7 &  run19 &  150    & 10    &  0.3    &  yes    &  yes    &  yes    &  no    &  low    &  100    &  no    &  -\\
 \rowcolor[cmyk]{1, 0, 0, 0} ML7 &  run20 &  150    & 10    &  0.3    &  yes    &  no    &  no    &  yes    &  low    &  100    &  yes    &  0.4\\
 \rowcolor[cmyk]{1, 0, 0, 0} ML7 &  run21 &  150    & 10    &  0.3    &  yes    &  yes    &  yes    &  yes    &  low    &  100    &  yes    &  0.1\\
 MH8 &  run22 &  150    & 10    &  0.3    &  yes    &  no    &  no    &  no    &  high    &  100    &  no    &  -\\
 MH8 &  run23 &  150    & 10    &  0.3    &  yes    &  yes    &  no    &  no    &  high    &  100    &  no    &  -\\
 \rowcolor[cmyk]{1, 0, 0, 0} ML9 &  run24 &  150    & 10    &  0.5    &  yes    &  no    &  no    &  no    &  low    &  100    &  yes    &  1.0\\
 \rowcolor[cmyk]{1, 0, 0, 0} MH10 &  run25 &  150    & 10    &  0.5    &  yes    &  no    &  no    &  no    &  high    &  100    &  yes    &  1.0\\
 \rowcolor[cmyk]{1, 0, 0, 0} MH10 &  run26 &  150    & 10    &  0.5    &  yes    &  yes    &  no    &  no    &  high    &  50    &  yes    &  --\\
 \rowcolor[cmyk]{1, 0, 0, 0} ML11 &  run27 &  180    & 6    &  0.5    &  yes    &  no    &  no    &  no    &  low    &  100    &  yes    &  1.0\\
 \rowcolor[cmyk]{1, 0, 0, 0} ML11 &  run28 &  180    & 6    &  0.5    &  no    &  no    &  no    &  no    &  low    &  100    &  yes    &  1.0\\
\hline
\end{tabular}
\caption{Scheme of the Simulations: columns 2-4 show again virial velocity, concentration and gas fraction of each models, for easier comparisons. Columns 6-9 indicate what kind of physics is turn on in the run considered: feedback (FB), Metal Cooling (MC), Metal Thermal Diffusion (MTD) or Geometric Density SPH (GDSPH, \citealt{K14}); column 10: resolution used in each run; column 11: value of the softening; column 12: whether in the selected run there is clump formation or not; column 13: duration of the clumpy phase in Gigayears (taking into account only the first Gyr after the relaxation phase).}
\label{table_runs}
\end{table*}
\begin{figure*}
 \includegraphics[width=0.28\textwidth]{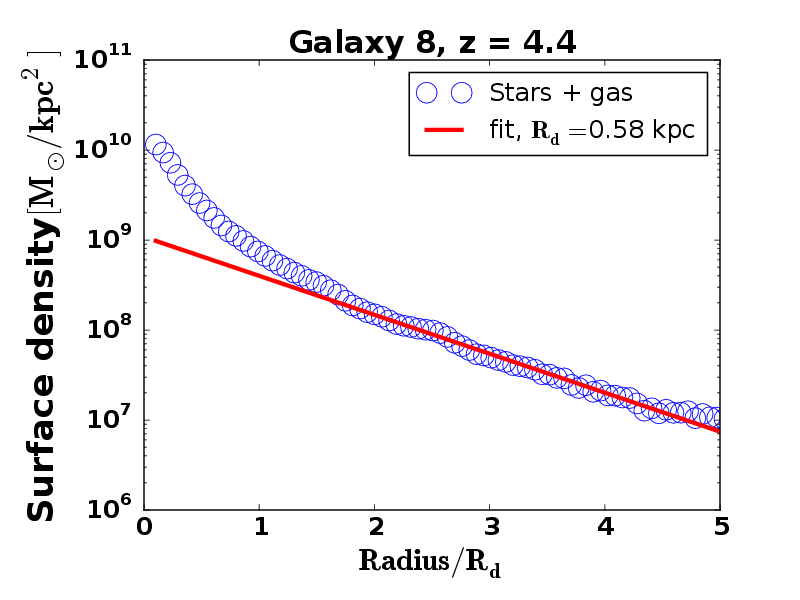}
 \includegraphics[width=0.28\textwidth]{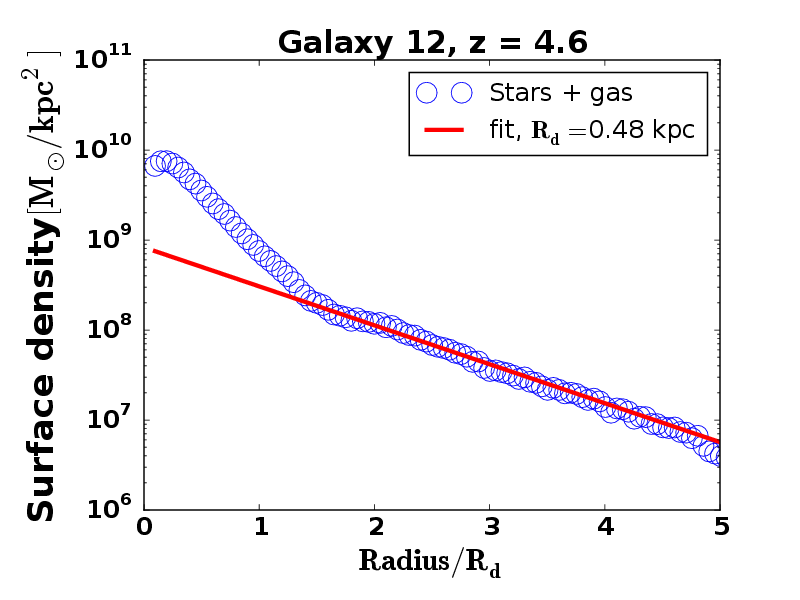}
 \includegraphics[width=0.28\textwidth]{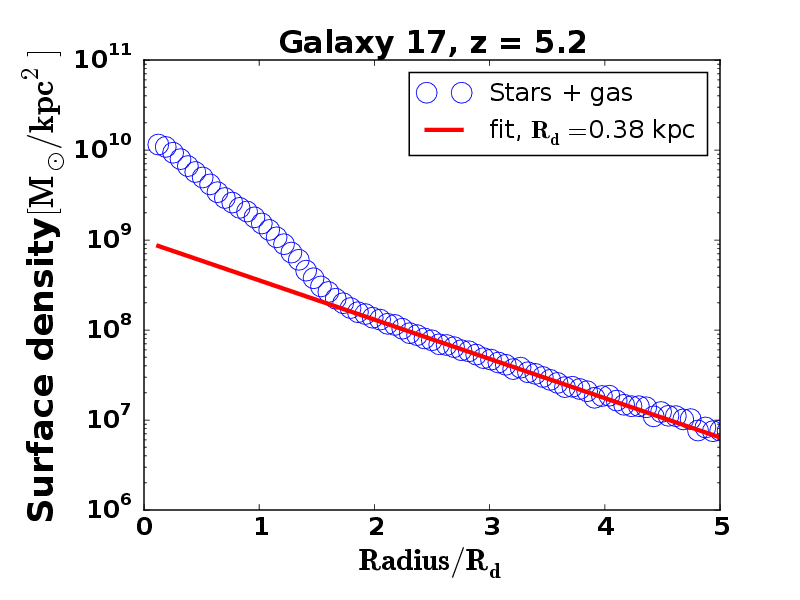}
\caption{The disc surface density profile (stars and gas) for three galaxies selected in the ARGO simulation (blue empty circles), including a fit with an exponential law (red solid line). On the x-axis we show the radius in units of the disc scale length ($R_d$) calculated for each galaxy, as shown in the legend.}
  \label{fig:surface_cosmo}
\end{figure*}
The models are built using the Hernquist technique (\citealt{H93}), which allows to construct multi-component galaxy models close to equilibrium.\\
More specifically, we analyse our cosmological galaxy sample in order to determine the most likely value of the exponential disc scale length, peak rotational velocity and gas fraction (see Figure \ref{fig:IC} for this last value). Our models have exponential baryonic surface density profiles (as described in \citet{MMW98}) in which the gas and stars have the same scale length $R_d$:
\begin{equation}
\Sigma(R) = \Sigma_0 exp\left(- R/R_d\right)
\end{equation}
where $\Sigma_0$ is the central surface density. We fit the stellar surface density profiles of the galaxies in ARGO simulation with an exponential fit, finding the scale length $R_d$. Note that most galaxies in ARGO, and certainly all the most massive ones, do not have a simple exponential profile (see Figure \ref{fig:surface_cosmo}) since bulge growth due to a variety of processes has already begun (\citealt{F15}). 
Nevertheless, we do not include a bulge component in the models, since this will tend to stabilise them. In addition, as we will discuss below, the surface density profile will become steeper, essentially producing naturally a small bulge, simply as a result of relaxation (see Figure \ref{fig:surface_density_IC} and Section \ref{Bulge}).\\
We measure the gas fraction inside a radius equal to $3 \times R_d$. The gas fraction (here after $f_{gas}$) is defined as:
\begin{equation}
f_{gas} = \frac{M_{gas}}{M_{gas} + M_{star}}
\end{equation}
We consider only cold gas, with a temperature below $3 \times 10^4$ K. We find that $f_{gas}$ in our sample of selected galaxies is in the range $0.2-0.4$, see Figure \ref{fig:IC}. We thus build the initial set of models using an intermediate value, $f_{gas} = 0.3$, but since observations of clumpy galaxies have found even higher gas fractions to be common, and since cosmological simulations may suffer from excessive star formation efficiency , we also constructed models with $f_{gas} = 0.5$ \citep{T10, T13, DZ14}.
\begin{figure}
   \includegraphics[width=.49\textwidth]{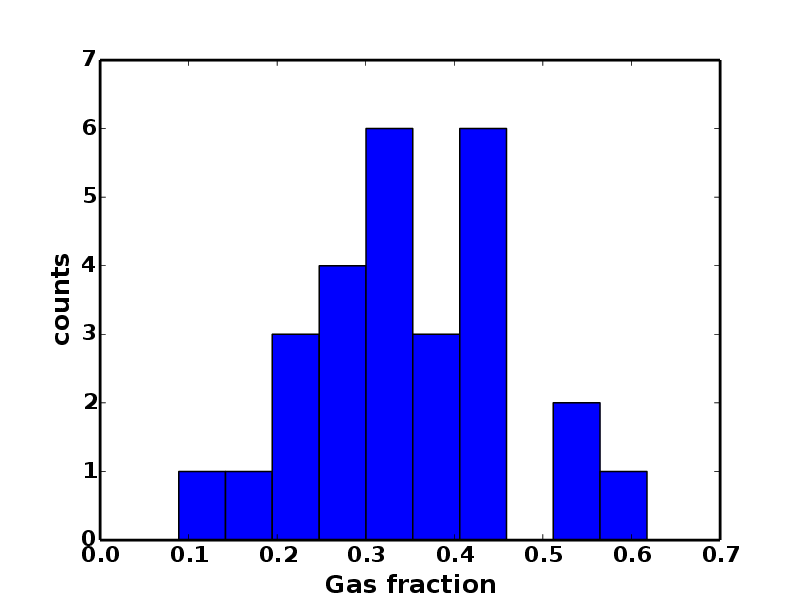}
   \caption{The histogram of the gas fraction, calculated as ratio between $M_{gas}$ and $M_{gas} + M_{stars}$ for all the galaxies selected in the ARGO simulation, showing as typical values 0.2-0.4.}
  \label{fig:IC}
\end{figure}
We caution that the gas fractions measured in the observations are determined using molecular gas rather than total mass of cold gas, while only the latter is accessible in the simulations. We will return on this point when we report our results in Section \ref{sec:structural_param}.\\
For our model set up we also need to define the velocity at the virial radius, $V_{vir}$, which, for an assumed cosmology (ours assumes $\Omega_0 = 0.27$, $\Omega_{\Lambda} = 0.73$ and $H_0 = 73$ km/s/Mpc as WMAP9), automatically determines the virial mass, $M_{vir}$, and the virial radius, $R_{vir}$, of the halo. We chose to use as halo virial velocity a value in the range between $100$ and $150$ km/s, which corresponds to the asymptotic total velocity of the most massive galaxies in the ARGO simulation, see Figure 4 in \citet{F15}.\\
For the halo we choose NFW density profiles \citep{NFW95, N97} with different halo concentrations, $c$, and spin parameter, $\lambda$.
The concentration is defined as $c = R_{vir}/r_s$, where $r_s$ is the halo scale radius. Basically the value of the concentration $c$ defines what fraction of the total mass of the halo is contained within its inner regions, where the baryonic disc lies. For our model we chose three values for the halo concentration, $c = 6, 10$ and $15$. 
For galaxies in the virial mass range $3.6 \times 10^{11}- 2.5 \times 10^{12} M_{\odot}$ considered here, this range of concentrations is indeed plausible (\citealt{Bullock01, Ludlow14}), but it is important to stress that for the most massive galaxies in our sample, which are closer in mass to those considered in previous simulation works, the choice $c = 6$ is most representative and should be considered as the reference case. 
Note that we do not measure the halo concentration in the cosmological galaxies, because baryons inevitably affect 
the halo potential, especially for the most massive galaxies \citep{F15}, making it unfeasible to determine the original concentration.\\
Finally the spin parameter is defined as $\lambda = J \mid E \mid ^{1/2} G^{-1} M_{vir}^{-5/2}$, where $J$ and $E$ are, respectively, the total angular momentum and the total energy of the halo. Also in this case, rather than measuring directly the spin parameter, we set it to a value that allows us to recover a range of disc scale lengths  similar to that of the original cosmological galaxies at $z \sim 3$ (we use $\lambda$ in the range 0.02-0.04). The disc scale lengths of the models are shown in Table \ref{table_resolution}.\\
\\
Following the procedure outlined above, we construct 11 disc models, varying virial velocity, concentration, gas fraction and resolution (see Table \ref{table_resolution}).
The spin parameter is adjusted in each model in order to obtain a sensible disc scale length, because the disc scale length depends directly or indirectly on the other structural parameters, see \citet{MMW98}. Three models have a mass resolution $6$ times higher, but the same gravitational softening as in the other models ($100$ pc). One of the high resolution models has been run with a softening reduced by a factor of 2 to explore the effect of increased spatial resolution (run 26 in Table \ref{table_runs}). Indeed, a vast literature on gravitational instability on protostellar and protoplanetary discs has shown that both mass and spatial resolution do affect the outcome of disc instability (\citealt{MW04, D07}), and similar studies on gaseous bars and spiral instabilities have shown that a smaller gravitational softening allows to better
capture the dynamics of non-axisymmetric modes, leading to stronger bars or spiral modes (e.g. \citealt{MW04,K07}). \\
The disc models are evolved adiabatically for $1$ Gyr, in order to allow the disc to relax and eliminate transient waves in the stellar and gaseous disc, which could artificially enhanced non-axisymmetric instabilities. Indeed we want to approach the instability from a stable state, in order to perform a systematic and controlled study of disc instability. Note that this is still a caution to be taken, even if we are using a pressure floor (see Section \ref{Code}); in fact the pressure floor enforces that density fluctuations, which lead to fragmentation, are well resolved, but does not ensure that numerical noise and transient perturbations in the initial conditions do not amplify those fluctuations to begin with (see \citealt{R14b}).
After 1 Gyr the models are fairly stable, with Toomre parameters well above unity everywhere (Figures \ref{fig:Toomre_low_high} and \ref{fig:Toomre_other_runs}, blue line) as in the ARGO galaxies (these however have even larger Toomre parameters, see Figure \ref{fig:Q_toomre_ARGO}). Note that transient perturbations have sustained a spiral pattern soon after the start of the simulation in the adiabatic phase, leading to some angular momentum transport in the disc and producing a small bulge-like central concentration. This shows up as a central steepening of the rotation curve inside 1 kpc. This is acceptable, since galaxies in the ARGO simulation also develop a moderate bulge component already at $z > 4$, as highlighted by the shape of the rotation curves in Figure 4 of \citet{F15} and by the stellar surface density maps in Figures 2 and 3 of \citet{F15}.\\
After the adiabatic relaxation phase, we turn on radiative cooling and run the models for 2 Gyr more, with varying sub-grid physics, namely switching on and off blast-wave supernovae feedback, metal-line cooling, thermal and metal diffusion, and using even the GDSPH in selected cases (see Table \ref{table_runs} for more details). Each run lasts 3 Gyr, but since the first Gyr covers only the initial relaxation phase, for our purpose we reset the clock to $t=0$ at the end of the relaxation phase, hence in the plots and Tables time will label only the disc evolution phase following the initial relaxation.
Note that, while the isolated disc simulations are not cosmological, assuming they are initialised to resemble galaxies at $z \sim 3$, the 1 Gyr time spanned by the simulations can be interpreted as evolving them to $z \sim 1.5-2$, which corresponds quite well to the typical redshift range of clumpy galaxies.\\
In Figures \ref{fig:velocities_IC} and \ref{fig:surface_density_IC} there are, respectively, the rotation curves and surface densities of all our simulated galaxies at the initial condition and after the relaxation time, 1 Gyr.\\
\begin{figure*}
\flushleft
 \includegraphics[width=0.33\textwidth]{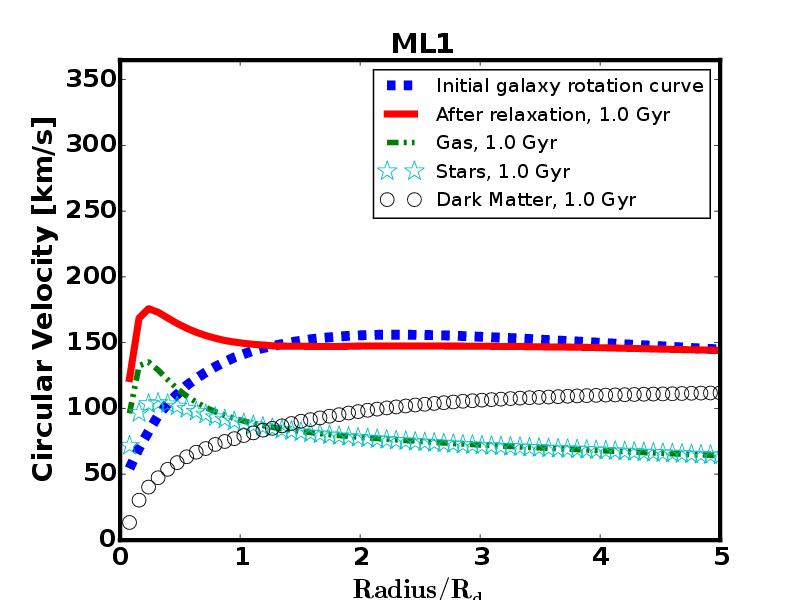}
 \includegraphics[width=0.33\textwidth]{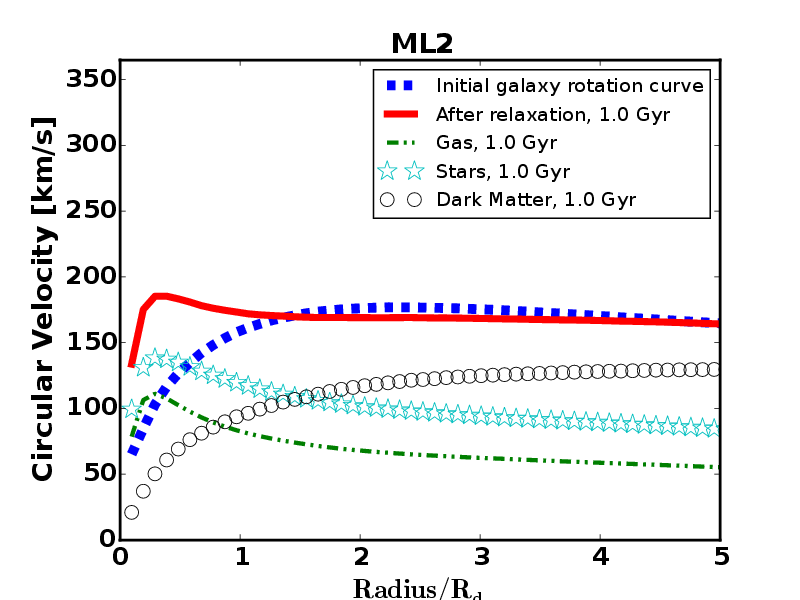}
\includegraphics[width=0.33\textwidth]{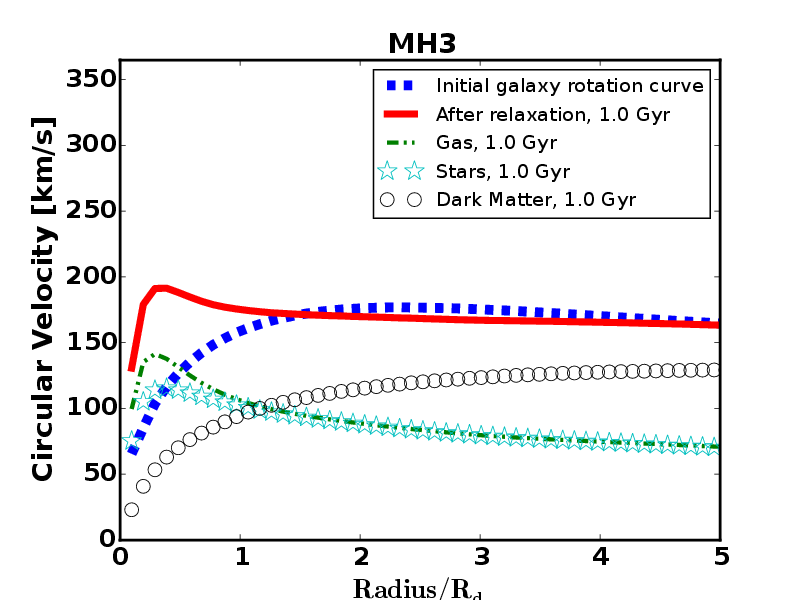}\\
 \includegraphics[width=0.33\textwidth]{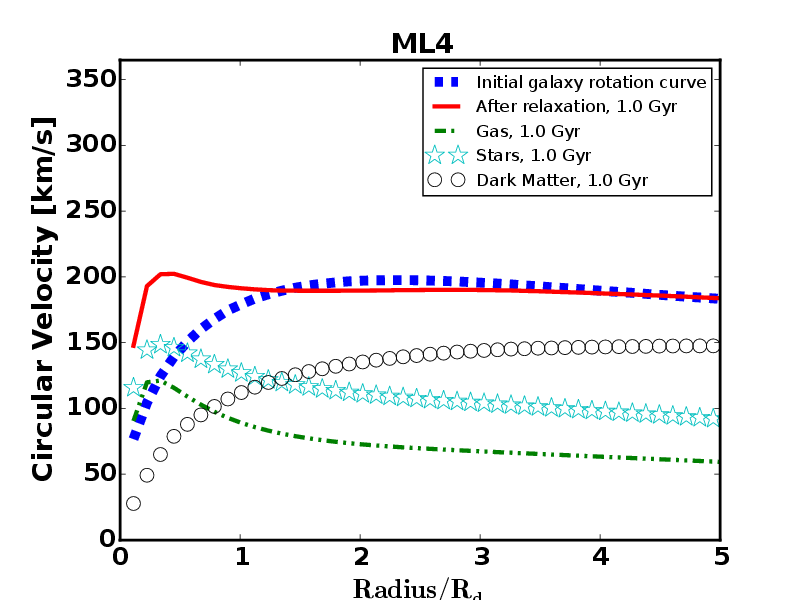}
 \includegraphics[width=0.33\textwidth]{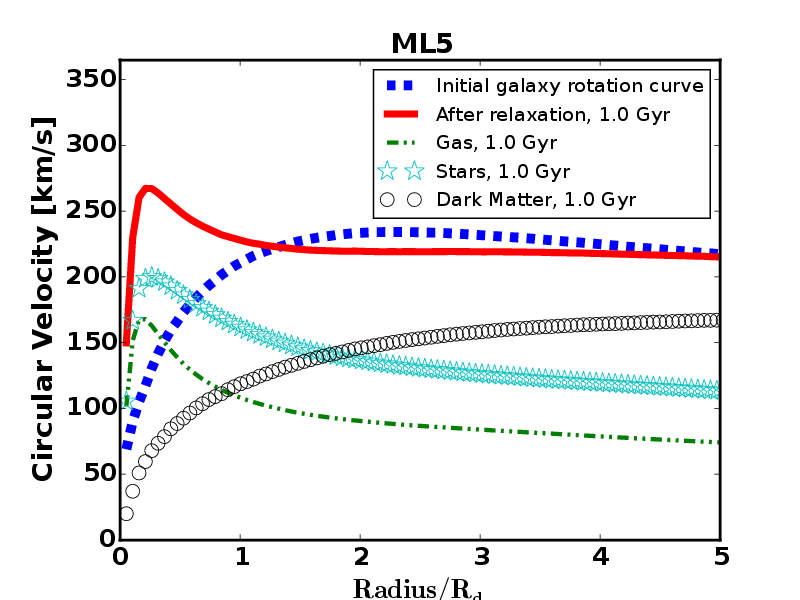}
 \includegraphics[width=0.33\textwidth]{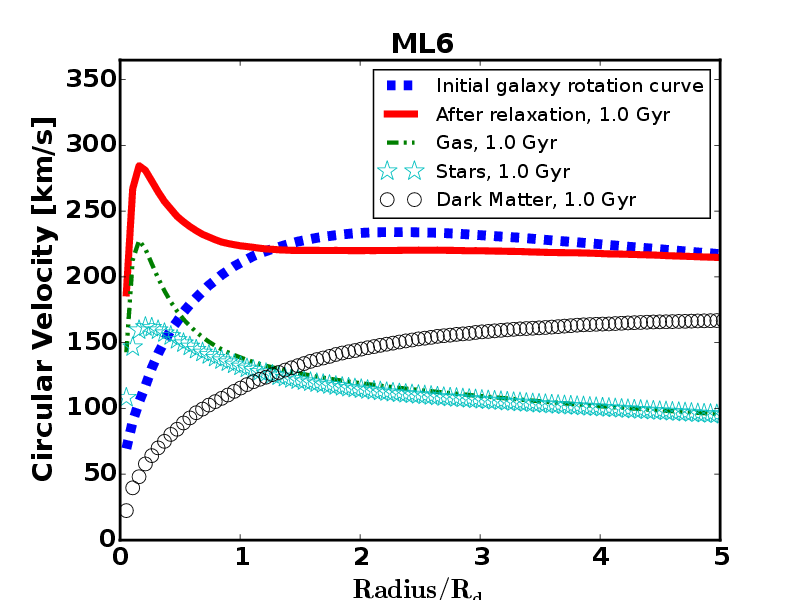}\\
 \includegraphics[width=0.33\textwidth]{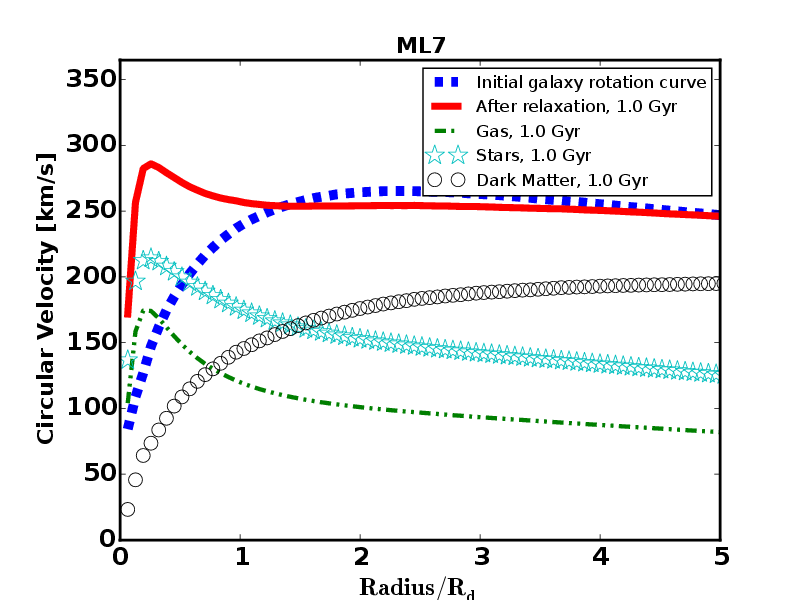} 
 \includegraphics[width=0.33\textwidth]{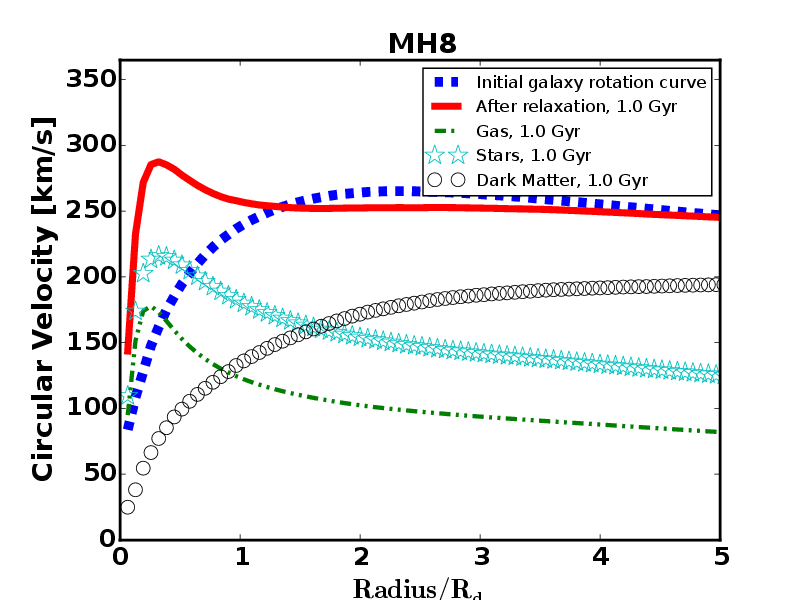}
 \includegraphics[width=0.33\textwidth]{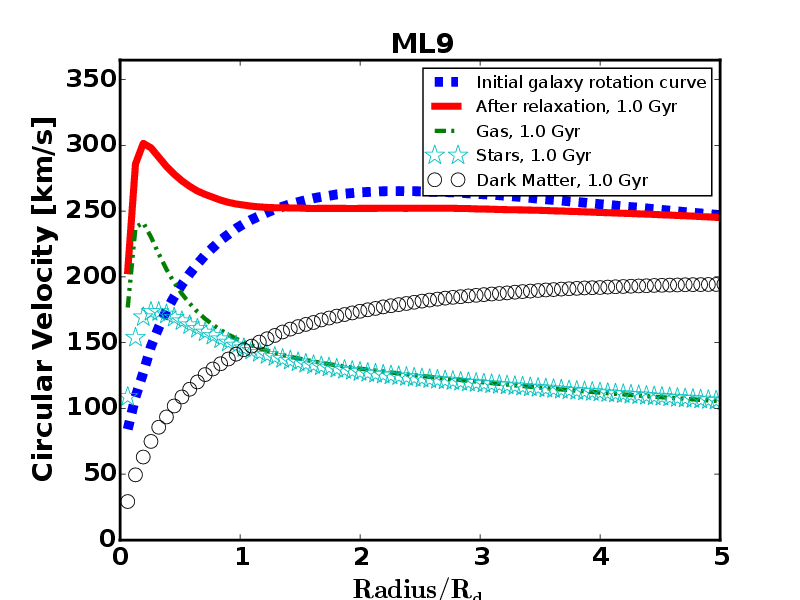}\\
 \includegraphics[width=0.33\textwidth]{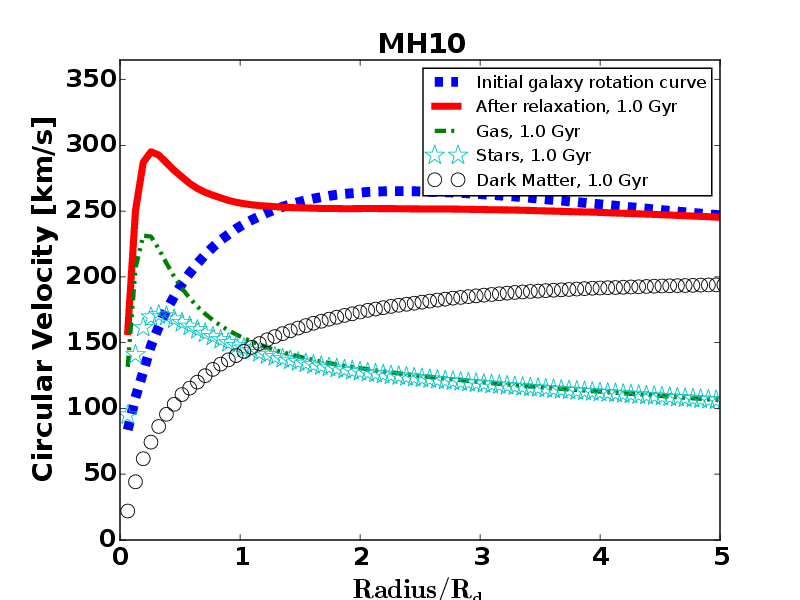}
 \includegraphics[width=0.33\textwidth]{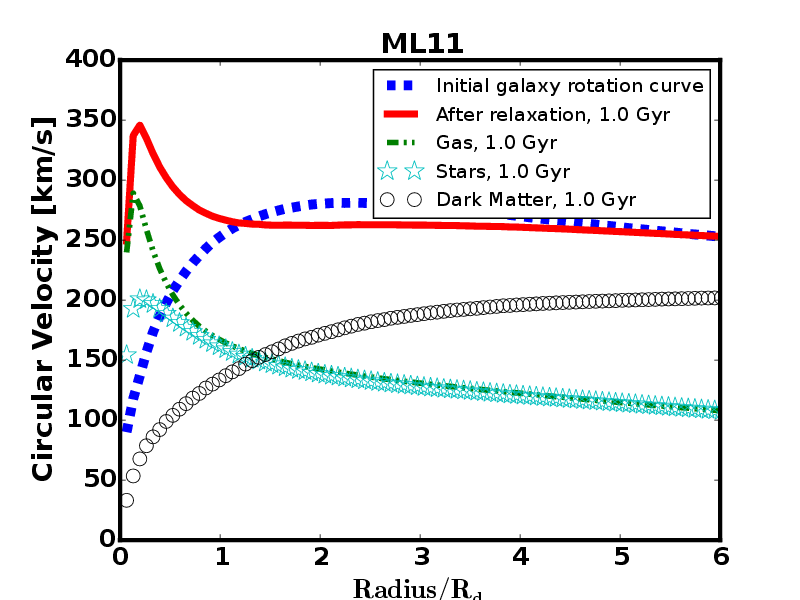}
\caption{\textbf{The rotation curves for all our galaxy models, before (dashed blue line) and after the relaxation phase , $t_0 = 1 $ Gyr. The last one (red solid line) is divided into its individual components, in particular gas, stars and dark matter are showed by green dash dot line, light blue stars and empty circle respectively, as explained in the legend.}}
\label{fig:velocities_IC}
\end{figure*}
\begin{figure*}
\flushleft
 \includegraphics[width=0.33\textwidth]{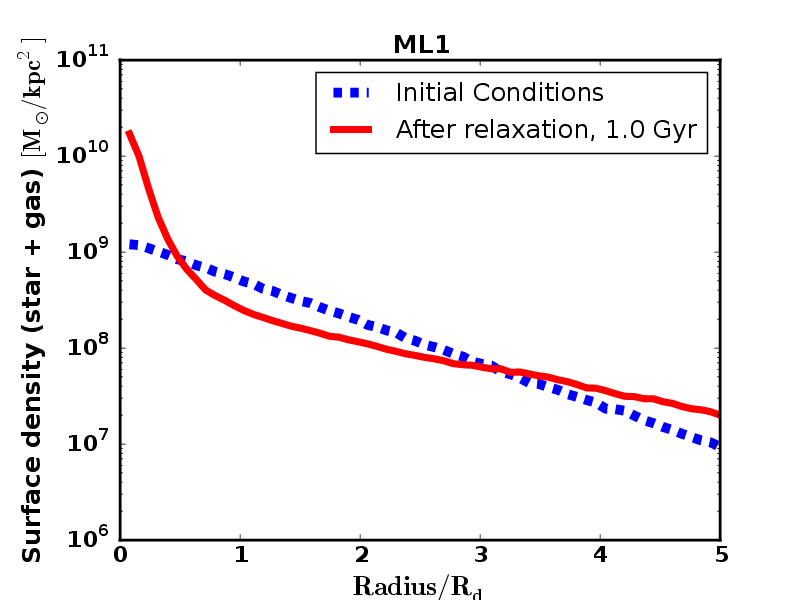}
 \includegraphics[width=0.33\textwidth]{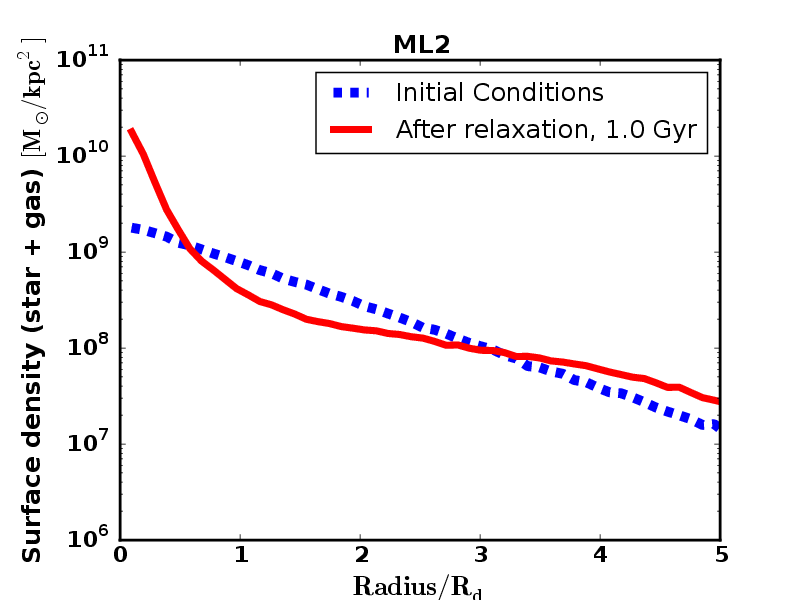}
 \includegraphics[width=0.33\textwidth]{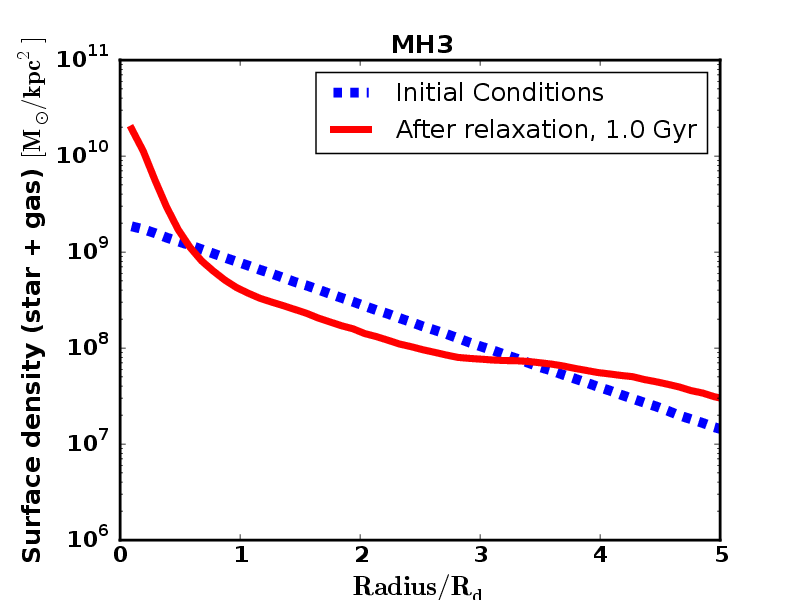}\\
 \includegraphics[width=0.33\textwidth]{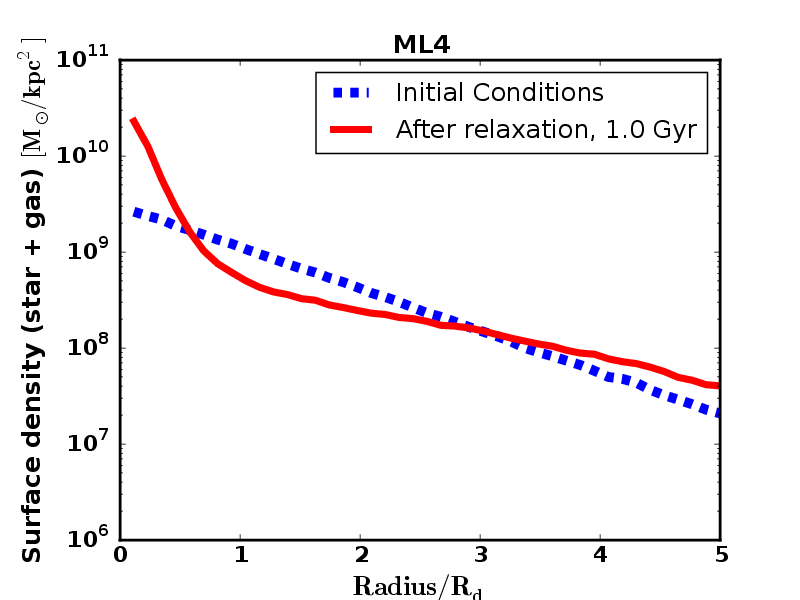}
 \includegraphics[width=0.33\textwidth]{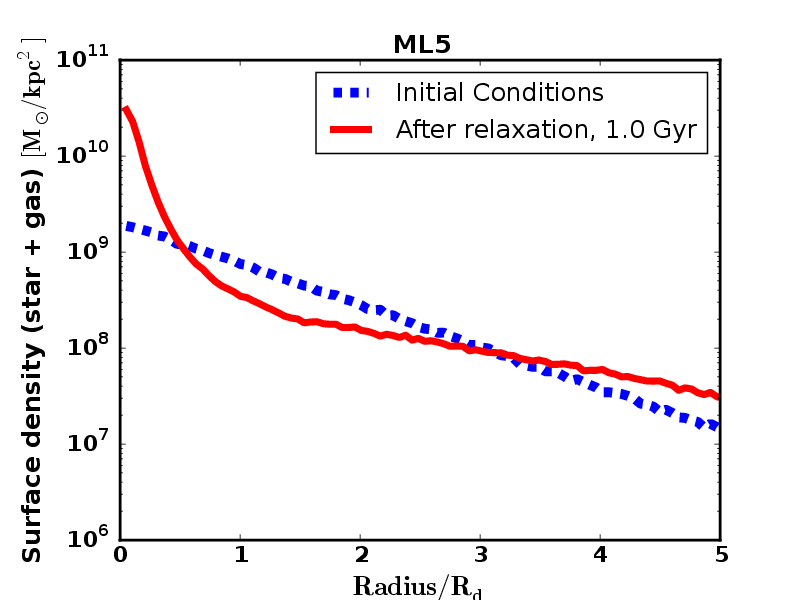}
 \includegraphics[width=0.33\textwidth]{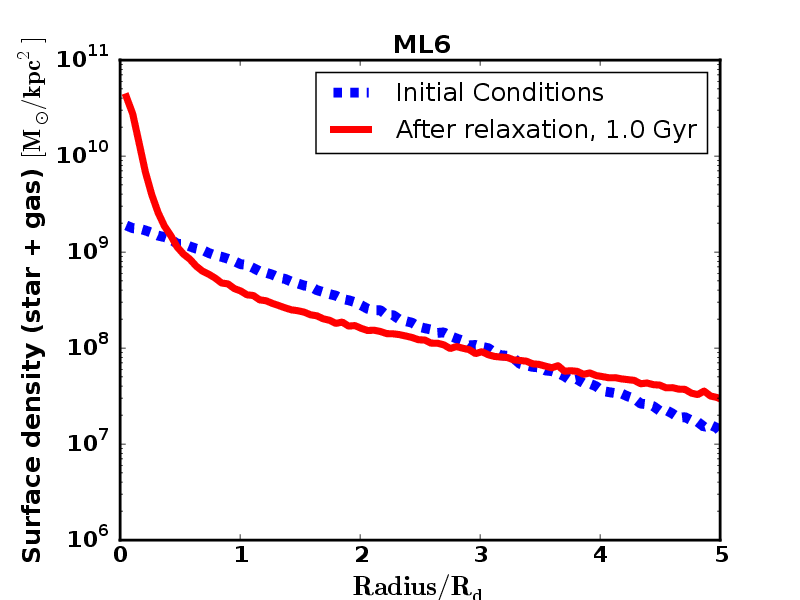}\\
 \includegraphics[width=0.33\textwidth]{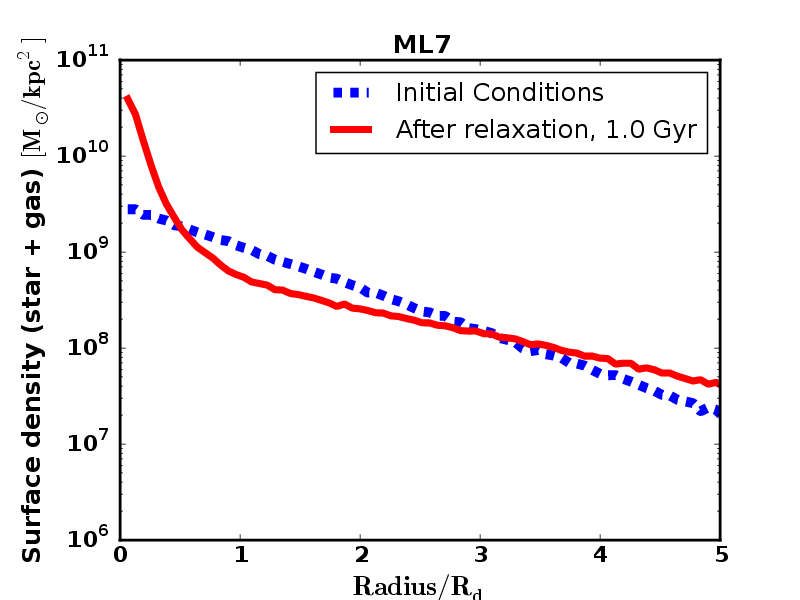} 
 \includegraphics[width=0.33\textwidth]{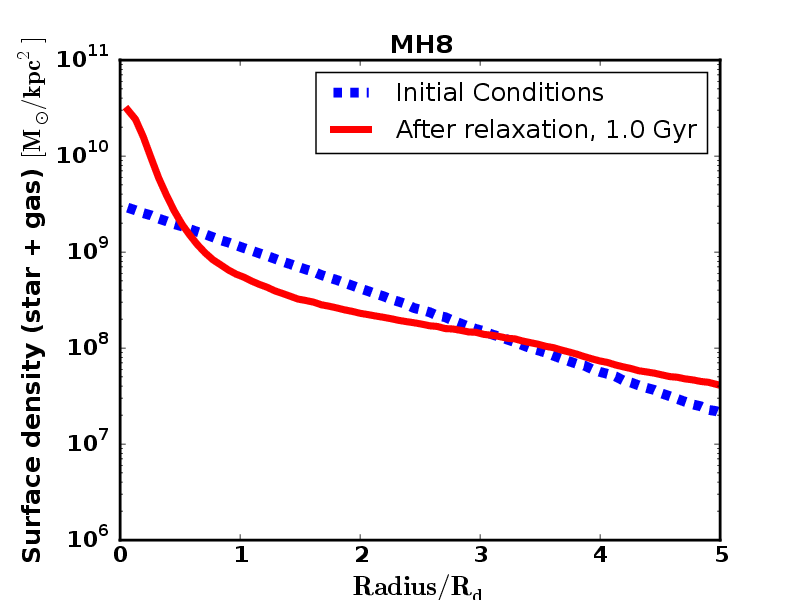}
 \includegraphics[width=0.33\textwidth]{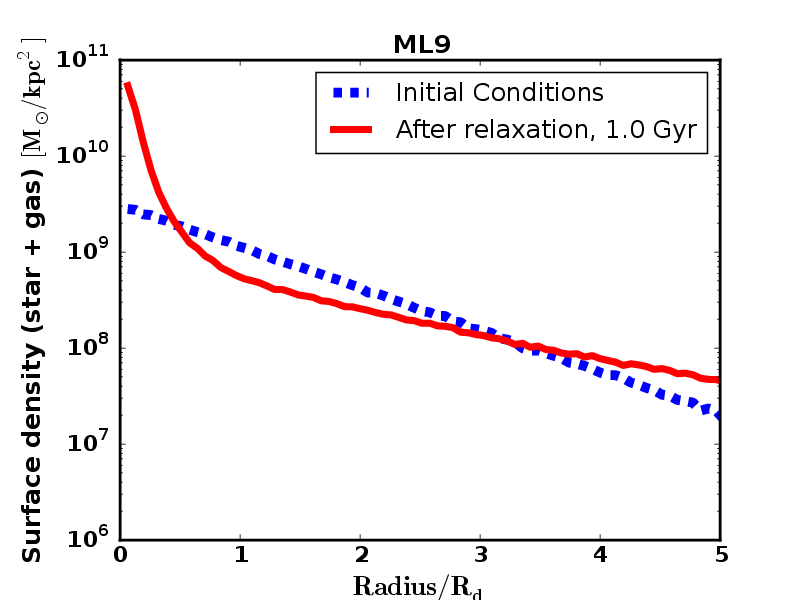}\\
 \includegraphics[width=0.33\textwidth]{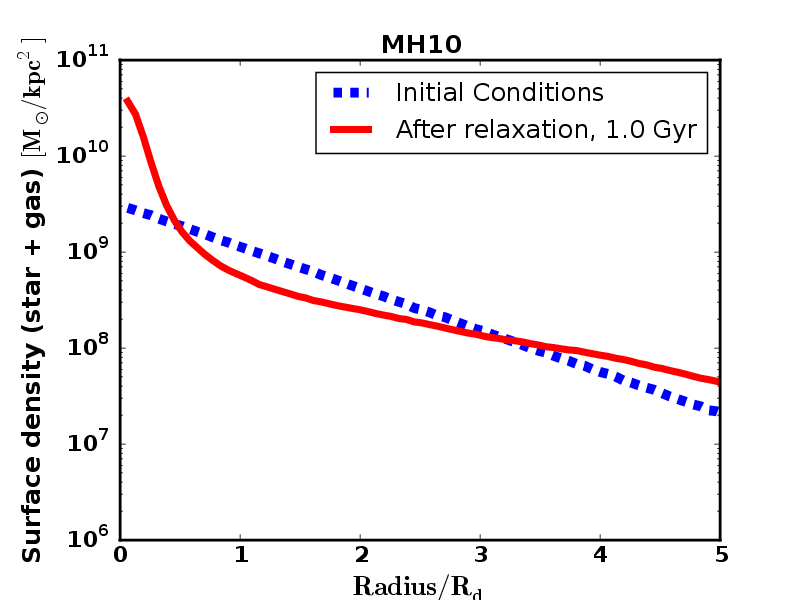}
  \includegraphics[width=0.33\textwidth]{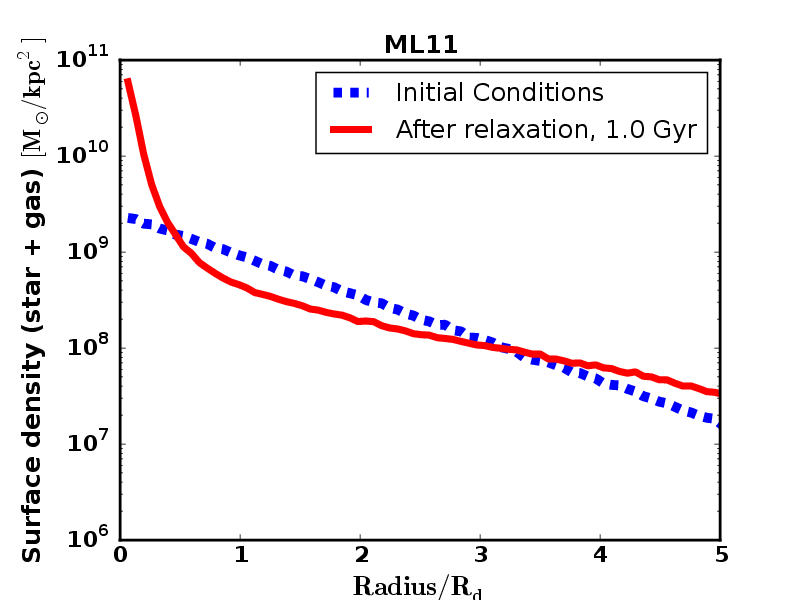}
\caption{Surface density profiles of the discs (including both stars and gas) for all our galaxy models. Surface densities for the initial conditions (blue dashed lines) and after the relaxation phase ($t_0 = 1 $ Gyr, red solid lines) are shown.}
\label{fig:surface_density_IC}
\end{figure*}
\section{RESULTS}\label{Results}
\begin{figure}
  \includegraphics[width=0.49\textwidth]{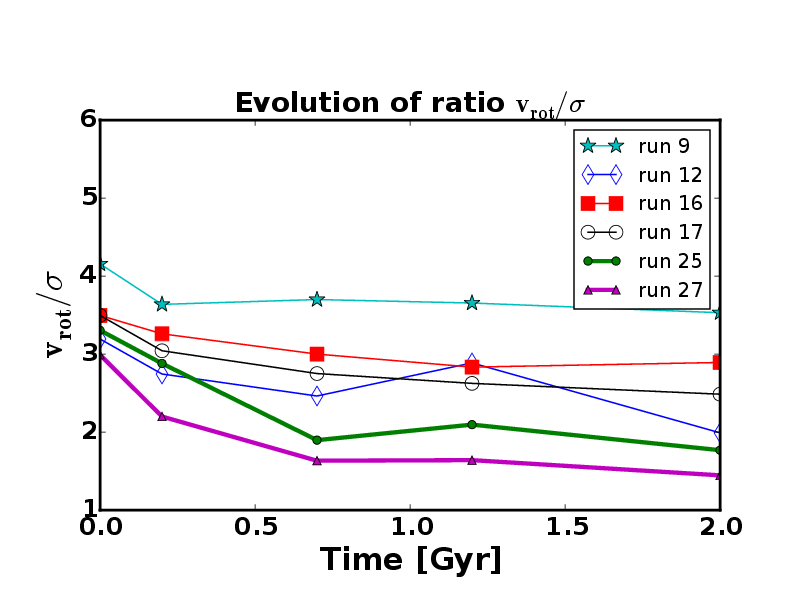}
  \caption{The evolution of the ratio $v_{rot}/ \sigma$ for some representative runs. We start from $t_0 = 0$, that corresponds to the end of the relaxation phase, when we switch on radiative cooling and other sub-grid physics, according to Table \ref{table_runs}, then we show the state at 0.2 and 0.7 Gyr, corresponding to the phase of clump formation, and finally at 1.2 and 2.0 Gyr, which signal the end of the clumpy phase, except for runs 12, 17, 25, 27 that continue fragmenting. Run 9 does not fragment and shows a ratio higher than the other cases, as expected.}
  \label{fig:ratio_v_s}
\end{figure}

As said before, isolated galaxy simulations allow us to reach higher resolution, compared to cosmological simulations, and to explore systematically the effect of varying sub-grid physics and structural parameters.\\
Table \ref{table_runs} summarizes the outcome of our large suite of simulations. Inspection of 
Table \ref{table_runs} highlights that the formation of long-lasting gravitationally bound gaseous clumps, due to gravitational instability, is a common occurrence for  massive galaxies (with virial velocities $150-180$ km/s), especially with low concentration and high gas fraction. 
This is consistent with previous simulations of clumpy galaxies (e.g. \citealt{C10, M14}). 
However, when we consider the complete sample of simulations, which includes lower mass galaxies, only half of the simulations show fragmentation into clumps, and of these only ten exhibit clumps until 1 Gyr of evolution (note, however, that the short lifetimes of clumps are in principle consistent with the young ages typically inferred for clumps in observed hi-z galaxies \citep{Adamo13}).
Long lived fragmentation lasting 1 Gyr occurs either in unrealistic setups, such as when feedback is not included in the simulation (runs 17 and 28), or in galaxy models with most vigorous instability because of largest disc mass, highest gas fractions and/or lowest halo concentration (ML6 and ML11). 
Note that model ML11, which has $V_{vir} = 180$ km/s and a disc mass $7.8 \times 10^10$ $M_{\odot}$, has a maximum circular velocity of $\sim 350 km/s$, see Figure \ref{fig:velocities_IC}, which is close to the tail of highest rotational velocities in the large high-z KMOS sample of \citealt{Wisnioski15} (see figure 7 in the quoted paper for comparison: our peak rotational velocity for the gas in ML11 is $\sim 300 $ km/s, while at redshift $1-2$ the typical value is $\sim 200$ km/s). \\
Finally, as we show in section \ref{Clumps} the typical mass and size of clumps in our sample is sensibly smaller than the
masses often reported in the literature of numerical simulations, reflecting both the wide range of galaxy masses
and structural properties considered and the different treatment of sub-grid physics.\\
We proceed now to have a closer look at the nature of fragmentation and the properties of clumps in our simulations. 
We recall that $t=0$ corresponds to the end of the 1 Gyr-long relaxation phase in our simulations, so that we study 
only the evolution past this point.
Fragmentation, when it happens, occurs along dense spiral arms, forming at a distance from the galactic centre usually comprised between $2$ and $5$ kpc, after roughly one orbital time (i.e. $\sim 100-150$ Myr), since this is the region that reaches the lowest Toomre parameters (see Figure e.g. \ref{fig:Toomre_low_high}). The exact value of the Toomre parameter near fragmentation depends on the way the Toomre parameter is defined and measured, which is not trivial in our case, since we are dealing with global instability in a thick, non-laminar, non-isothermal disc, hence with all the assumptions of linear Toomre analysis breaking down. We have measured the Toomre parameter in different ways, using both the two-fluid and one-fluid definition (see Section \ref{Intro}), finding that the azimuthally averaged value of the parameter for the cold gas phase only (single fluid approximation) yields a result well in line with the consensus reached in the vast literature on protostellar and protoplanetary gaseous discs, using a variety of numerical hydrodynamical techniques, namely that fragmentation in 3D gaseous discs requires a minimum Toomre parameter $Q< 1.4-1.5$ (e.g. \citealt{D07}). Despite this, for completeness we have preferred to use the two-fluid Toomre parameter (see Equation \ref{eq:Q_total}), hence including both gas and stars in our analysis. The fragmentation will then occur if the local cooling time is not much longer than the local orbital time, a condition that is always satisfied in galactic discs (but not necessarily in astrophysical discs at smaller scales, see \citep{D07}).\\
In our simulations, the fragmentation phase lasts from few orbital times up to 1 Gyr. Clumps, which we identify as gravitationally bound gaseous objects (see Section \ref{Clumps}) form with a diameter between $150$ and $400$ pc and gas masses typically comprised between $10^7 M_{\odot}$ and $10^8 M_{\odot}$ (Figure \ref{fig:hist_clumps_evolution}), undergoing rapid star formation in the first 100-200 Myr (Figure \ref{fig:SFR_FB_vs_MC}). They can then reach higher total masses (including stars and gas), up to $\sim 10^9 M_{\odot}$ as they accrete mass and merge with other clumps. The initial masses and evolution of the mass function is thoroughly addressed in section \ref{Clumps}.
Their orbits decay towards the galactic nucleus due to dynamical friction (\citealt{N99}), but, as we will show, their contribution to growing the mass of the bulge is very modest in our simulations (see section \ref{Bulge}), a reflection of their moderate masses relative to previous works, claiming bulge formation via clumps (\citealt{N99, IS12, P13}).\\
We caution from the beginning that our simulations can only capture gravitationally bound clouds above our resolution limit.
Assuming conservatively that 1-2 SPH kernels yield the minimum cloud mass that can be reliably captured by the simulations (see e.g. \citealt{BB97}), based on the mass of our gas particles (see Table \ref{table_resolution}) such mass is of order $10^6$ $M_{\odot}$ in most of our simulations, since in GASOLINE2 the SPH kernel contains 32 particles (in the high resolution simulations the same limiting mass would be of order $10^5$ $M_{\odot}$).
This means we can hardly resolve fragmentation on the typical mass scale of molecular clouds, but this is not an issue for our aim of resolving giant clumps with masses well exceeding $10^6$ $M_{\odot}$.
Furthermore, the final masses of clumps could be overestimated, irrespective of the initial fragmentation scales, since we lack feedback from massive stars (radiative heating, radiation pressure, protostellar outflows and jets) interior to the clumps that may cause outflows and dissipate them on shorter timescales (\citealt{B10}). We will extensively discuss the origin of clump masses and sizes in Section \ref{Clumps}.\\
The effects of sub-grid physics on the fragmentation scale and clump evolution are often non-trivial, as we detail below. 
Before we move on with a detailed account of how fragmentation is affected by sub-grid physics and galaxy structural parameters, it is important to verify that the global properties of galactic discs, during the fragmentation phase, are consistent with observational constraints on high redshift clumpy galaxies.
An important diagnostic is disc kinematics. As we said in the introduction, observations show that clumpy galaxies have a ratio $v_{rot}/ \sigma \simeq 1-7$, therefore showing a strong evidence for the fact that clumpy discs are rotationally supported \citep{G06, FS06, S08, FS09, Wisnioski15}. In Figure \ref{fig:ratio_v_s} we show the evolution of the ratio $v_{rot}/ \sigma$, calculating the values at half mass radius, using the $\sigma_{1D}$, as in the observations. The figure shows that the ratio remains in the range $v_{rot}/ \sigma \sim 2-4$, as expected from the fact that discs is relatively undisturbed during clump formation, because clumps comprise a very small fraction of the disc mass. This is true even for runs in which fragmentation is more vigorous, such as run 17, that has no feedback, or run 25, that has a high disc mass and a high gas fraction. This is an important self-consistency check that was not always carried out in past works. Indeed, the more prominent fragmentation is, the lower $v_{rot}/ \sigma$ should be (as shown in Figure \ref{fig:ratio_v_s}), because the galaxy disc becomes increasingly gravitoturbulent, making it more difficult to explain why observed high redshift discs are clearly rotationally supported.\\
\\
In the following  we report on  the effect of varying sub-physics and the structural parameters on clump formation and evolution.
\subsection{Effects of different sub-grid physics}
\subsubsection{Feedback}
\begin{figure*}
  \includegraphics[width=0.36\textwidth]{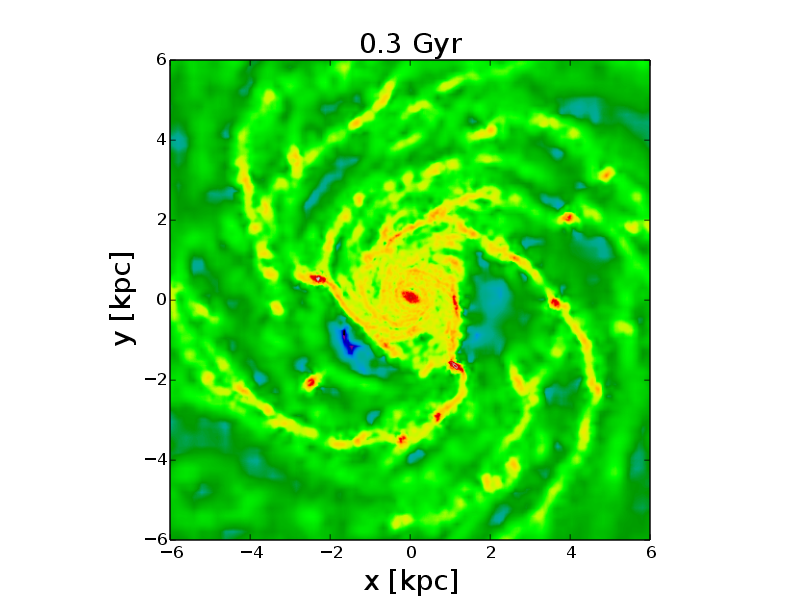}
  \hspace{-1.2cm}
  \includegraphics[width=0.36\textwidth]{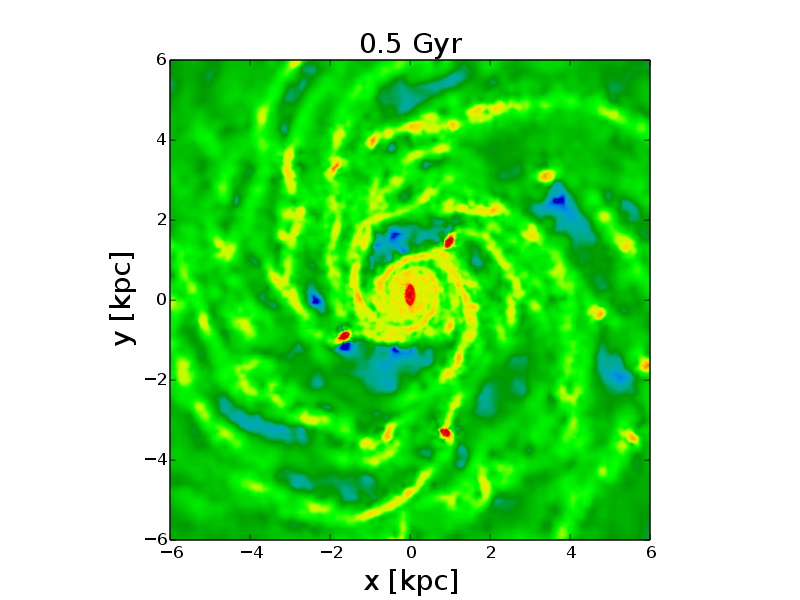}
  \hspace{-1.2cm}
  \includegraphics[width=0.36\textwidth]{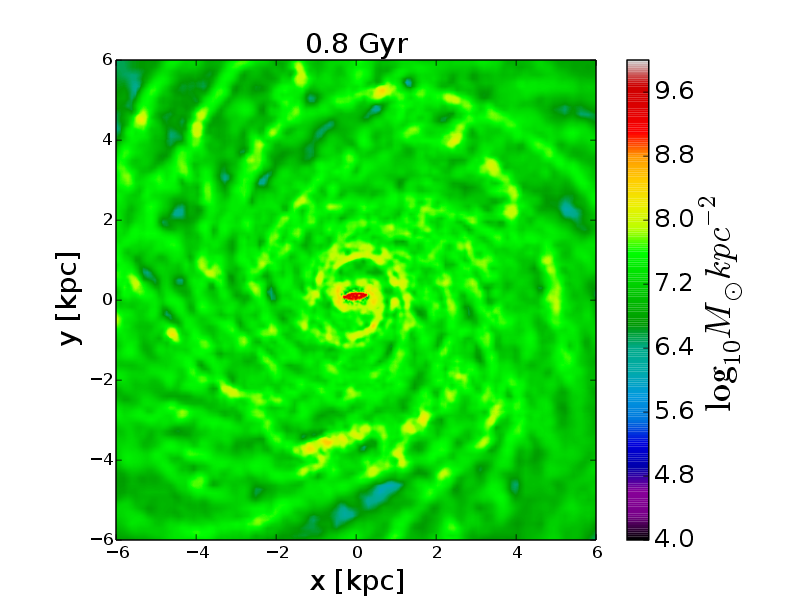}
  \includegraphics[width=0.36\textwidth]{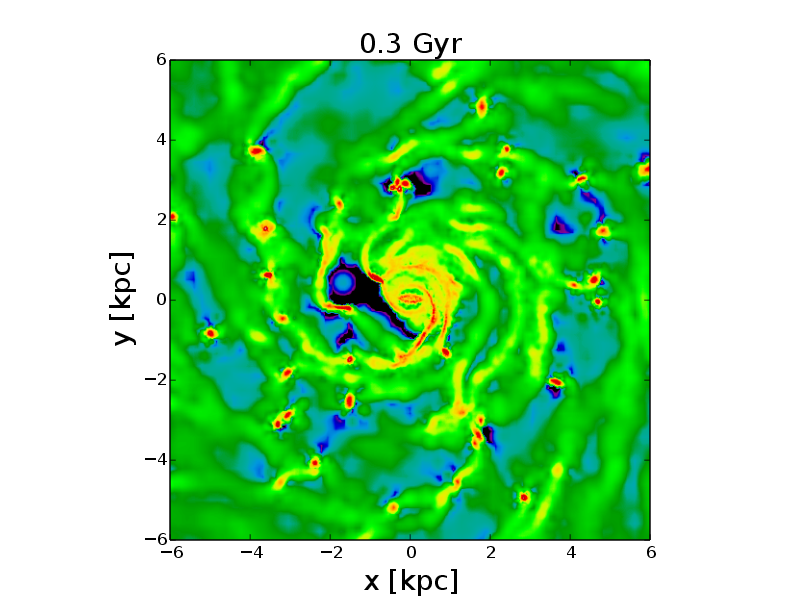}
  \hspace{-1.2cm}
  \includegraphics[width=0.36\textwidth]{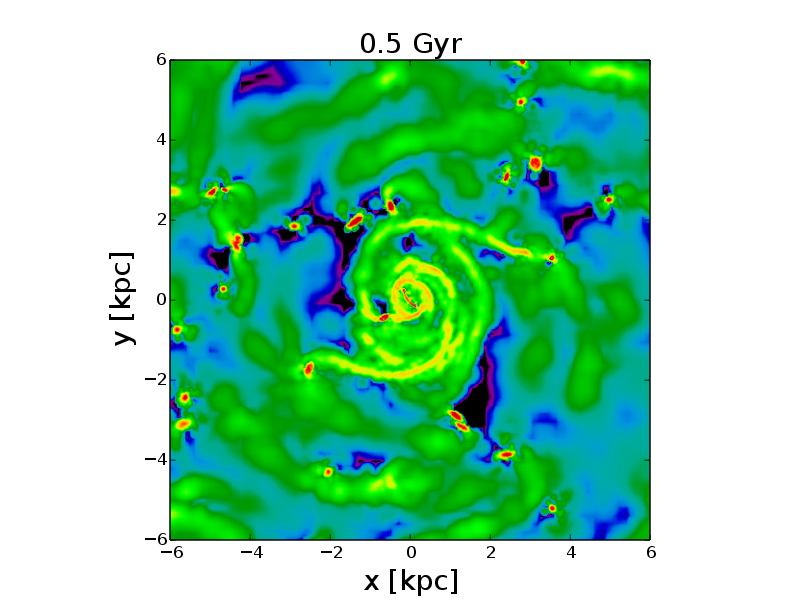}
  \hspace{-1.2cm}
  \includegraphics[width=0.36\textwidth]{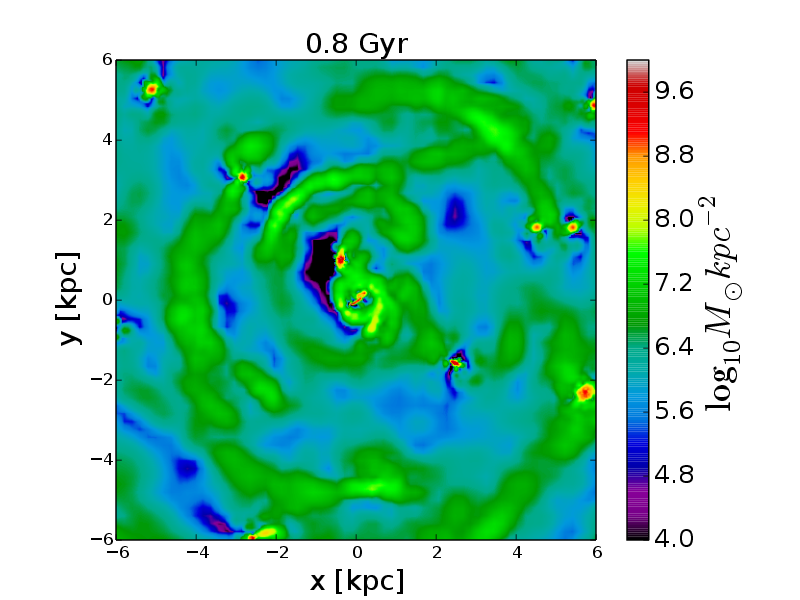}
  \caption{The evolution of the gas density maps (from left 0.3, 0.5 and 0.8 Gyr) for two runs with the same structural configuration: $c = 10$, $V_{vir}=150$ km/s, $f_g = 0.3$. Top panels show the galaxy in run 16, where feedback is included, while bottom panels show run 17, the extreme case without feedback. In run 16 we can observe the formation of some clumps, but these disappear in $500$ Myr and the galaxy continues evolving smoothly afterwards. In run 17, instead, clumps form vigorously and the galaxy completely fragments.}
  \label{fig:gas_density_noFB_FB}
\end{figure*}

\begin{figure*}
  \includegraphics[width=0.32\textwidth]{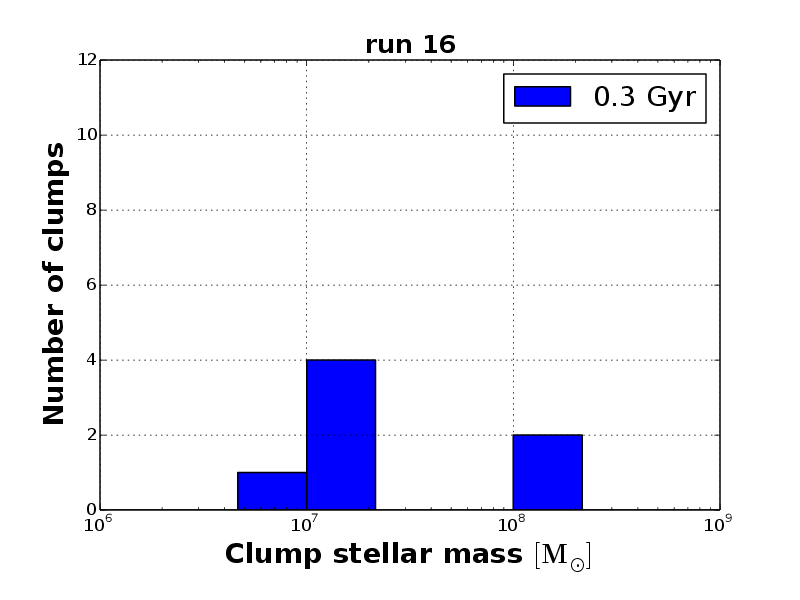}
  \includegraphics[width=0.32\textwidth]{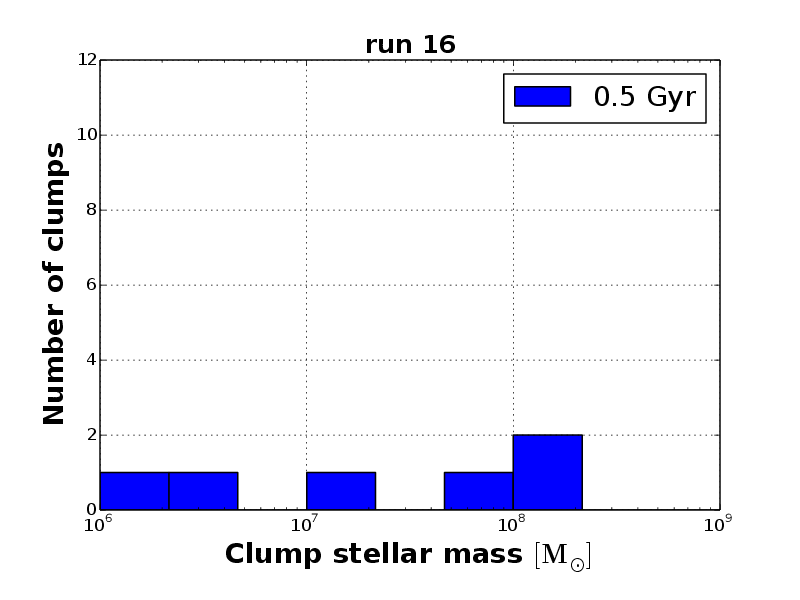}
  \includegraphics[width=0.32\textwidth]{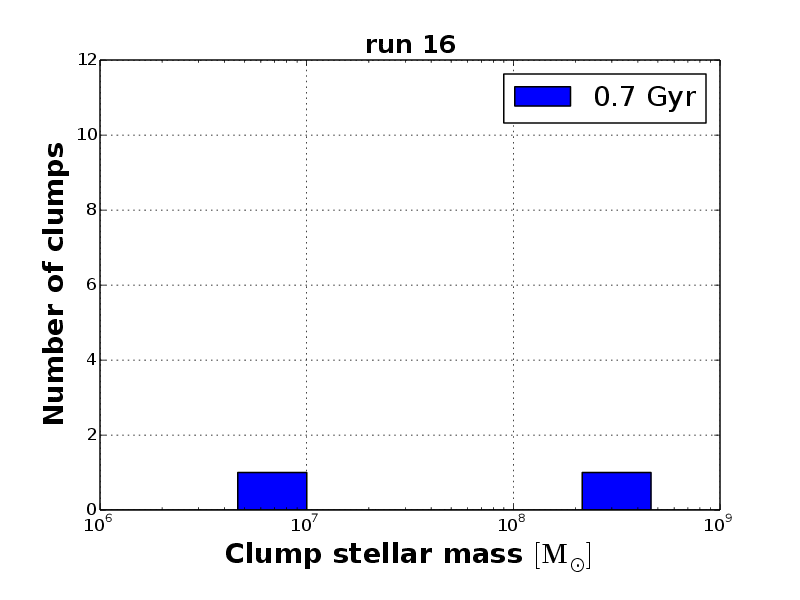}
  \includegraphics[width=0.32\textwidth]{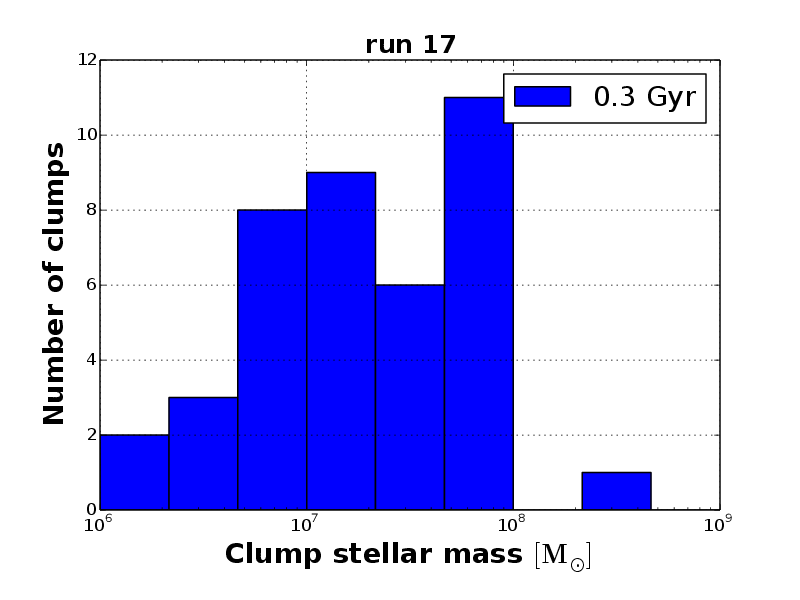}
  \includegraphics[width=0.32\textwidth]{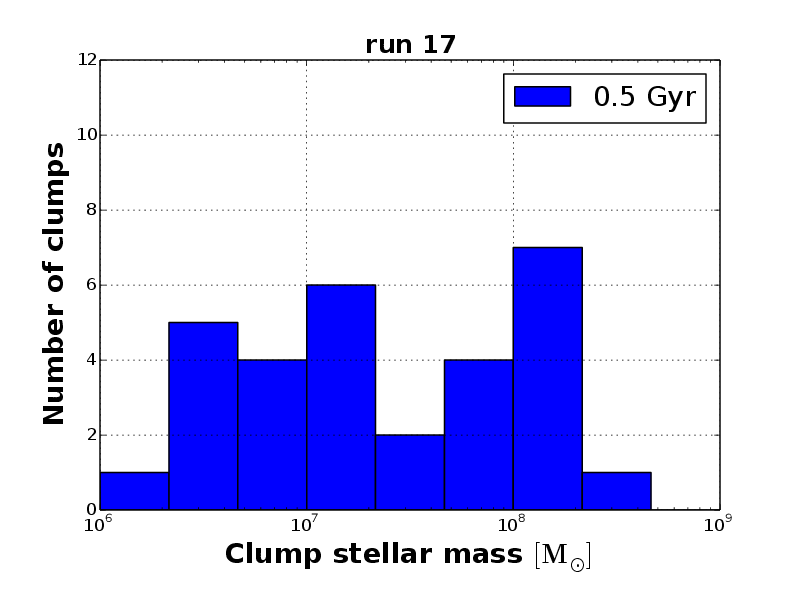}
  \includegraphics[width=0.32\textwidth]{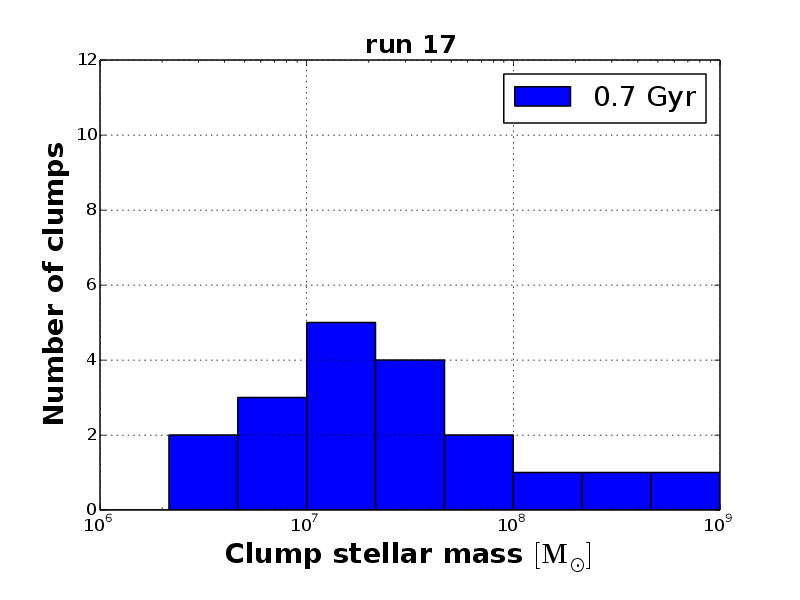}
  \caption{The histograms show the clump mass distribution for run 16 (top panels) and run 17 (bottom panels). It is clear that feedback suppresses clumps formation and that massive clumps can be formed only via clump-clump mergers, especially in the run without feedback, where more clumps are produced and can survive until they reach mass larger than $10^8$ $M_{\odot}$.}
  \label{fig:comparison_FB_noFB}
\end{figure*}

Feedback has a key role for allowing/suppressing fragmentation. To better understand this point, we focus on two runs, 16 and 17, which differ only by the presence of the feedback. In both the simulations the clumps form after $0.2$ Gyr and they are well visible at $0.3$ Gyr. When the blast-wave feedback is included (see top panels of Figure \ref{fig:gas_density_noFB_FB} for the gas density map evolution at some time steps), like in run 16, clumps form, but they are much fewer than in run 17 (bottom panels) and are dissolved quickly, in about $500$ Myr. 
The extreme case of run 17 shows that, when the galaxy evolves without any mechanism to heat and stir the gas, which would act to stabilised the disc against fragmentation, clumps formation is favoured. The galaxy appears already fragmented soon after $0.3$ Gyr, even if the clumps are smaller ($\sim 200-400$ pc in diameter) than the ones shown in previous papers using simulations with either no feedback or thermal feedback without any delayed cooling, which  is known to result in much weaker feedback compared  to our blastwave feedback prescription (e.g. in \citealt{C10} clump sizes are $\sim 1$ kpc).\\
The comparison between the two runs in Figure \ref{fig:gas_density_noFB_FB} shows clearly that feedback not only reduces the amount of fragmentation that happens in the disc but also stops the fragmentation earlier. Indeed in run 16 there are no more clumps after $0.8$ Gyr, while in run 17 there are still some. Note that in run 16 the number of clumps that are forming at time $0.2$ Gyr (not showed since they are still not well detectable) is the same that we observe at time $0.3$ Gyr, meaning that they are not disrupted in $0.1$ Gyr nor they migrate faster to the centre, but there is simply less clump formation relative to run 17. Comparing the mass histograms for these runs at different times makes the point even clearer (see Figure \ref{fig:comparison_FB_noFB}). 
For completeness we investigate the effect of feedback looking also at runs 27-28 that have a configuration more similar to the clumpy galaxies at high redshift. The same analysis confirmed us that clumps are fewer at all times with feedback. In this last case, however, fragmentation continues for longer due to higher disc mass also in the feedback case (i.e. run 27). This is at variance with previous theory \citep{DK13} and other simulations \citep{Bo14, Mandelker14}, in which clumps do form nearly as efficiently when the feedback is strong but they disrupt later. This is likely because the main effect of our feedback model is to help stabilize the disc on a global scale rather than to affect individual clumps. 
We also note that feedback has almost no effect on the high mass clumps, in the sense that the largest masses reached by clumps with and without feedback are essentially the same, which agrees with previous works (\citealt{M14, Bo14, Mandelker14}). Indeed in both runs 16 and 17 clumps can reach up to $\sim 3 \times 10^8$ $M_{\odot}$ (look at Figure \ref{fig:comparison_FB_noFB}, especially the middle panels). The reason is that the high mass tail of clumps (right panels in Figure \ref{fig:comparison_FB_noFB}) does not arise from fragmentation, so it does not depend on the presence/absence of feedback affecting fragmentation directly, rather they originate from clump-clump mergers that happen after $200-400$ Myr. This was verified by tracking back in time the particles of individual clumps. Once they are formed, these clumps are massive enough to avoid being affected by (our) feedback. 
We remark that small-scale feedback from massive stars inside clumps themselves, or radiation pressure, could have an 
important additional effect and will have to be explored in the future. Internal feedback could possibly affect the maximum
clump mass that can be achieved.
Finally, while all feedback models considered in the literature, including ours, are purely phenomenological and very approximate, we note that with ours the massive galaxies in our sample are consistent with abundance matching (Figure \ref{fig:abundance_matching}), which again reflects the large-scale self-regulation operated by feedback.
\subsubsection{Metal cooling}
\begin{figure*}
  \includegraphics[width=0.36\textwidth]{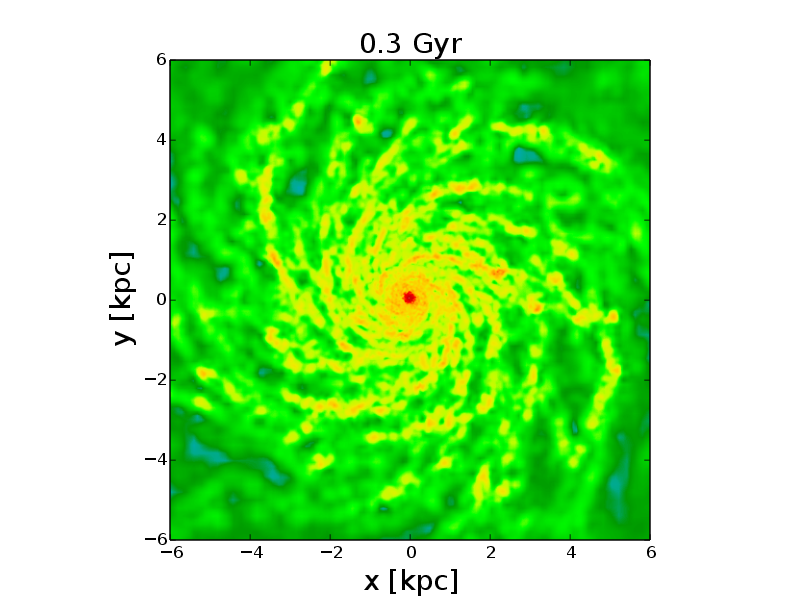}
  \hspace{-1.2cm}
  \includegraphics[width=0.36\textwidth]{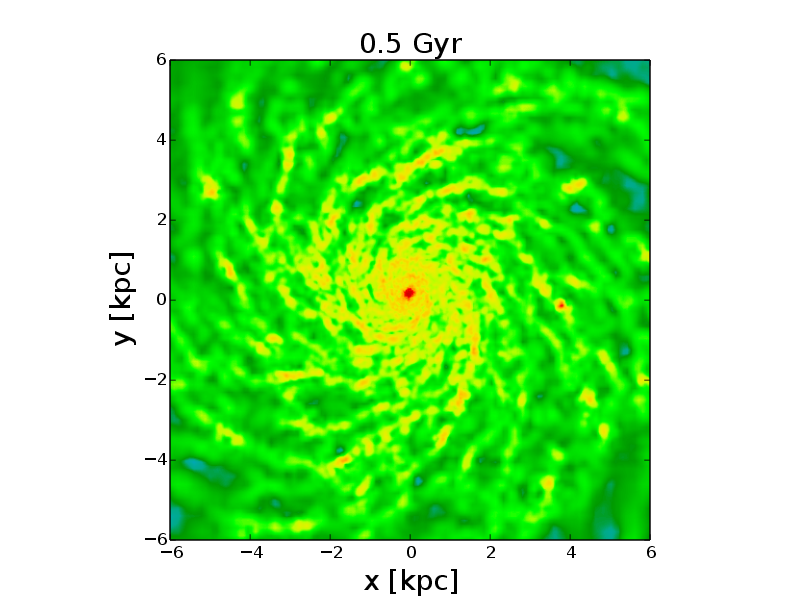}
  \hspace{-1.2cm}
  \includegraphics[width=0.36\textwidth]{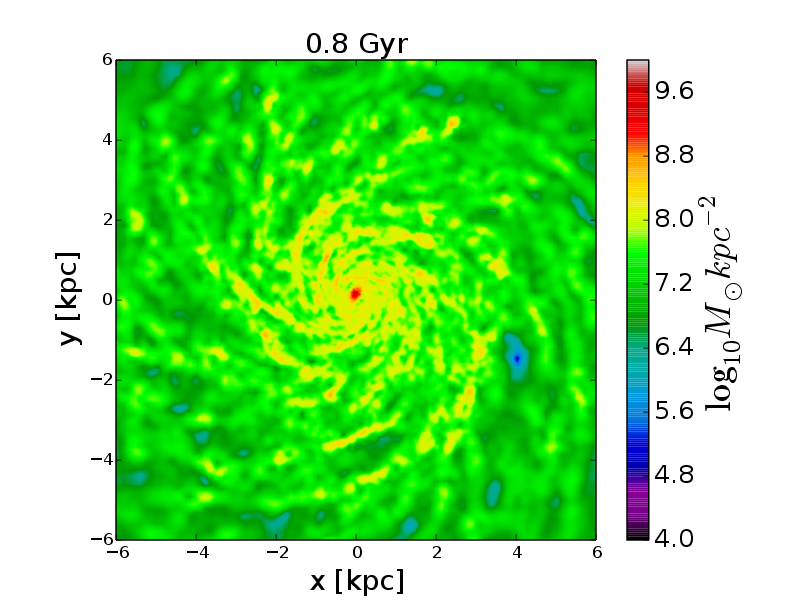}
  \caption{The evolution of the gas density map (from left 0.3, 0.5 and 0.8 Gyr) for the galaxy in run 18, having the same structural configuration of run 16 ($c = 10$, $V_{vir}=150$ km/s, $f_g = 0.3$) shown in Figure \ref{fig:gas_density_noFB_FB}, top panel. In this case we also switch on metal cooling and no clumps form (see text in the Results section).}
  \label{fig:gas_density_MClow}
\end{figure*}

\begin{figure}
\includegraphics[width=.49\textwidth]{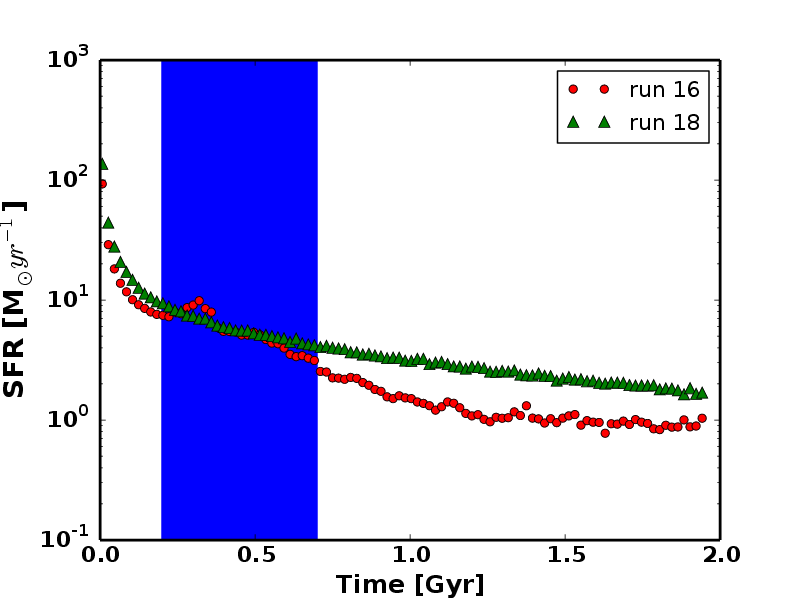}
   \caption{The SFR for run 16 (without metal cooling) and for run 18 (with metal cooling). The blue band highlights the clumpy phase for run 16. As one can see, the SFR is not significantly higher in the run with metal cooling, especially at the beginning, when clumps used to form (after the first $\sim 100-200$ Myr).}
  \label{fig:SFR_FB_vs_MC}
\end{figure}

We have explored the effect of varying the radiative cooling prescription on fragmentation. In particular, while all runs account for cooling of atomic hydrogen and helium, only in a subset of the runs we include cooling by metal lines. Galaxies reach metallicities up to  $0.3$  Z$_{\odot}$ by the end of the simulations as a result of star formation and subsequent metal yields from supernovae. Note that this is at the low end of the metallicities found in galaxies at $z \sim 2-3$, which are in the range $0.1-1$  Z$_{\odot}$, hence any difference we find between runs with and without metal cooling should be regarded as conservative. 
When metal-line cooling is included, not only the cooling at $T > 10^4$ K is enhanced in gas with non-zero metallicity, but, most importantly, gas can also cool to much lower temperature ($\sim 100-1000$ K).
Comparing run 16 and run 18, which are otherwise identical, shows that adding metal-line cooling (run 18) suppresses fragmentation at the scales that we resolve here (see Figure \ref{fig:gas_density_MClow}).\\
One could think that the reason for this lower fragmentation would be due to the more efficient cooling, which could induce faster SFR (so faster gas consumption), which also leads to stronger feedback, and thus to more efficient gas removal from clumps. To be sure that this is not the case, we check the SFR for both run 16 and run 18 and we found that a more efficient cooling by metals does not induce significantly higher or more rapidly rising star formation, see Figure \ref{fig:SFR_FB_vs_MC}, especially at early times (i.e. $100-300$ Myr), when clumps begin to form.
The reason is probably that the Toomre wavelength, which, as we explain below, expresses an upper limit on the fragmentation length scale (see Section \ref{Clumps}) becomes of order of one hundred parsecs, which is comparable to the resolution limit set by the gravitational softening. Indeed the Toomre wavelength is proportional to $c_s^{2}$, hence to the temperature T, which implies it drops from about 1 kpc to 100 pc or lower as the temperature drops from $ \sim 10^4$ K to $< 1000$ $K$ when we add metal-line cooling.
This means, however, that the structures that we do not resolve are on scales as small as those of typical molecular
clouds, (sizes $< 100$ pc, masses $< 10^6 M_{\odot}$). One could argue that in a gravitoturbulent system  the scale imposed by the thermal sound speed is not relevant, since gas pressure is dominated by turbulence. However, the latter argument applies to  the global instability of the disc, whereas locally collapse will be possible only in regions where turbulent support becomes negligible, in which $Q_T$ approaches unity. In such regions thermal pressure support becomes important again, analogously to what happens in cores of familiar molecular clouds.
Enhanced cooling with metal-lines then simply shrinks the scale of collapse in the latter regions (see e.g. \citealt{Elmegreen11}, calling for more demanding resolution requirements).
In runs with low dark matter halo concentration, the shallower shear profile, as measured by $ \kappa (r)$, drives the fragmentation scale up, counteracting the effect of enhanced cooling and allowing clump formation on resolved scales. This can be understood from the definition of the Toomre wavelength as well as  from that of the modified local fragmentation length, introduced in Section \ref{Clumps}. This happens for runs 2 and 15 (also in the first one the mass resolution is higher than in the second case, see Table \ref{table_resolution}), which also have the highest gas fraction.
\subsection{Effects of structural parameters}\label{sec:structural_param}
In what follows, we want to investigate what happens when we change the structural parameter of our galaxy.\\
\subsubsection{Virial Velocity}
\begin{figure*}
  \includegraphics[width=0.36\textwidth]{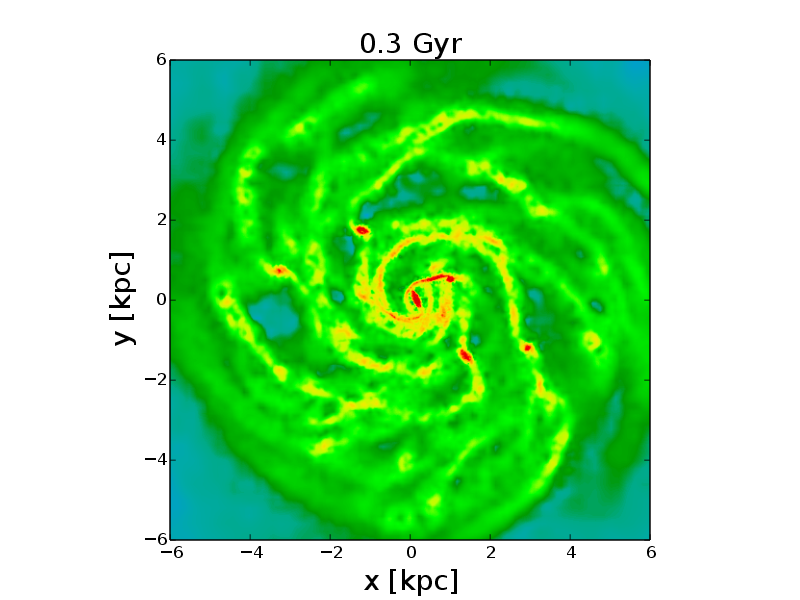}
  \hspace{-1.2cm}
  \includegraphics[width=0.36\textwidth]{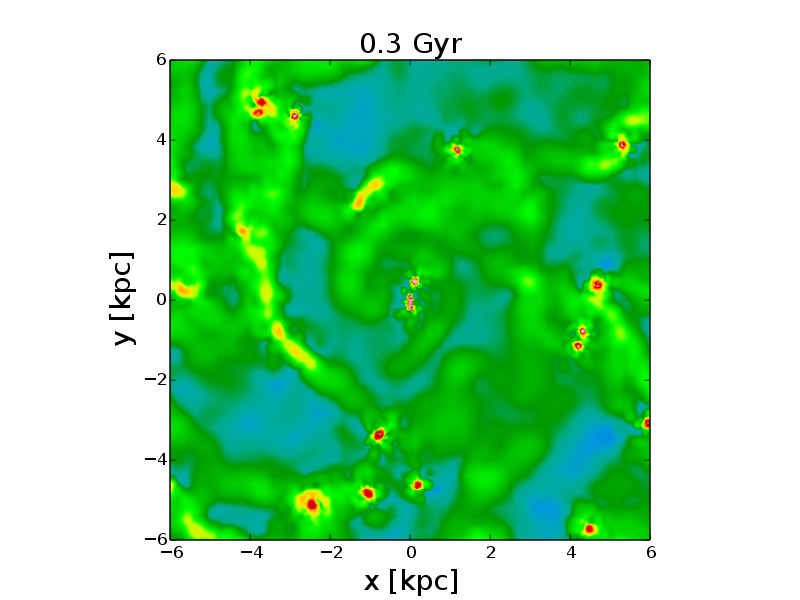}
  \hspace{-1.2cm}
  \includegraphics[width=0.36\textwidth]{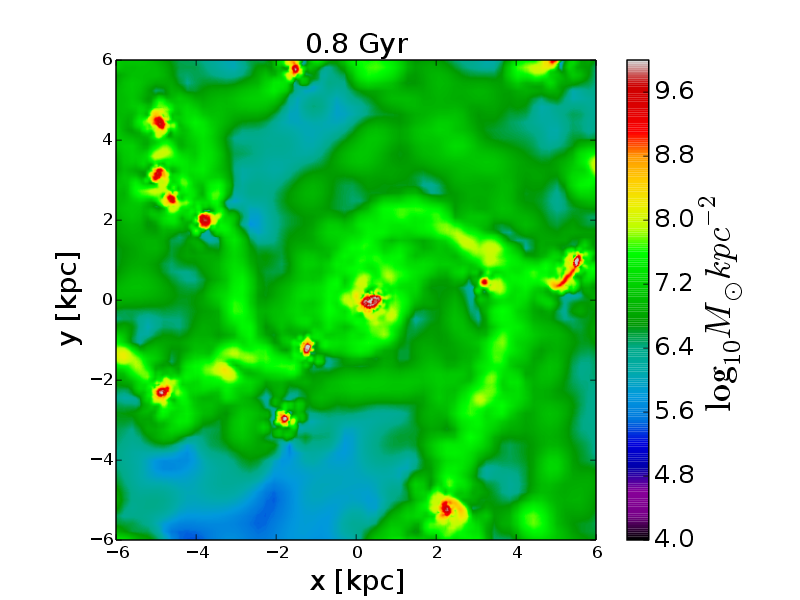}
  \caption{The gas density maps for galaxies in run 1 (left), in run 14 (middle) and in run 27 (right) at time 0.3 Gyr, when the clumps become detectable. The three galaxies evolve with feedback and with the same parameters, except for the virial velocity that ``increases" from left to right. All the galaxies fragment, but note how fragmentation increases with the virial velocity (i.e. with the mass) of the galaxy. Note also that clumps are shown 100 Myr after the beginning of fragmentation and many of them are very close each other and will merge in the next 200 Myr.}
  \label{fig:comparison_v_vir}
\end{figure*}

\begin{figure*}
  \includegraphics[width=0.36\textwidth]{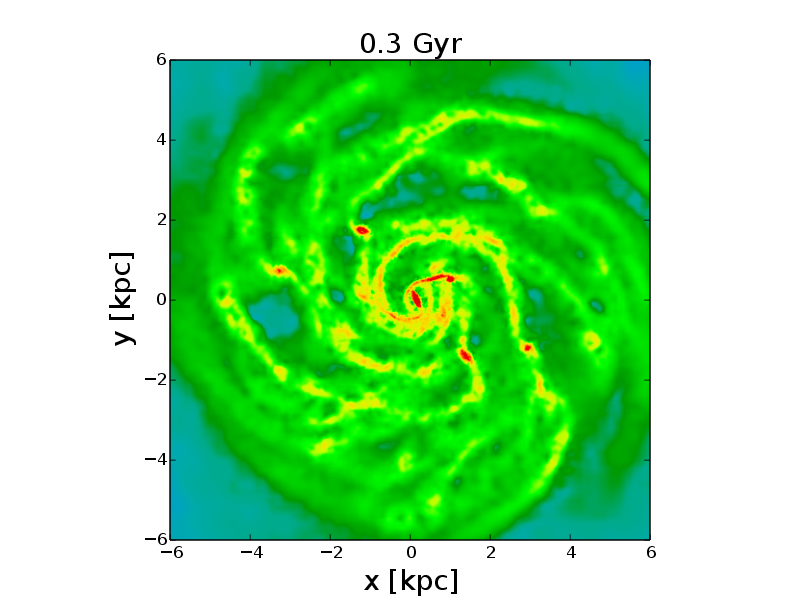}
  \hspace{-1.2cm}
  \includegraphics[width=0.36\textwidth]{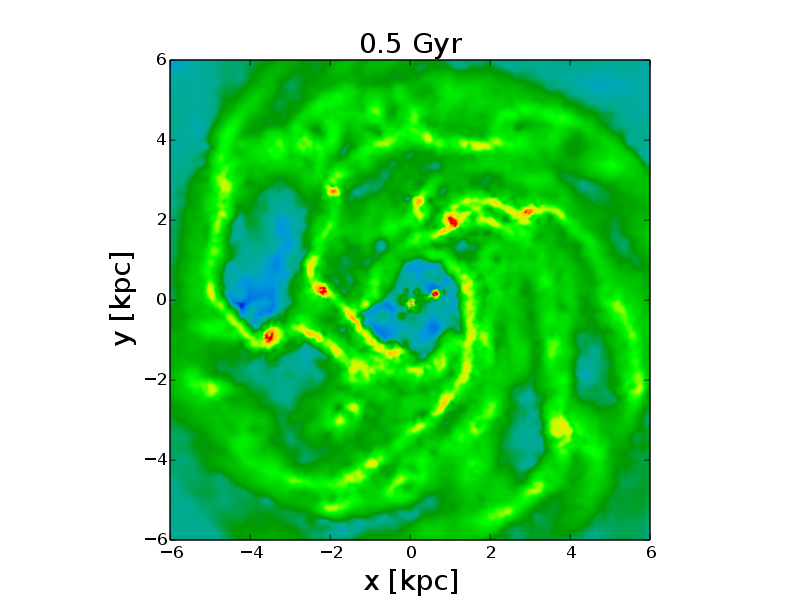}
  \hspace{-1.2cm}
  \includegraphics[width=0.36\textwidth]{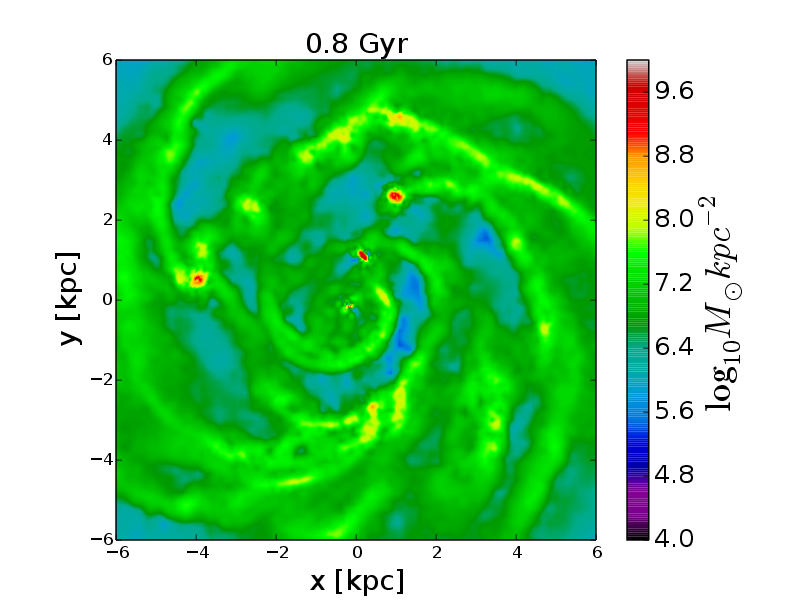}
  \includegraphics[width=0.36\textwidth]{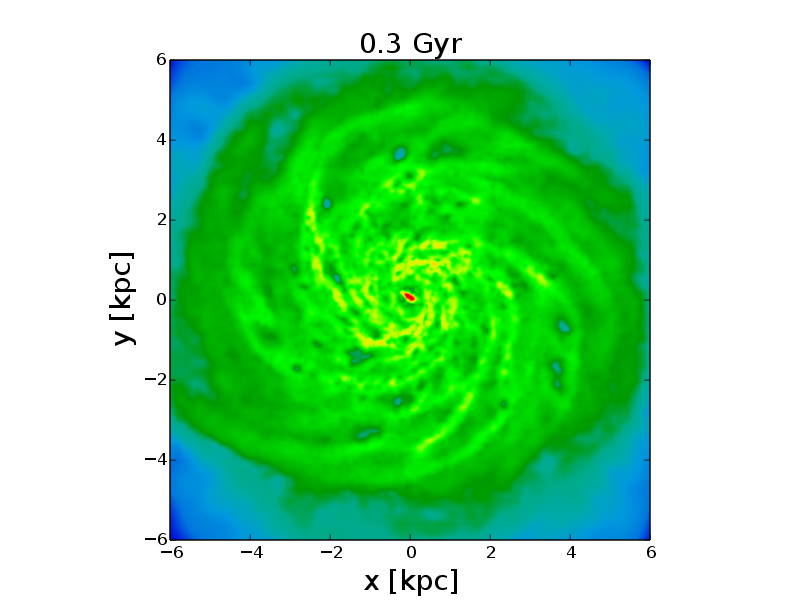}
  \hspace{-1.2cm}
  \includegraphics[width=0.36\textwidth]{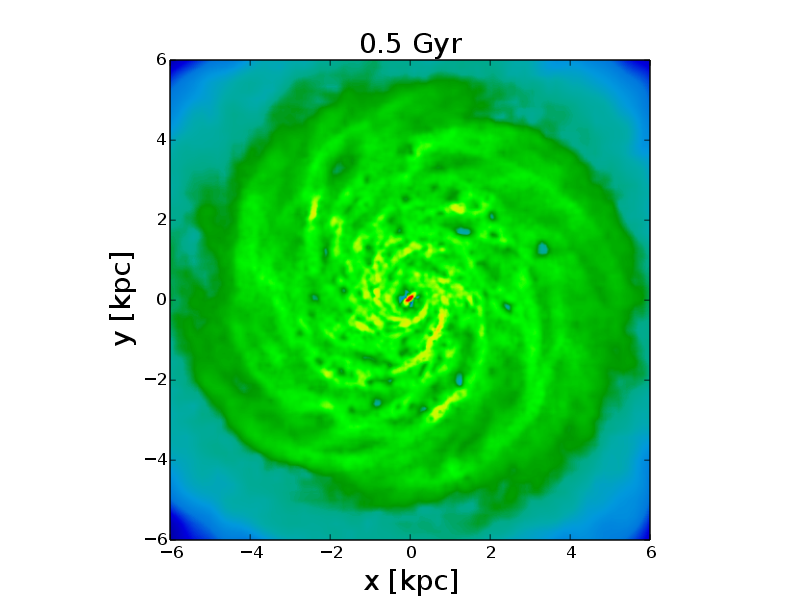}
  \hspace{-1.2cm}
  \includegraphics[width=0.36\textwidth]{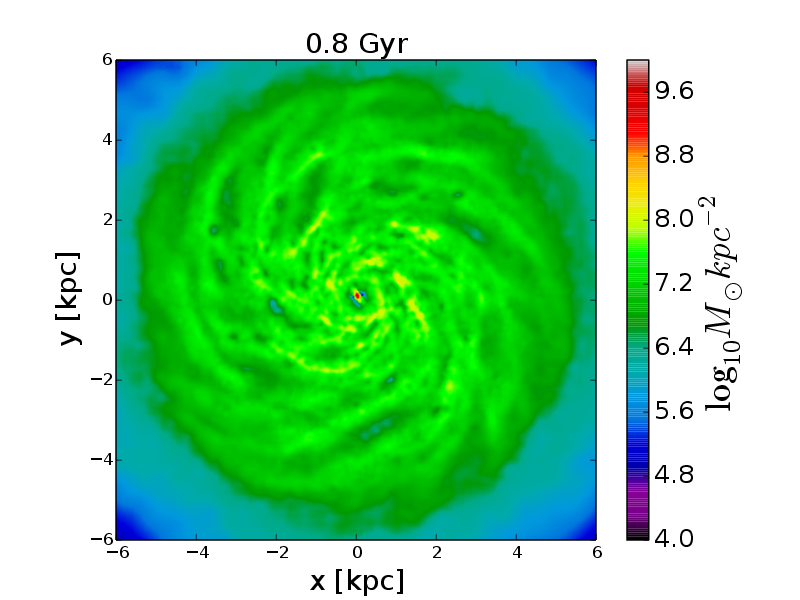}
  \caption{The evolution of the gas density map (ifrom left 0.3, 0.5, 0.8 Gyr) for the galaxy in run 1 (top panels) and for the galaxy in run 4 (bottom panels). Both evolve with feedback and have $V_{vir} = 100$ km/s, but with this low circular velocity, the galaxy requires also a high gas fraction, $f_g = 0.5$, and a low concentration, $c = 6$, in order to fragment, as in run 
1, top panels. The galaxy model in run 4, instead, does not fragment owing to a lower gas fraction, $f_g = 0.3$, and a higher concentration, $c = 10$. See, for comparison, also the top panels of Figure \ref{fig:gas_density_noFB_FB}: run 16, which fragments, was run with the same sub-grid physics and code setup as in run 4, and structural parameters in the galaxy model were the same, except for the higher circular velocity.}
  \label{fig:v_vir100}
\end{figure*}

We use virial velocity rather than virial mass to label our galaxy models because rotational velocities, or at least line-widths, are observable parameters in real galaxies, while total mass is not. 
Our models have $V_{vir} = 100, 150, 180$ km/s. Since we constructed models that obey the cold dark matter halos scaling between mass and velocity, $M_{vir} \sim {V_{vir}^3} $ \citep{MMW98}, a difference of a factor 1.8 in mass corresponds to a difference of about 5.8 in velocity. Based on our model building technique, the disc mass also scales as the total virial mass, while the radius scales as the velocity (again see \citealt{MMW98}). As a result, the disc surface density, both stellar and gaseous, varies by a factor $\sim {V_{vir}^3}/{V_{vir}^2} \sim V_{vir} \sim 1.8$. The Toomre parameter and wavelength are inversely proportional to the surface density $\Sigma$, suggesting thus that the disc should be more prone to fragmentation in the models with the highest $V_{vir}$ (see Figure \ref{fig:comparison_v_vir}). Table \ref{table_runs} shows that indeed this is what happens: the vast majority of runs with $V_{vir} = 150-180$ km/s does fragment, while the reverse is true for the models with $V_{vir} = 100$ km/s, which produce clumps only if they have low concentration and high gas fraction (see Figure \ref{fig:v_vir100}).
Moreover the most massive clumps are found in the most massive galaxy (run 27) or in runs without feedback after $500-700$ Myr. This reflects the fact that, as we have already explained, mergers can drive up the mass of clumps further simply because more clumps form in more massive discs (with higher $V_{vir}$). It does not reflect an intrinsic scaling of clump mass with the ratio of disc mass to virial velocity. Indeed we find no systematic variation in the initial mass of clumps as we vary $V_{vir}$ (see Section \ref{Clumps}).
\subsubsection{Concentration}
\begin{figure*}
  \includegraphics[width=0.36\textwidth]{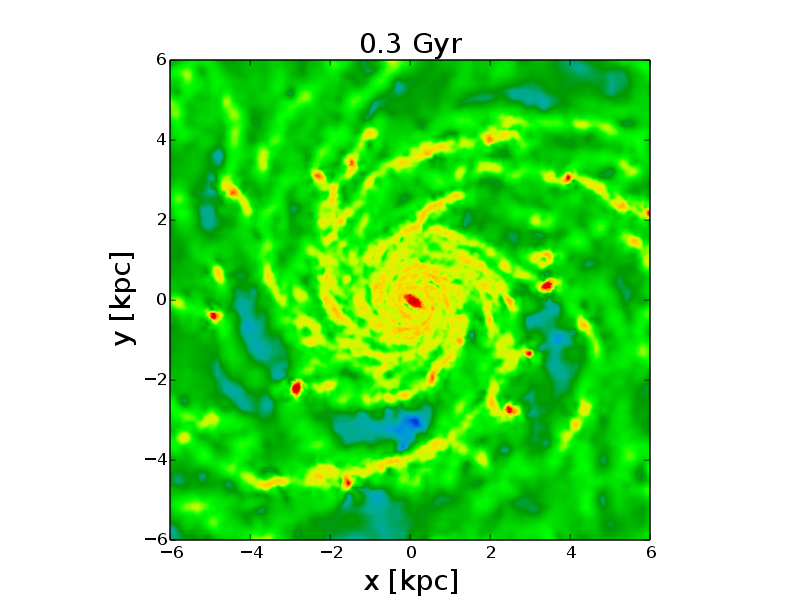}
  \hspace{-1.2cm}
  \includegraphics[width=0.36\textwidth]{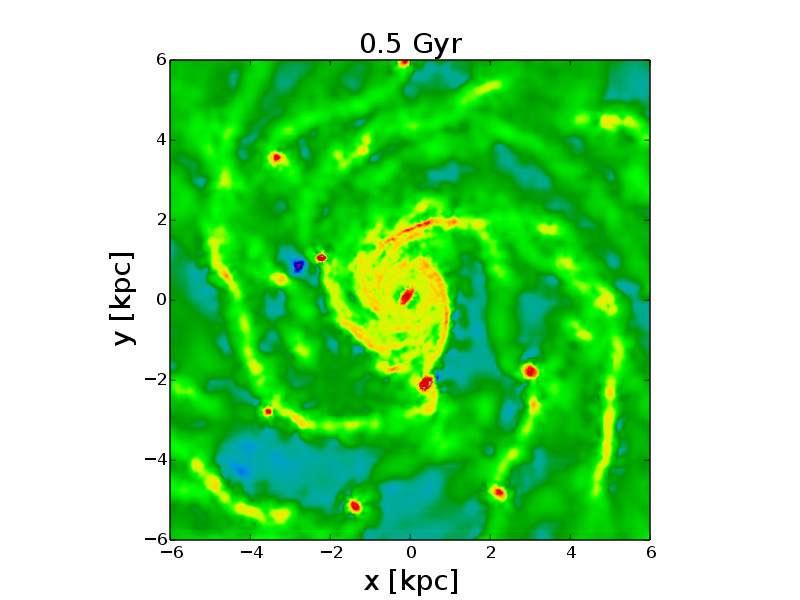}
  \hspace{-1.2cm}
  \includegraphics[width=0.36\textwidth]{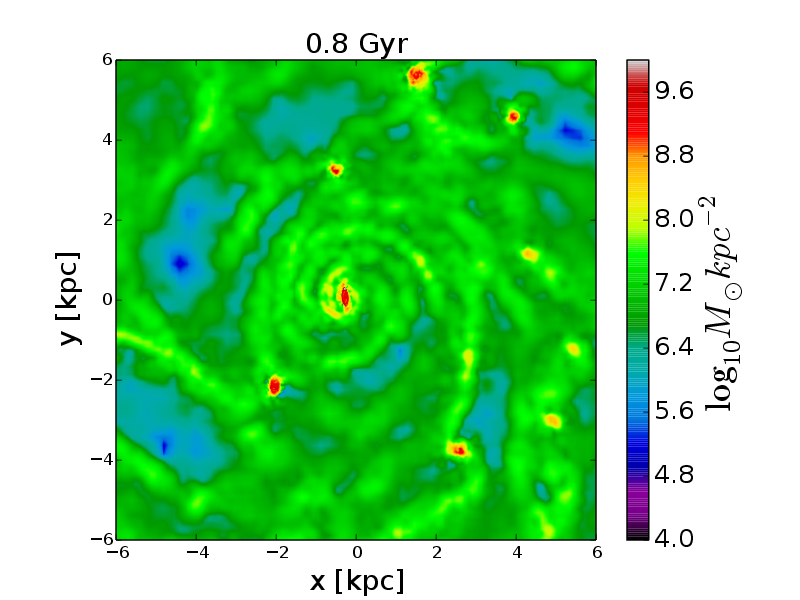}
  \caption{The evolution of the gas density map (from left 0.3, 0.5, 0.8 Gyr) for galaxy in run 12. This run evolves with feedback and employes a model with $V_{vir} = 150$ km/s and $f_g = 0.3$ as in run 16, but with a lower concentration, $c=6$. Here more clumps form and they survive for a longer time (see for comparison the top panels of Figure \ref{fig:gas_density_noFB_FB}, especially time $0.8$ Gyr).}
  \label{fig:gas_density_galaxy_cc6_FB}
\end{figure*}

Halo concentration modifies the Toomre parameter by affecting the velocity/shear profile of the disc, namely the radial profile of $\kappa$. In particular, with all other parameters being equal, lowering the concentration leads to a shallower velocity profile towards the centre, hence a lower $\kappa$ and thus a lower Toomre parameter at fixed radius.
For instance, as a result of this effect, galaxies are more prone to bar instabilities in lower concentration halos 
(see e.g. \citealt{MW04}). Therefore we expect models with lower concentration to be more susceptible to fragmentation. This is indeed what we find (see Table \ref{table_runs}) and, for example, it is clear by  comparing run 16 with run 12, see Figures \ref{fig:gas_density_noFB_FB} (top panels) and \ref {fig:gas_density_galaxy_cc6_FB}.

The two runs adopt the same sub-grid physics (radiative cooling, star formation and feedback), but in run 12 the concentration is lower ($c=6$ versus $c=10$ in run 16). As shown in Figure \ref {fig:gas_density_galaxy_cc6_FB}, decreasing the concentration produces more clumps than in run 16 (again, see the top panels in Figure \ref{fig:gas_density_noFB_FB}) and they survive for a longer period (see, in particular, times 0.8 Gyr to compare). They are completely dissolved after long enough time though (0.6 Gyr after their formation).
We stress that models with low concentration are most representative of the massive high redshift galaxies usually  
considered in the literature on giant clumps, but our goal in this paper is to carry out a comprehensive parameter study which applies also to less massive galaxies ($V_{vir} \sim 100$ km/s), such as those studied in \citet{Livermore15}.\\
 \subsubsection{Gas fraction}
\begin{figure*}
  \includegraphics[width=0.36\textwidth]{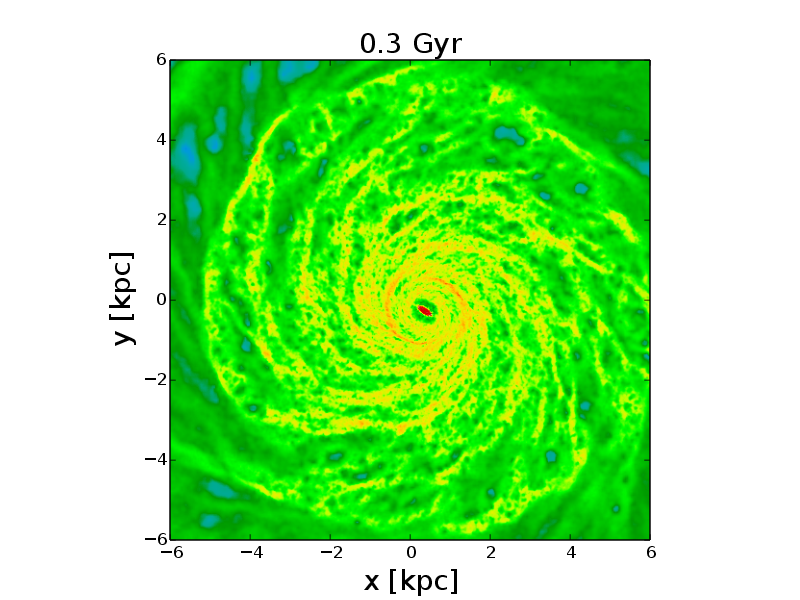}
  \hspace{-1.2cm}
  \includegraphics[width=0.36\textwidth]{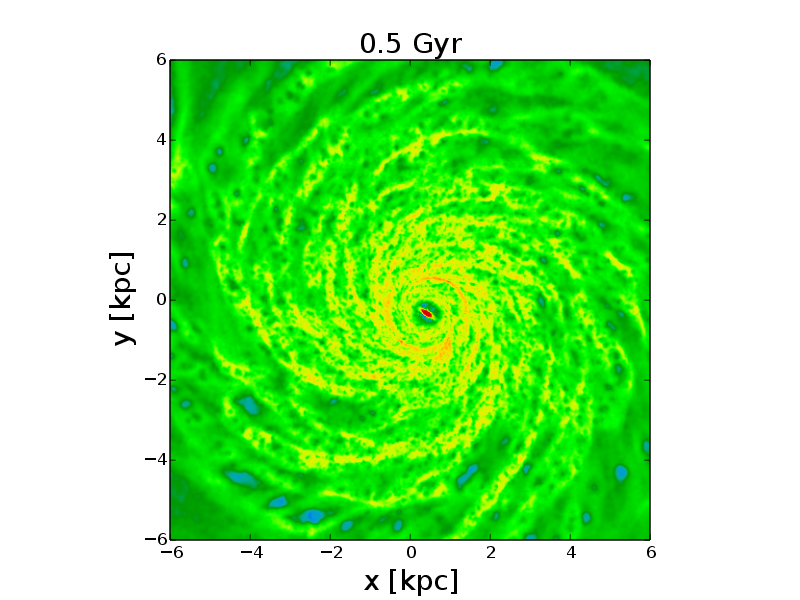}
  \hspace{-1.2cm}
  \includegraphics[width=0.36\textwidth]{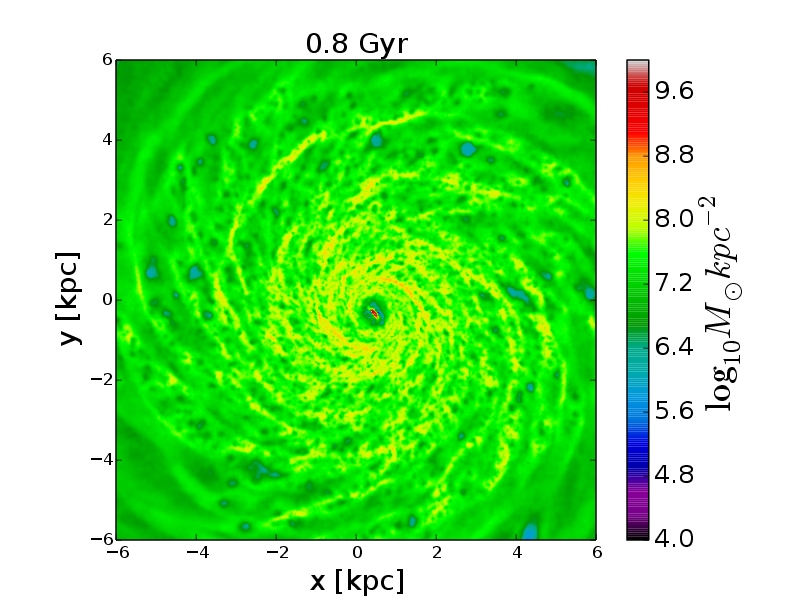}  
  \includegraphics[width=0.36\textwidth]{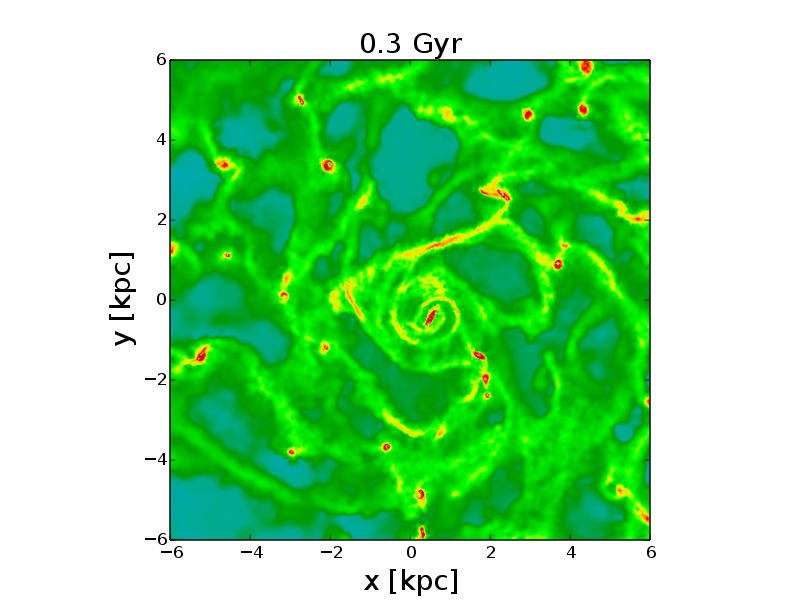}
  \hspace{-1.2cm}
  \includegraphics[width=0.36\textwidth]{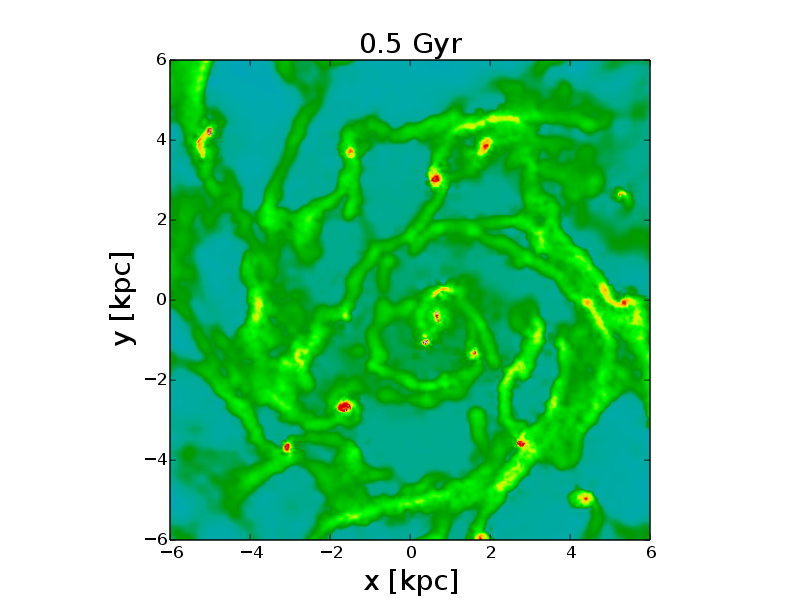}
  \hspace{-1.2cm}
  \includegraphics[width=0.36\textwidth]{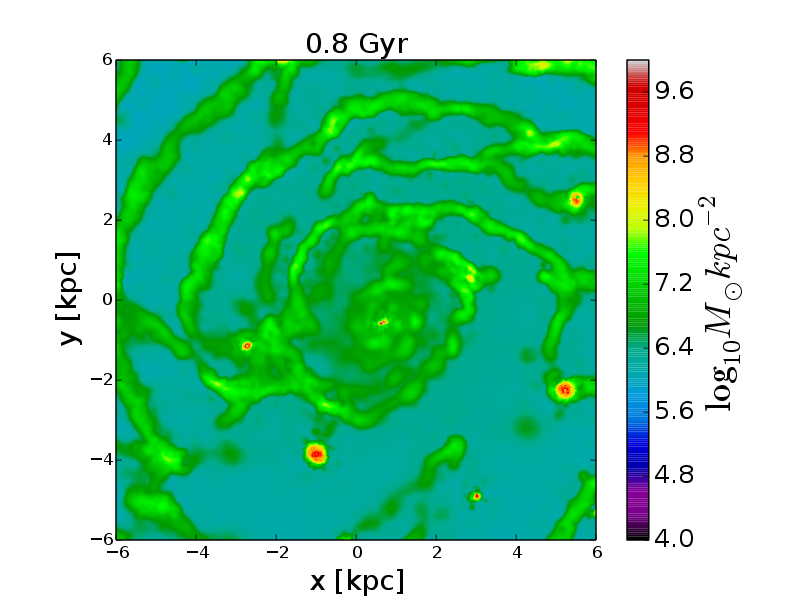}
  \caption{The evolution of the gas density map (from left 0.3, 0.5, 0.8 Gyr) for the galaxies in runs 22 (top panels) and 25 (bottom panels). Both runs have high resolution, the same concentration ($c=10$) and virial velocity ($V_{vir} = 150$ km/s), and both are evolved with feedback. Run 22 (top panel) has a gas fraction of $f_g= 0.3$. Note that this  case is almost identical to run 16 (which is at low resolution), but here no clumps form. In run 25, instead (shown in the bottom panels), the gas fraction is $f_g = 0.5$. More clumps form and they survive for a longer time (see for comparison also Figure \ref{fig:gas_density_noFB_FB}, top panels, time $0.8$ Gyr).}
  \label{fig:gas_density_g03_g05_high_FB}
\end{figure*}

Since observations have shown that clumpy high redshift galaxies have a range of gas fractions $f_g = 0.4 - 0.6$ \citep{DZ14, Livermore15}, we study the effect of increasing the gas fraction in the discs. We note that in the simulations gas fractions account for the total gas content, while in observations only the molecular gas is detected. A posteriori we verified that, once radiative cooling is switched on, gas at and above the threshold density for star formation, $n_{SF} =10$ atoms/cm$^3$, typically accounts for $70-80 \%$ of the total gas content in the disc. At such high densities gas would be mostly molecular if the simulations included the phase transition between atomic and molecular gas, hence this shows that the quoted values of the gas fractions can be meaningfully compared with the observational estimates.
Inspection of Table \ref{table_runs} clearly shows that the number of runs, that lead to clumps, increases in the subset that adopts a higher gas fraction (only 3 out of 12 runs with $f_g=0.5$ do not fragment, while only 5 out of 16 runs with $f_g = 0.3$ do fragment).
As an example, both run 22 and run 25 employ high resolution, have the same concentration, the same virial velocity and evolve with the same physics (radiative cooling, star formation and feedback), they differ only by the gas fraction ($0.3$ versus $0.5$, respectively). In Figure \ref{fig:gas_density_g03_g05_high_FB} we can see that clumps form only in the case with higher gas fraction (bottom panels). This is probably due to the fact that a higher gas fraction produces a lower Q Toomre parameter for the gas by increasing the surface density. Note that even in the two-fluid case the Toomre parameter would decrease somewhat, because the kinematically colder component of the two, namely the gas, increases its contribution to the disc mass with increasing gas fraction (see Equation (\ref{eq:Q_total}) in Section \ref{Intro}), so the disc is more unstable.
\subsection{Effects of resolution}\label{resolution}
\begin{figure*}
  \includegraphics[width=0.49\textwidth]{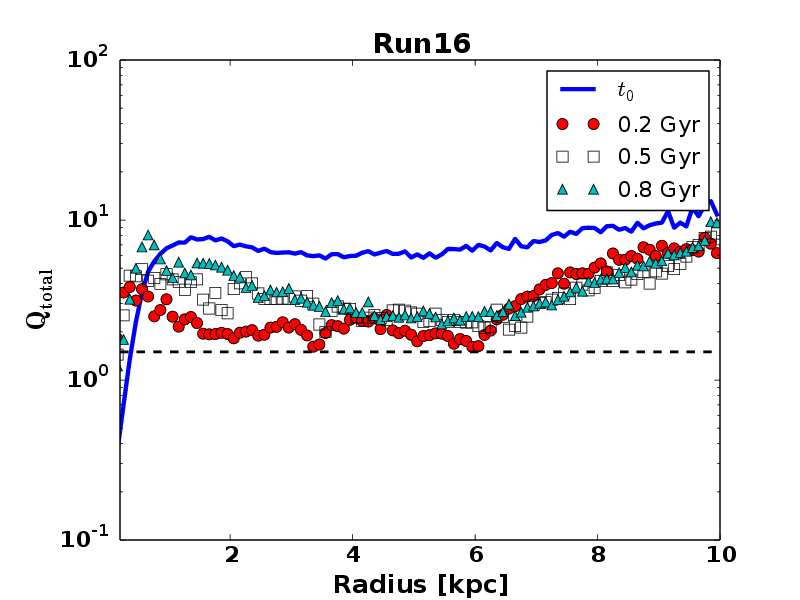}
  \includegraphics[width=0.49\textwidth]{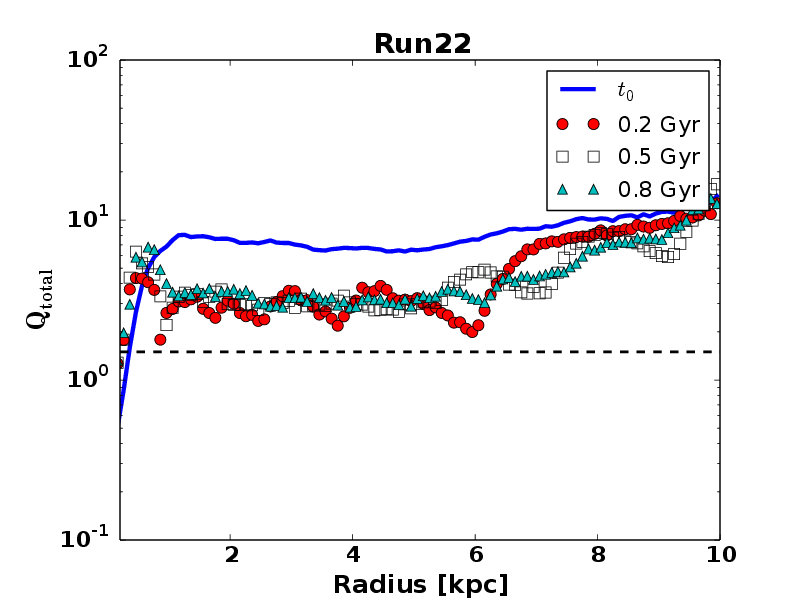}
  \caption{The evolution of the $Q_{total}$ Toomre parameter in run 16 (i.e. low resolution) on the left and run 22 (i.e. high resolution) on the right. At $t_0$ the adiabatic relaxation phase ends and, in the case with high resolution, the profile is more stable and smoother than in the case with low resolution. For run 16 the time steps 0.2, 0.5.  and
0.8 Gyr correspond to the beginning of the clump formation phase, the middle of the clumpy phase and the end of it, respectively. Horizontal dashed line marks the threshold 1.5 for the stability of a disc. Note that for run 16 at $t = 0.2$ Gyr the sudden drop around $3.5$ kpc corresponds to the radius at which clumps form.}
  \label{fig:Toomre_low_high}
\end{figure*}

\begin{figure*}
 \includegraphics[width=0.49\textwidth]{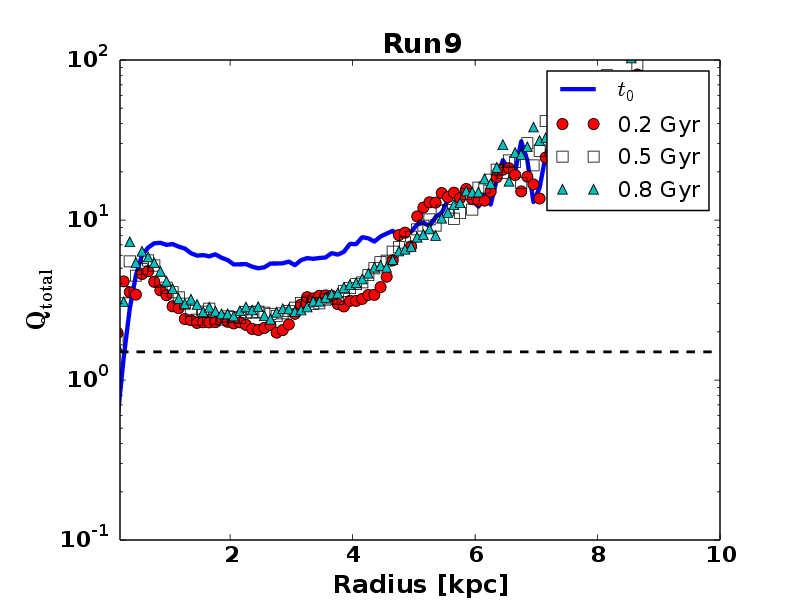}
 \includegraphics[width=0.49\textwidth]{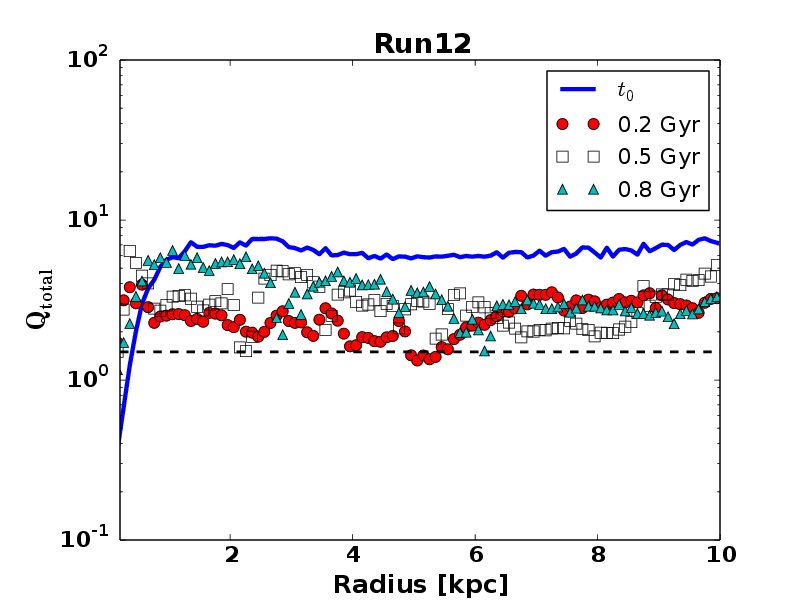}
 \includegraphics[width=0.49\textwidth]{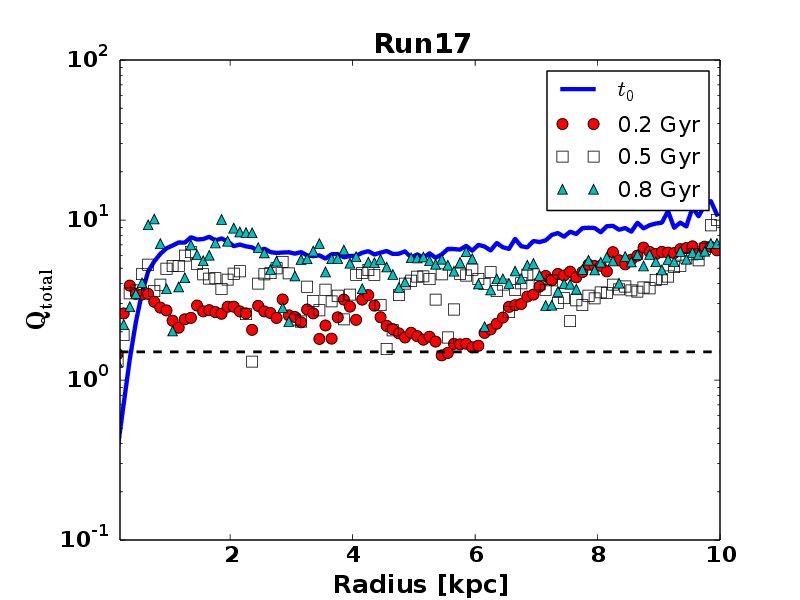}
\includegraphics[width=0.49\textwidth]{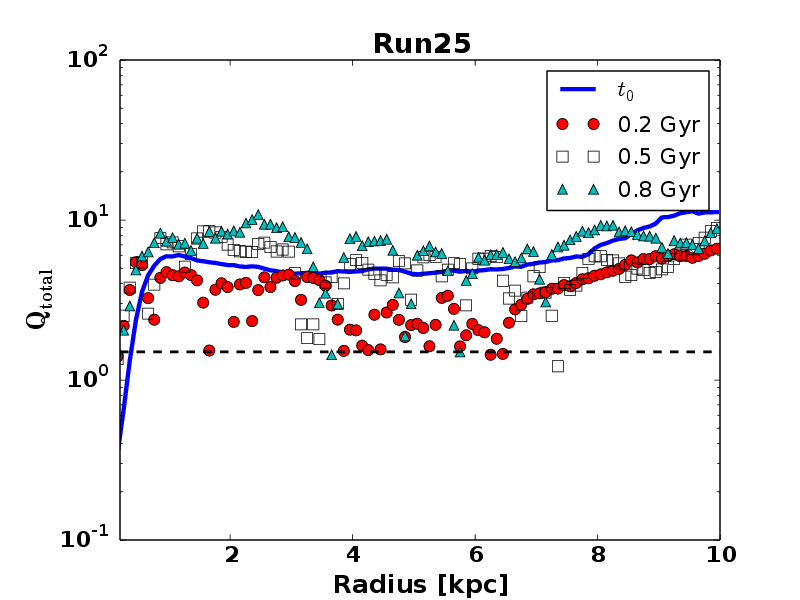}
\caption{The evolution of the $Q_{total}$ Toomre parameter in other representative runs. Run 9 is the only one among these that does not fragment, so the Toomre parameter is always well above the critical value of 1.5  (horizontal dashed line). Runs 12, 17 and 25, instead, fragment and show a lower and more disturbed value of Q, because of the presence of clumps during  the whole duration of the  simulation.}  
\label{fig:Toomre_other_runs}
\end{figure*}

Mass and spatial resolution also have an effect on the outcome of the instability. In our simulations we see that, with higher resolution, hence with initial conditions that have lower particle noise, the disc relaxes to a more stable state during the adiabatic evolution, which tends to suppress fragmentation. This shows the importance of relaxation and highlights the importance of minimizing the effects of built-in numerical perturbations, when studying disc instability.
Run 16 and run 22 are identical regarding the chosen structural parameters (i.e. concentration $10$, virial velocity $150$ km/s, gas fraction $0.3$) and the adopted sub-grid physics (radiative cooling with no metal lines, feedback, no thermal and metal diffusion), they only differ by the resolution.
The results are clear in the top panels of both Figure \ref{fig:gas_density_noFB_FB} and Figure \ref{fig:gas_density_g03_g05_high_FB}, for run 16 and 22 respectively. 
In the low resolution case, clumps form and disappear in $\sim 500$ Myr, while in the high resolution case no clumps form.
Inspection of the Q Toomre parameter profile for both runs, see Figure \ref{fig:Toomre_low_high}, clearly shows that the state of the disc emerging from the initial relaxation phase is different: in both cases the value of $Q_T$ is similar, but in the low resolution simulation, run 16, the disc is noisier at the beginning and the galaxy fragments immediately, while in the high resolution case, run 22, the Toomre parameter is smoother. The increased noise in the density field has been
shown to easily trigger fragmentation in self-gravitating discs that are just close to the Q threshold (e.g. \citealt{Mayer04}). This has been shown robustly by imposing a small, percent level perturbation of the density field in both SPH and AMR protostellar disc models (e.g. \citealt{MG08}).\\
We have checked differences in other quantities that may affect stability, such as star formation, but we found those to be negligible. However it is important to note that high resolution does not prevent fragmentation, indeed in runs 25 and 26 (which have also higher gas fraction), we still observe clumps.
Unfortunately high resolution simulations require a long calculation time (note that we stopped run 26 before the end of the simulation, due to its high computational cost), so we could not perform as many high resolution simulations as low resolution ones to make a systematic comparison.\\
For more Toomre parameter profile evolution, look also at Figure \ref{fig:Toomre_other_runs}.
\subsection{Effect of  SPH implementation}\label{GDSPH}
\begin{figure*}
  \includegraphics[width=0.36\textwidth]{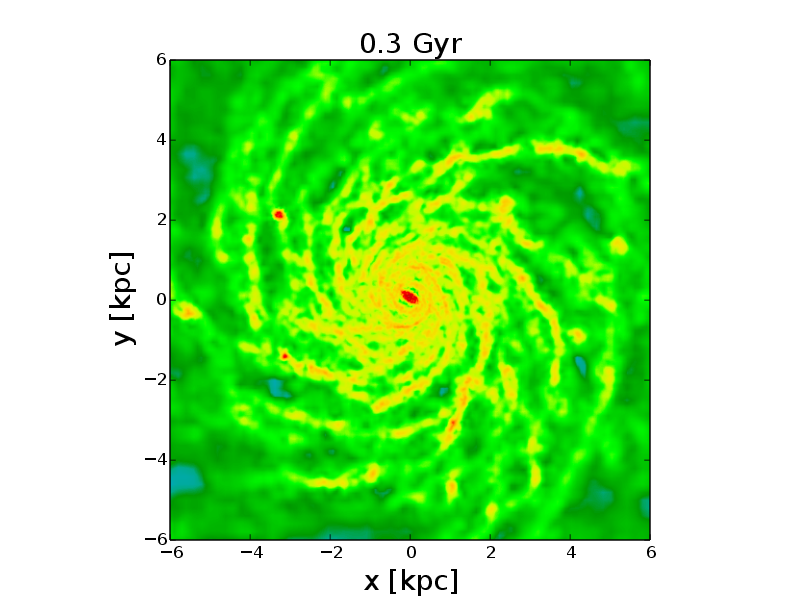}
  \hspace{-1.2cm}
  \includegraphics[width=0.36\textwidth]{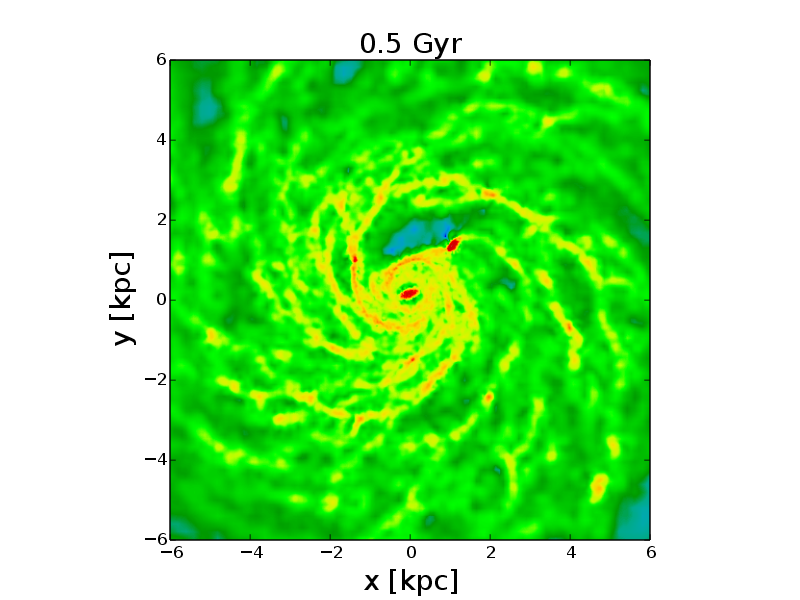}
  \hspace{-1.2cm}
  \includegraphics[width=0.36\textwidth]{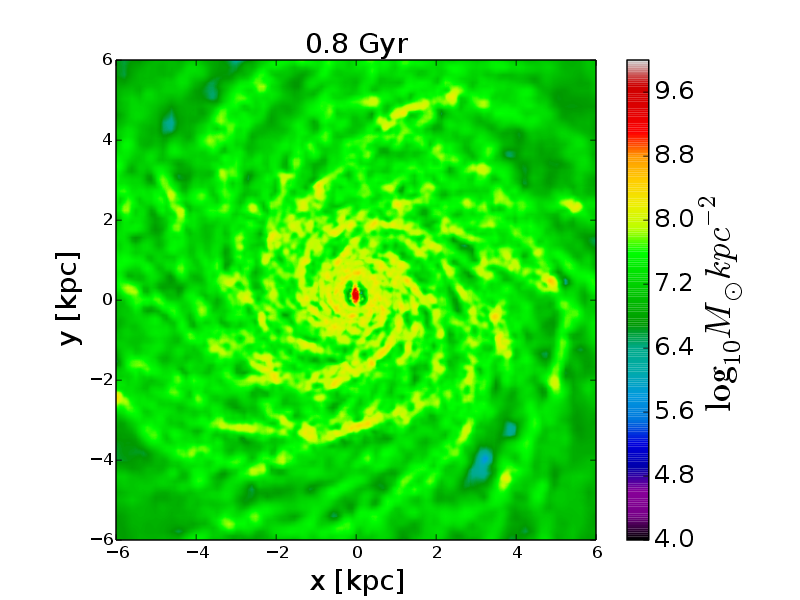}
  \caption{The evolution of the gas density map (from left 0.3, 0.5, 0.8 Gyr) for the galaxy in run 20. It has the same structural parameters ($c=10$, $v = 150$ km/s and $f_g = 0.3$) of the galaxy in run 16 and it evolves with feedback in the same way, but using GDSPH rather than standard SPH. Less clumps form and they survive for only $0.4$ Gyr (see for comparison the top 
panels of Figure \ref{fig:gas_density_noFB_FB}, especially times $0.3$ and $0.5$ Gyr, to compare the number of clumps).}
  \label{fig:newSPH}
\end{figure*}

Some of the runs were repeated with and without metal thermal diffusion (MTD), others with the new GDSPH method. Diffusion and GDSPH are designed to avoid artificial surface tension in SPH, leading to a smoother, less noisy flow, which allows mixing of fluid phases and suppresses spurious clumps generated by numerical artefacts rather than gravitational instability (\citealt{H13}, \citealt{P12a} and \citealt{P12b}). The results of such runs should thus be regarded as more accurate relative to standard ``vanilla'' SPH. First of all, inspection of Table \ref{table_runs} shows that both MTD and GDSPH lead to a weakening of the fragmentation. Clumps do form, but they disappear really quickly, sometimes in only one orbit ($\sim 0.1$ Gyr), becoming a transient phenomenon that has a negligible effect on disc evolution. In particular we run three simulations (runs 3, 20 and 21) with  GDSPH and we found that in two cases we have clumps formation (in a low number, 1-2), even with metal cooling switched on, while in the case with only feedback, run 20, we still have the same results of run 16, even if here clumpy phase lasts 0.1 Gyr less and the number of clumps is lower than in run 16 (see, for comparison, the top panels of Figure \ref{fig:gas_density_noFB_FB} and Figure \ref{fig:newSPH}).
We observe in general more moderate fragmentation as fewer clumps form and the disc stabilises sooner, while fragmentation happens even in the presence of metal cooling. Hence we conclude that the effect of GDSPH on fragmentation
does not show a systematic trend. There is only a slight tendency to have longer lived clumps in standard SPH, which
can be an effect of the artificial surface tension present in that case.
We have not carried out a high resolution version of these runs, but the discussion of the previous subsection leads us to think that, at higher resolution, fragmentation could be reduced further.
%
%
%
\subsection{Analysis and interpretation of clump masses and sizes}\label{Clumps}
\begin{figure*}
 \includegraphics[width=0.32\textwidth]{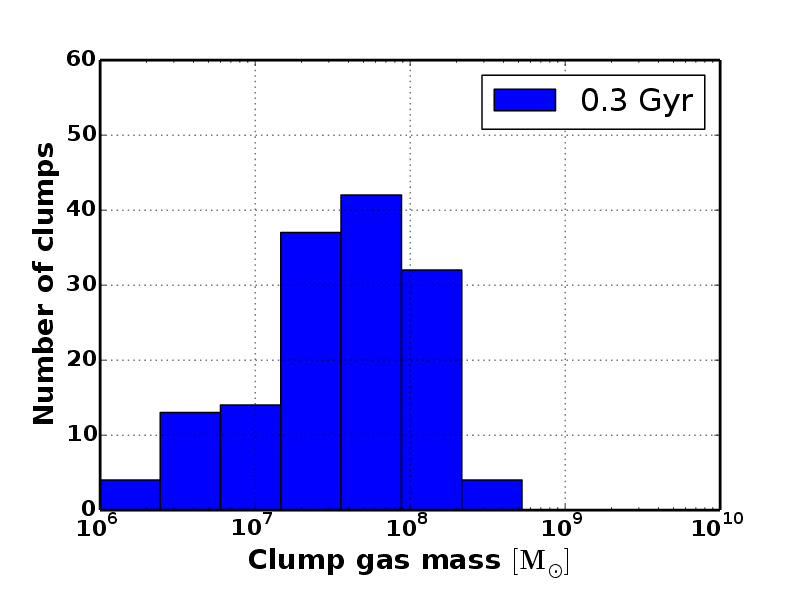}
 \includegraphics[width=0.32\textwidth]{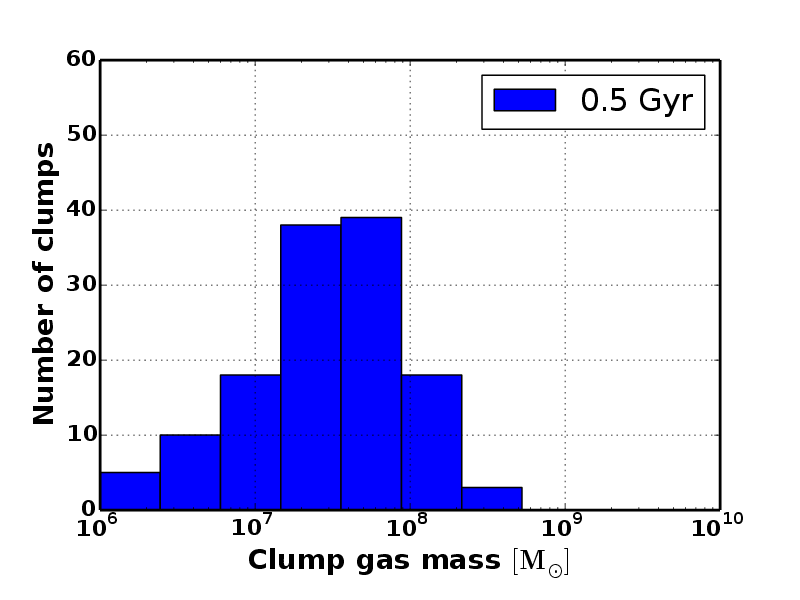}
 \includegraphics[width=0.32\textwidth]{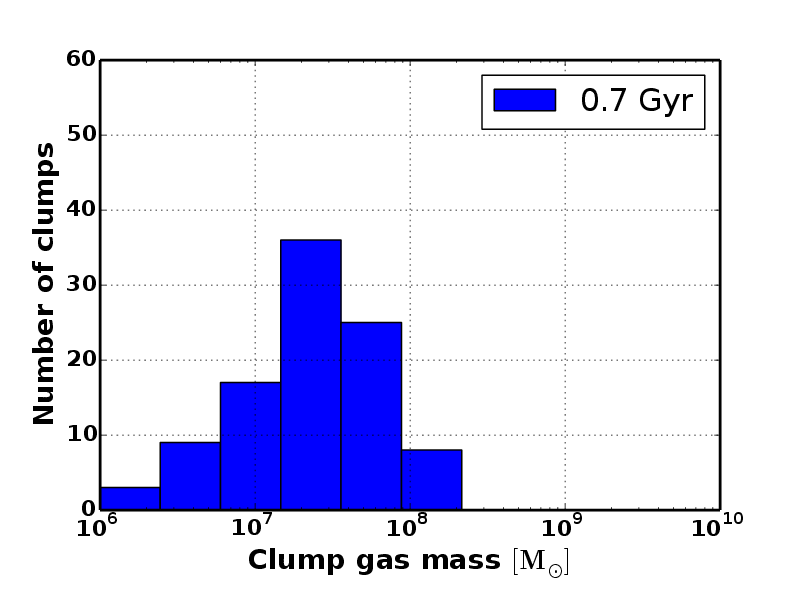}
 \includegraphics[width=0.32\textwidth]{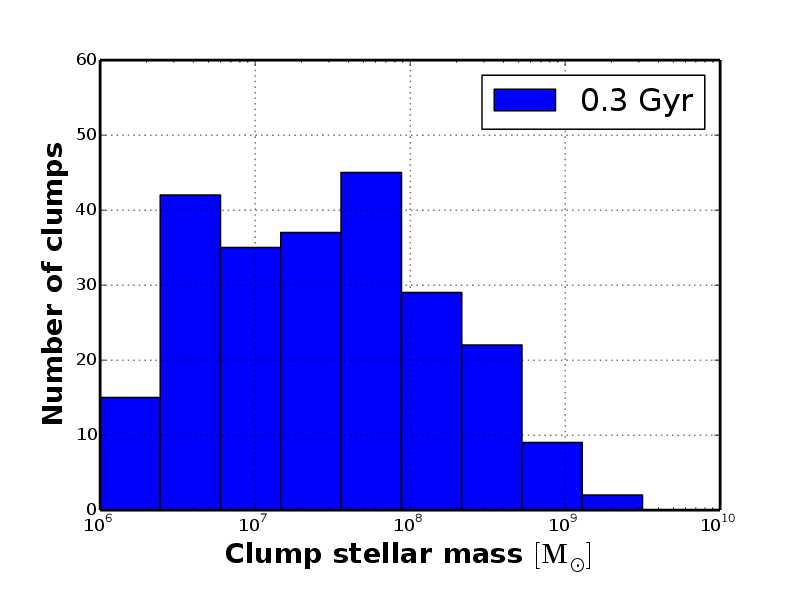}
 \includegraphics[width=0.32\textwidth]{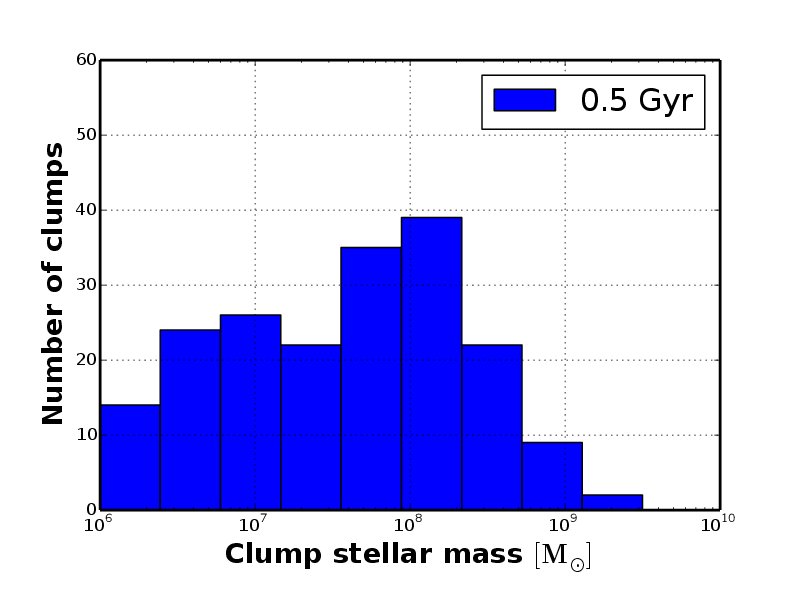}
 \includegraphics[width=0.32\textwidth]{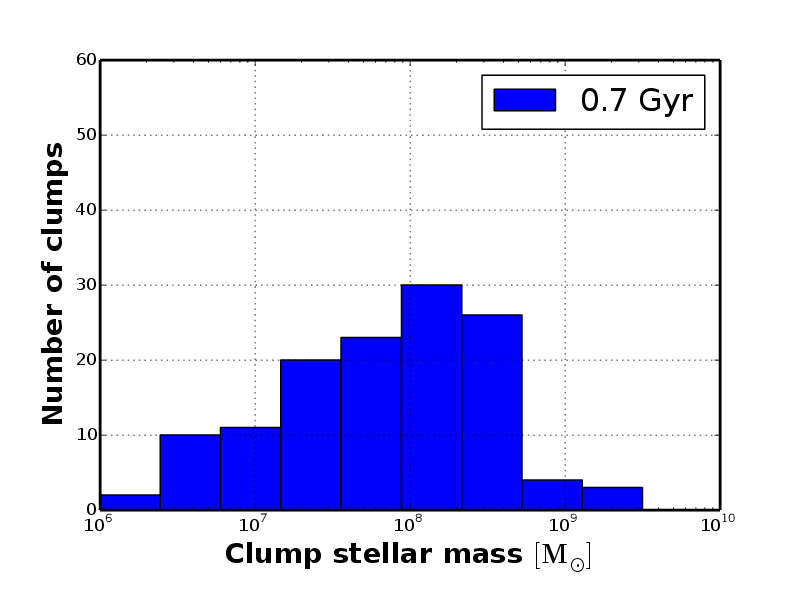}
  \caption{The histograms of the clump mass distribution as a function of time. In the top panels the gas mass is shown, while in the bottom panels there is the stellar mass. Clump masses are computed by stacking the results of all the simulations in order to have a significant statistical sample. $0.3$ Gyr corresponds to the time at which clumps are well detectable in all runs. As one can see looking at the top panels, most of the clumps form with mass $\sim 5 \times 10^7 M_{\odot}$. Note that in some simulations clumps formation happens earlier, already at $150-200$ Myr, but we chose to show $0.3$ Gyr since at this point clumps have formed in all runs that fragment, hence one can consider the values shown at $0.3$ Gyr as an upper limit for the mass at the formation time, because in some runs clumps have already grown via mergers. Looking at the bottom panels, one can see that many clumps are already made only by stars, probably due to the rapid conversion of gas to stars.}
  \label{fig:hist_clumps_evolution}
\end{figure*}

\begin{figure}
 \includegraphics[width=0.49\textwidth]{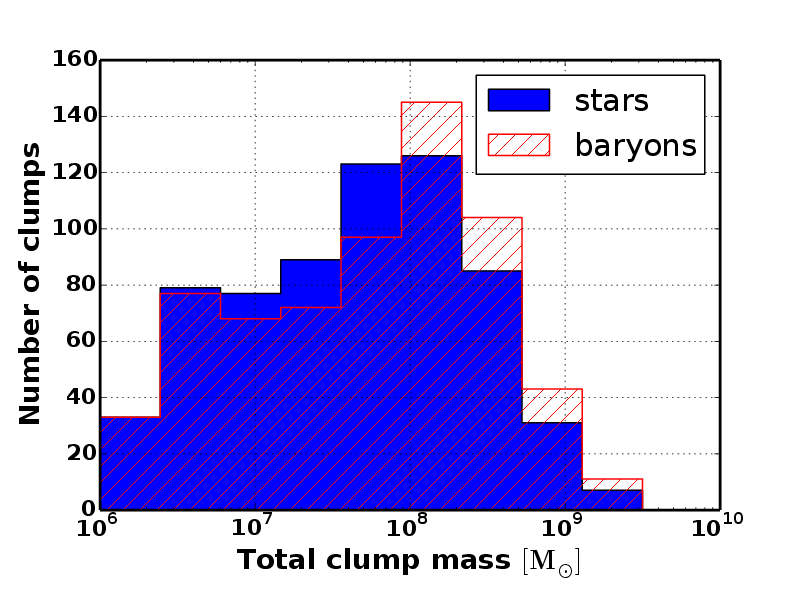}
\caption{The total clump mass (gas and stars) distribution obtained by further stacking of the distribution obtained at different times (see Figure \ref{fig:hist_clumps_evolution}). This should be interpreted as a typical mass distribution for galaxies undergoing fragmentation until 1 Gyr.}
  \label{fig:hist_clumps_total}
\end{figure}

\begin{figure}
\includegraphics[width=0.49\textwidth]{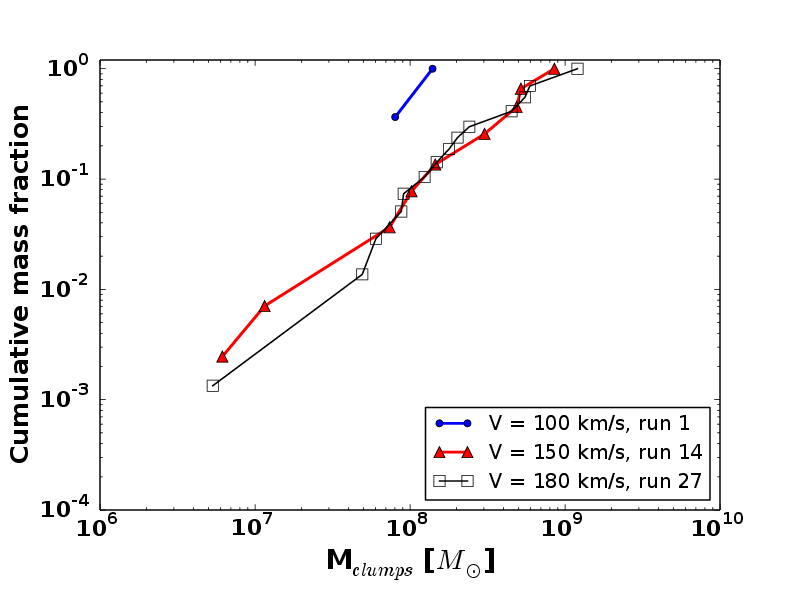}
\caption{The cumulative clump stellar mass (in y-axis the ratio of cumulative clump mass to total clump mass) computed at 1 Gyr for three significant runs (1, 14, 27), that differ only by the virial velocity (i.e. the total mass), shown in the legend.}
  \label{fig:cumulative_mass}
\end{figure}

In this work we have studied in detail the masses and sizes of the clumps, obtained in our runs, and have applied results previously derived for protostellar and protoplanetary discs, in order to understand our results beyond the simple framework of the conventional Toomre instability.\\
We begin by noting that previous simulation work has claimed that clumps in massive high redshift galaxies are as large as
$\sim 0.5-1$ kpc and have typical masses in the range $\sim 10^8-10^9$ $M_{\odot}$, hence too massive and dense to be disrupted by winds generated by feedback \citep{C10, M14}. As it will be evinced by the analysis presented in this section, in our simulations clumps have typically smaller sizes, their diameter being $200-500$ pc, and their typical gaseous and stellar mass is in the range $10^7 - 10^8 M_{\odot}$, although some clumps can reach stellar masses of $\sim 10^9$ $M_{\odot}$ after nearly 1 Gyr of evolution, before being accreted onto the central bulge (Figure \ref{fig:hist_clumps_evolution}).
Note that the mass resolution is $\sim 10^4-10^5$ $M_{\odot}$ as shown in Table \ref{table_resolution}, hence clumps with mass $\sim 10^7-10^8$ $M_{\odot}$ are extremely well resolved (by several SPH kernels as one kernel contains 32
particles).
We decided to divide the analysis and consider separately the gaseous and stellar clump masses since this help to understand the physical processes affecting clump formation and evolution, and because most previous observational and theoretical work has focused on one of the two (predominantly the stellar mass). 
In particular, since our galaxies do not accrete gas from the environment and the gas is depleted by star formation, we consider the gaseous clump mass at the beginning of our simulations and the stellar clump mass at the end, when clumps accreted mass by merging each other.
Since the average number of massive clumps (above $10^8$ $M_{\odot}$) per galaxy is too small (10 in the case with more fragmentation, run 27) to make significant statements, we compute the mass distribution by stacking the results from all the simulations in order to have a significant statistical sample. \\
In order to obtain the histograms shown in Figure \ref{fig:hist_clumps_evolution}, we had to identify the clumps. We used the group finder SKID \citep{S01}, that identifies gravitationally bound groups in N-body simulations. It basically groups particles that satisfy a certain cut criterion, i.e. to have density above a minimum value of $500$ atoms/cm$^3$ and temperature below a maximum value, $T=10^5$ K.\\ 
The mass distribution resulting from stacking is shown in Figure \ref{fig:hist_clumps_evolution} at three different times. First of all, at 0.3 Gyr, when clumps can first be robustly identified with SKID in all the fragmenting discs, at 0.5 Gyr, when there is still enough gas to form new clumps, and at 0.7 Gyr, when more than $70\%$ of the gas has been already consumed in star formation and the mass budget of clumps is clearly dominated by stars rather than gas.
Note that clumps start to condense out of the gas phase as early as $150$ Myr in some simulations. This, combined with the rapid initial star formation burst (Figure \ref{fig:SFR_FB_vs_MC}), explains why at $0.3$ Gyr there are already massive stellar clumps present in the mass distribution of Figure \ref{fig:hist_clumps_evolution}. Another important feature that can be evinced by inspection of Figure \ref{fig:hist_clumps_evolution} is that the gaseous mass of clumps peaks at $3-5 \times 10^7$ $M_{\odot}$, which suggests this is a characteristic mass of clumps set by the fragmentation process itself (see discussion of characteristic clump mass later in this section). \\
We also find (Figure \ref{fig:hist_clumps_evolution}) that this characteristic fragmentation mass scale is roughly independent on the mass of the disc as it is determined by the local conditions inside overdense spiral arms, a result that is discussed below in the context of a definition of the initial fragmentation mass that is alternative to the conventional Toomre mass. Instead, the final of the clumps does depend on the disc mass, since it is determined by evolutionary process that are enhanced at higher disc mass.\\
In particular, by tracing the particles back in time we have also checked that the clumps in the high mass tail of the distribution at $0.3$ Gyr, especially those already dominated by stars, are those that have undergone already mergers and that formed as early as $100$ Myr. The bottom panels show, indeed, in contrast to the gas mass,  the mean stellar clump mass increases with time leading to massive clumps exceeding $3 \times 10^8$ $M_{\odot}$ with a tail around $10^9$ $M_{\odot}$. Using particle tracing to follow individual clumps, we verified that the latter high mass tail is mainly the result of multiple clump-clump mergers. Hence it does not reflect the mass scale set by the fragmentation process, rather is a product of dynamical evolution. The most massive clumps, with masses well above $10^8 M_{\odot}$, which would classify as giant clumps according to definitions used in the literature, arise in
the most massive discs simply because more clumps form in those, thus increasing the likelihood of multiple mergers, and also because a larger gas reservoir is available for accretion.\\
We caution, though, that such high mass tail is not statistically very significant due to the low number of clumps belonging to it and because it develops relatively late, while observed  clumps are ``young" (estimated ages $1-400$ Myr, see \citealt{Adamo13}).
However, by mass the largest clumps can dominate the overall mass budget of clumps as a fraction of the disc mass, especially in the most massive discs (see Figure \ref{fig:cumulative_mass}). We have also measured the maximum contribution in stellar mass that massive clumps give at any time over the total stellar mass of the disc and we have found that this is at most $20\% $ in the most massive discs (typically around $5-12 \%$ in the whole simulation sample).
In Figure \ref{fig:hist_clumps_total} we show the total clump mass (gas and stars) distribution got by further stacking the results obtained at different times (until 1 Gyr), which should be interpreted as a typical time-averaged mass distribution. Note that the peak value is $\sim 2 \times 10^8 M_{\odot}$, in agreement with what found in cosmological simulations \citep{M14} and observations \citep{Swinbank09, Livermore15}. Finally, we have checked that the distribution of clumps size and mass is not affected by varying resolution.\\
Since the interpretation of the masses and sizes of clumps in previous works have always been conducted in the context of standard Toomre instability, hence treating clumps as linear perturbations, we will now try to answer these two basic key questions:
\begin{enumerate}
	\item What is the clump mass we expect from the Toomre instability analysis?
	\item Is the mass derived via the Toomre analysis a sensible prediction for the clump mass, and do we understand the characteristic masses of the clumps found in the  simulations in light of Figure \ref{fig:hist_clumps_evolution}? 
\end{enumerate}
These questions of course concern the initial fragmentation phase, so they are relevant to understand the {\it initial masses} of the clumps. Their masses and sizes can still evolve via mass accretion from the surrounding disc, mass-losing tidal effects from disc tides and clump-clump gravitational encounters, and clump-clump mergers. All these effects are of course taken into account automatically in our analysis shown in Figures \ref{fig:hist_clumps_evolution}, \ref{fig:hist_clumps_total} and \ref{fig:cumulative_mass}, but will not be taken into account in the following discussion, since they concern a later evolutionary phase that is characterised by strongly nonlinear effects.\\
\\
The Toomre mass is defined (e.g. \citealt{N06}) as:
\begin{equation}
M_T = \pi \Sigma \left(\frac{\lambda_T}{2}\right)^2
\end{equation}
where $\lambda_T$, the Toomre wavelength, is the maximum unstable radial wavelength for a given local sound speed $c_s$ and (axisymmetric) background surface density $\Sigma$ and can be written as:
\begin{equation}
\lambda_T = 2 \frac{c_s^2}{G\Sigma}
\end{equation}
assuming $Q = 1$ and the epicyclic frequency $k = \sqrt{2} \Omega$, with $\Omega$ angular velocity (we checked that this is a good approximation in our simulations). This wavelength should measure the maximum scale of fragmentation, hence should be comparable to the maximum mass of clumps soon after they form. The most unstable Toomre wavelength, which is in Equation (\ref{eq:most_unstable}), should instead be close to the characteristic clump mass soon after the onset of fragmentation, which is $\sim 5 \times 10^7$ $M_{\odot}$, based on the shape of the mass function shown in Figure \ref{fig:hist_clumps_evolution}. \\
\begin{figure*}
 \includegraphics[width=0.49\textwidth]{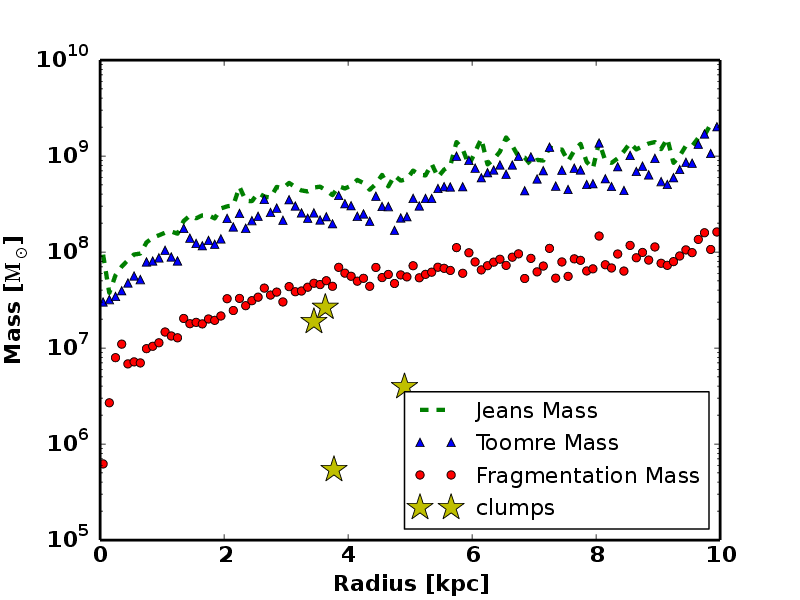}
 \includegraphics[width=0.49\textwidth]{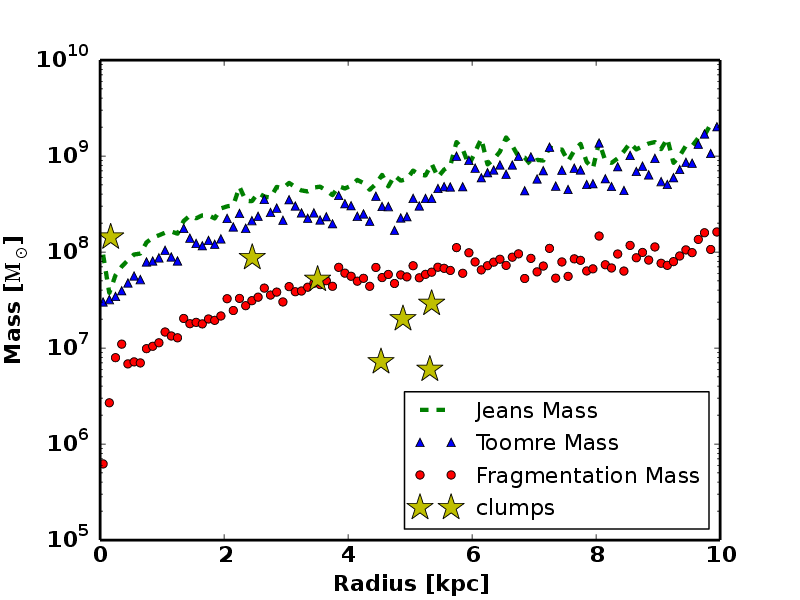}
 \includegraphics[width=0.49\textwidth]{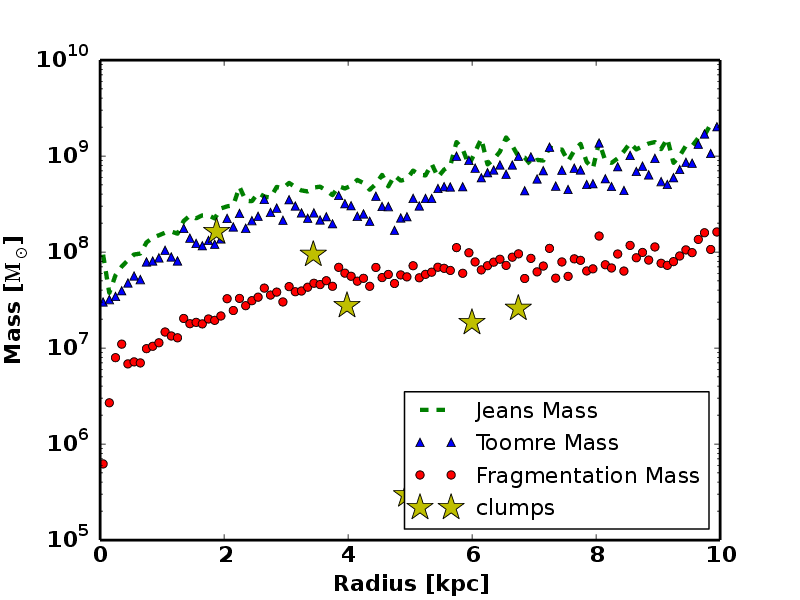}
 \includegraphics[width=0.49\textwidth]{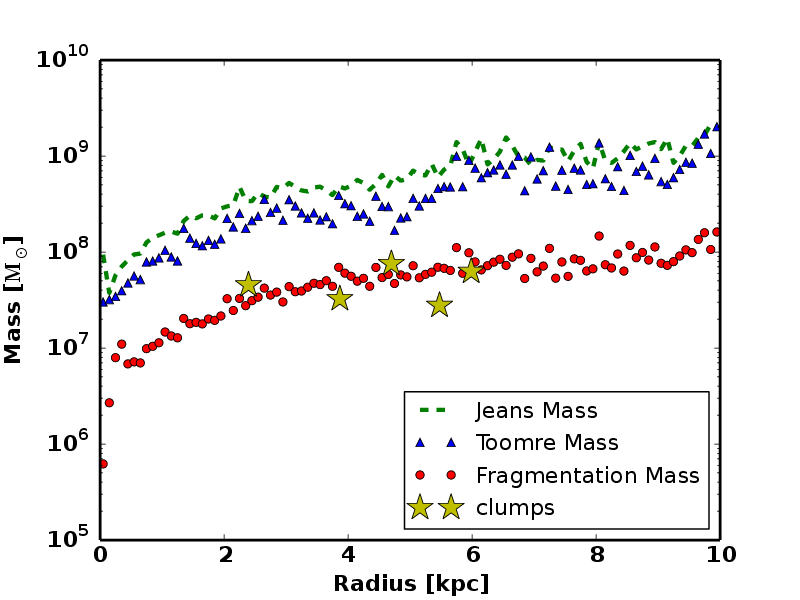}
  \caption{Masses of clumps around $300$ Myr, when they form, and their later evolution for run 12. Note, looking at the figure on the top left panel (initial time for the clumpy phase), that clumps form with mass $\sim 2-4 \cdot 10^7 M_{\odot}$ and then, looking at next time steps, they can increase their mass until $\sim 2 \cdot 10^8 M_{\odot}$ 
only by merging with other clumps or accreting gas. We show the different characteristic mass scales discussed in Section \ref{Clumps}. Note the very good match to the initial clump masses provided by the local fragmentation mass defined in Section \ref{Clumps}}
  \label{fig:comparing_mass_cc6_v150_gf03}
\end{figure*}
The surface density is usually computed from the Toomre parameter using $\Sigma = c_s \Omega / (\pi G \langle Q\rangle)$, with $ \langle Q \rangle \sim 1$, where $ \langle Q \rangle$ in this case is the azimuthally average value of $Q$ for the gas only, which, at the onset of fragmentation, when the disc is still gas-rich, is very similar to the value obtained using the two-fluid approach. Note $c_s$ is strictly the thermal sound speed for the remainder of this discussion, since regions that do fragment are by construction those in which the turbulent support is locally negligible (see also Section \ref{Intro}).\\ 
We recall that the Toomre mass and wavelength are obtained using linear perturbation theory for local axisymmetric perturbations in an isothermal thin disc, while in our case discs are thick, the instability is global and non-axisymmetric (spiral modes form first, then clumps collapse inside the arms), and the fluid enters the nonlinear regime before clumps form. We will not attempt to account for all these limitations, since the simulation already does that by construction, but we will apply a model that overcomes at least partially the limitations of the Toomre analysis while remaining still simple and providing an analytical estimate of the characteristic fragmentation mass.
An important aspect of non-linearity is that in the Toomre mass definition only the azimuthally averaged velocity of the unperturbed disc appears, even if the velocity is not the same everywhere in spiral unstable disc, i.e. spiral wave regions are already disturbed and have velocity gradients. It would thus seem that fragmentation is better described assuming the background gas is traversed by a spiral density wave, as in \citealt{B10}. These authors, who studied fragmentation in protoplanetary discs, use length scales and surface density perturbations that are appropriate for spiral arms. The radial extent of the fragmenting region in the non-axisymmetric conditions can be estimated using the results of \citealt{D08}, who used the virial theorem to show that a disc, under isothermal conditions, is most susceptible to fragmentation within a region $\delta r$ from the corotation of a spiral wave.
In this treatment the velocity is not the same everywhere as in Toomre theory since, assuming the gas is cold and therefore pressure is negligible, it is the velocity gradient that prevents the gas from collapsing. This should define the size of clumps at the onset of fragmentation. There will be pressure gradients developing as the clump begins to collapse: in fact it may resist the collapse to some extent, but if the gas can cool efficiently, as it is the case in galaxy discs even more than in protoplanetary discs, this should be a secondary effect. Indeed, using this approach \citet{B10} could predict quite well the masses and sizes of clumps in their protoplanetary disc simulations, which were severely overestimated by the Toomre wavelength. Here we follow the same approach.\\
We use the formula for the initial clump mass given in \citet{B10}: 
\begin{equation}\label{eq:Mf}
M_f = 2 \cdot \lambda_T \frac{\Sigma c_s}{k f_{grav}}
\end{equation}
where $k$ is the epicyclic frequency (and, as we have already said, we checked that a good approximation in our case is $k = \sqrt{2} \Omega$, with $\Omega$ angular velocity) and $f_{grav}$ is a free parameter, taking into account self gravity effect of all the system, stars and gas (in all our cases it is equal to unity). Note that in the \citealt{B10} framework the region that will collapse is independent on the disc surface density (substituting Equation (\ref{eq:lambda}) in Equation (\ref{eq:Mf}), $\Sigma$ clearly cancels out), because it is assumed that this was already high enough to trigger the instability (in other words, the disc has already reached a low Toomre parameter value), instead it is only the material with locally coherent velocity field that continues to collapse, while the rest is sheared away. \\
\\
Now we use the Equation (\ref{eq:Mf}) to compare the resulting mass, which we simply call ``local fragmentation mass'', with the expected local Toomre mass and with the mass of our clumps. In Figure \ref{fig:comparing_mass_cc6_v150_gf03} we show a typical example of such comparison from run 12, done at different times, reporting also the local Jeans mass for completeness.\\
One can notice two important facts. First, our local fragmentation mass estimate can predict fairly well the masses of the clumps at formation time, in contrast to the 
conventional Toomre mass, which in the best case overestimates the masses of clumps by a factor of 5-6. This satisfactorily confirms earlier results obtained for fragmentation in protoplanetary discs \citep{B10}. Clearly this mass estimate captures much better the physics of the fragmentation process.
Second, the conventional Toomre mass yields an estimate in quite good agreement with the mass of the clumps at later times, but at this point clumps have grown in mass through processes unrelated to fragmentation such as gas  accretion and mergers, hence the match is purely coincidental. For fairness, we note that if one uses the most unstable wavelength, namely the wavelength of the fastest growing mode, to define the Toomre mass, an estimate for the clump mass more in agreement with ours can be obtained.  \\
\subsection{Clumps and bulge formation}\label{Bulge}
\begin{figure}
 \includegraphics[width=0.49\textwidth]{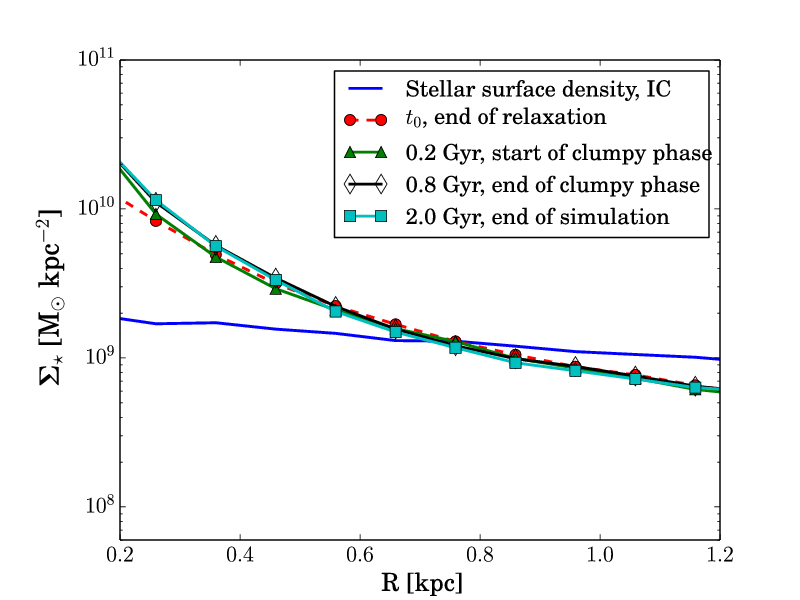} 
 \caption{The stellar density evolution inside $2$ kpc for run 16, highlighting that a small bulge is already assembled before the clumpy phase, during which the central surface density grows only by a factor $\sim 2$.}
 \label{fig:bulge_ok}
\end{figure}
\begin{figure*}
 \includegraphics[width=0.49\textwidth]{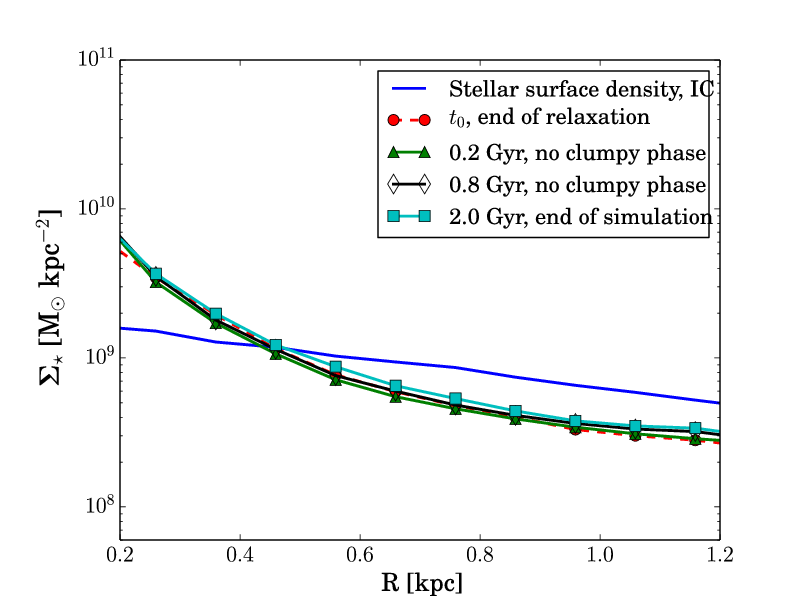}
 \includegraphics[width=0.49\textwidth]{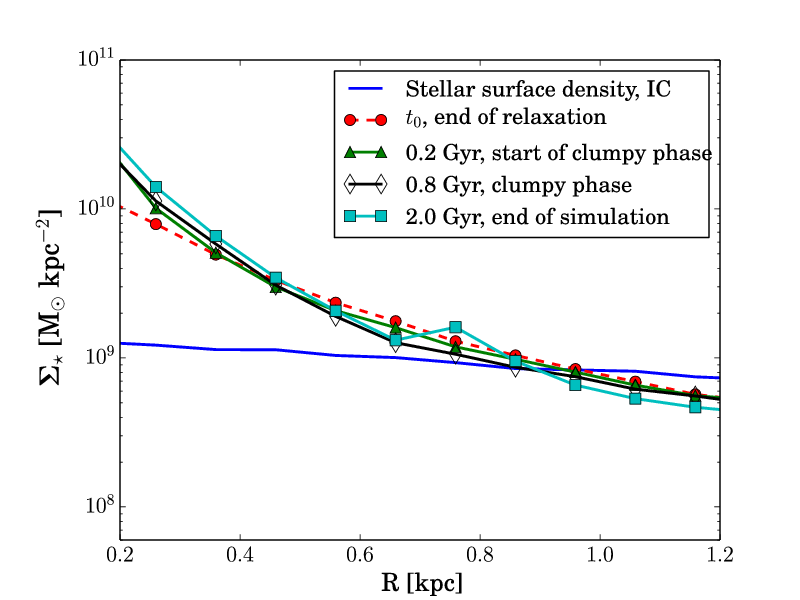}
 \includegraphics[width=0.49\textwidth]{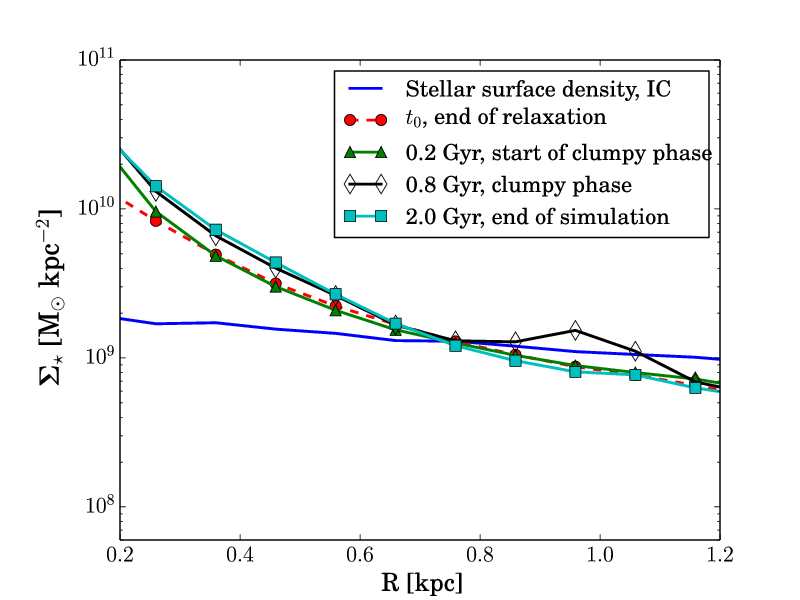}
 \includegraphics[width=0.49\textwidth]{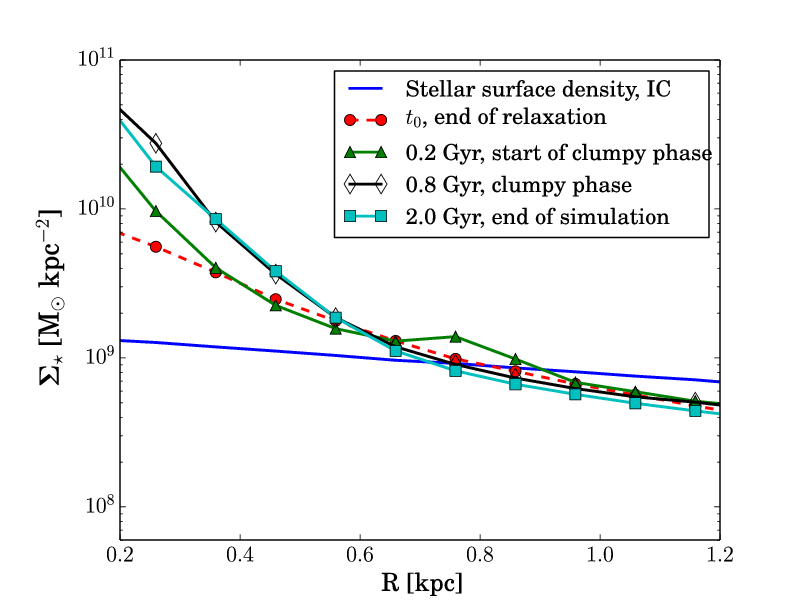}
 \caption{The stellar density evolution for runs 9 (left, top panels), 12 (right, top panels), 17 (left, bottom panels) and 25 (right, bottom panels). In all the runs it is clear that bulge grows more before the clumpy phase, 
except in run 25, that yields a particularly vigorous clump formation phase (see the text for more details) owing to the combination of both higher circular velocity and high gas fraction. Note also that run 9 never form clumps, yet a central bulge grows anyway.}
 \label{fig:bulge}
\end{figure*}
\begin{figure*}
 \includegraphics[width=0.49\textwidth]{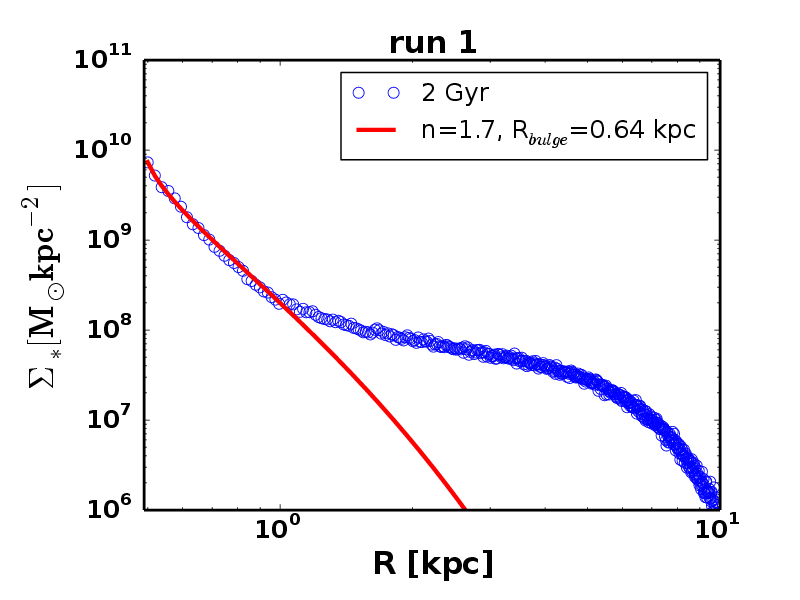}
 \includegraphics[width=0.49\textwidth]{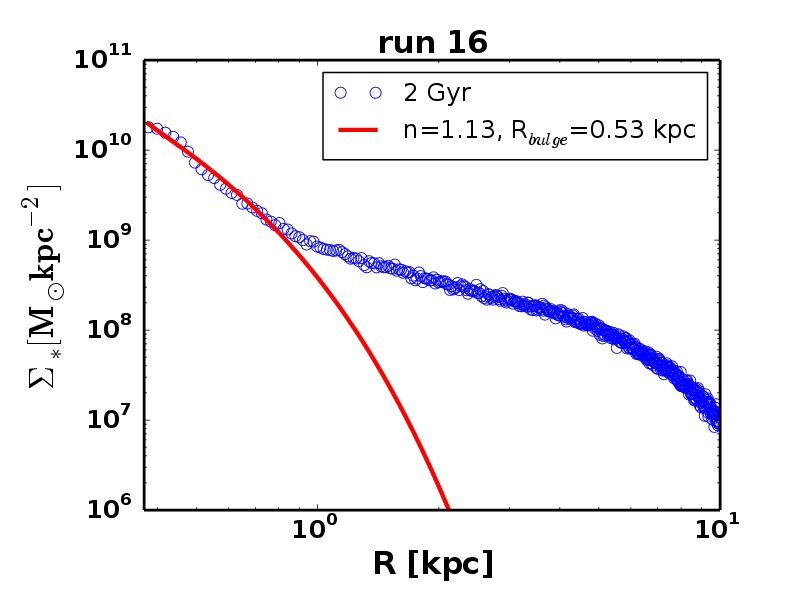}
 \includegraphics[width=0.49\textwidth]{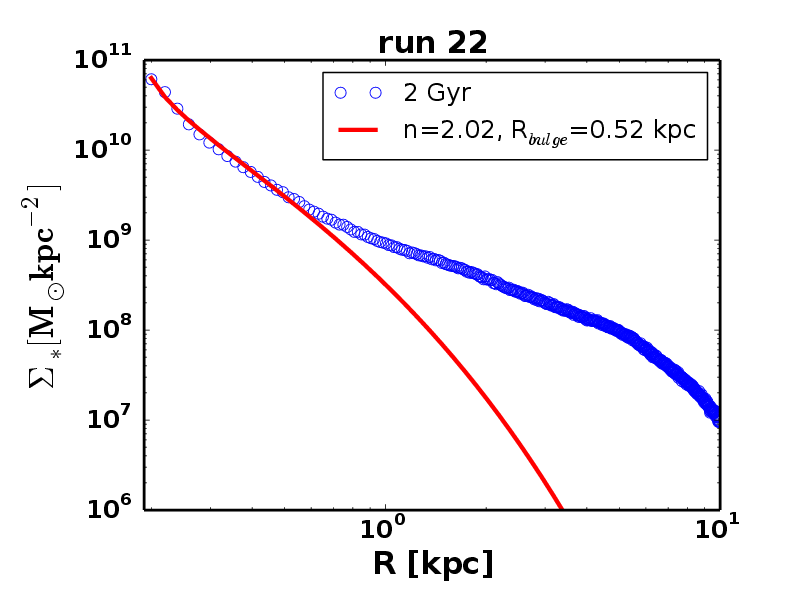}
\includegraphics[width=0.49\textwidth]{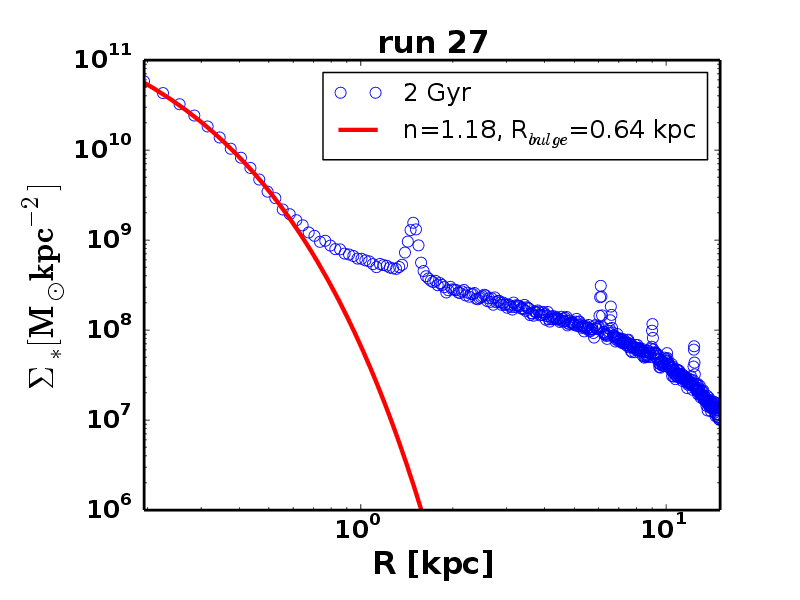}
 \caption{The stellar density evolution (blue circles) at the final step (2 Gyr) for runs 1, 16, 22 and 27 (from left up panel, to right down panel), fitted with a Sersic law (solid red line). The resulting Sersic index, $n$, and the size 
of the bulge, $R_{bulge}$, are shown for each case in the legend.}
 \label{fig:sersic_index}
\end{figure*}

In this section we discuss how clumps can contribute to the growth of a central stellar bulge in our simulations.\\
Bulges can form via mergers at high redshift, via secular evolution, due to the presence of a stellar bar bringing gas to the centre and turning into a spheroid via the buckling instability or resonant thickening \citep{R91, D05, Deb06}, or via gravitational instability, producing giant gas clumps that spiral-in towards the centre via dynamical friction \citep{N99, C10, IS12}. Recently it has been shown that disc galaxies forming in cosmological simulations
can grow a bulge via bar instabilities, that are dynamically excited by satellites and minor mergers at high redshift, essentially an accelerated and more dynamical version of the picture proposed in \citet{Deb06}.\\
Depending on their formation path, bulges develop different characteristics. Classical bulges probably form via mergers and have high Sersic index ($n>2$) profiles. Pseudobulges, instead, have disc-like density profiles and kinematics, low Sersic index ($n<2$) and probably originate by internal non-axisymmetric instabilities (secular or dynamical).
Finally, classical bulges have older stellar populations than pseudobulge.
Analysis of the galaxies in the ARGO simulation supports this picture, but also shows that even bulges from minor mergers can have low Sersic indexes as those classified as pseudobulges. In particular,  at $z \sim 3$ the central galaxy in ARGO, that is the most massive and has undergone a number of major mergers, is the only one that shows a Sersic index high enough to be compatible with a classical bulge \citep{F15}. However, \citet{Cev15} find that classical bulge can form by both violent disc instabilities and mergers.
Bulges arising from collective mergers and spiral-in of clumps in simulations have been shown to have common characteristics to classical and pseudobulges \citep{IS12}. However, such previous works \citep{C10, IS12} produced giant clumps in their simulations ($M > 10^8 M_{\odot}$), and concluded that the clumpy phase can be a major stage of bulge formation. 
Not surprisingly, in our simulations this does not happen, because clumps have moderate masses for the most part, and contribute only $10-15\%$ of the total mass budget of the disc in the most massive simulations.
For simplicity we do not attempt to measure bulge-to-disc ratios, rather we use the increase of central stellar surface density within the inner 0.5-1 kpc, where the stellar profile becomes significantly steeper (see Figures \ref{fig:bulge_ok} and \ref{fig:bulge}) as a measure of the bulge growth. In this sense a moderate bulge forms by the end of the relaxation phase in all simulations, via angular momentum transport from weak spiral waves. However, it evolves little after fragmentation begins, in the sense that the increase of surface density is modest.
In fact, inspection of Figures \ref{fig:bulge_ok} and \ref{fig:bulge} shows that the surface density grows by a factor of $\sim 6 $ in absence of clumps, while after fragmentation ensues it increases by less than a factor of 4 until the end of the simulations. 
The late mass accumulation at the centre is also not only provided by sinking clumps since simultaneously a strong spiral pattern persists, which would surely contribute to the mass inflow by torquing the gas. Overall there is indeed quite some theoretical consensus, supported by a variety of simulation work, that gas inflows are the crucial ingredient for bulge formation \citep{Krumholz10, Cacciato12, Dekel13, Forbes14, Dekel14, Zolotov15}.
A peculiar case is that of run 25 (right, bottom panels in Figure \ref{fig:bulge}), in which the central surface density grows before the clumpy phase by a factor $\sim 5$, but it also grows during the clumpy phase by a similar factor. However this is an especially favourable case for clump formation among our runs, due to both high gas fraction and high virial velocity.
Yet we note that, when compared with observations of clumpy galaxies, which are biased towards massive and very gas rich galaxies, the structural properties of this galaxy model are not extreme \citep{Wisnioski15}.\\
We emphasise that the presence of a small bulge already before fragmentation should not be regarded as an artifact of our simulations. 
Spiral or bar instabilities will easily be excited during the chaotic early stages of galaxy formation (see e.g. \citealt{G13} and \citealt{F15}) via minor mergers and tidal interactions, and will inevitably drive gas inflows likely before the galaxy becomes massive enough to be prone to disc instability.
The strength of non-axisymmetric instabilities will depend on the self-gravity of the discs, hence we expect it to be stronger in more massive discs. This is indeed what we see in the ARGO simulation, 
where massive galaxies ($V_{max} > 150$ km/s) acquire a steeper inner stellar profile early (at $z = 4-5$), while low mass galaxies ($V_{max} < 100$ km/s) remain pure exponential discs with no bulge for much longer \citep{F15}. Hence, for massive galaxies such as those studied here it is actually more realistic to start from a configuration which already includes a small bulge.
Finally, it is worth noting that at the end of the simulation our galaxies exhibit a Sersic index $n\leq 2$ (see Figure \ref{fig:sersic_index}), typical of pseudo-bulges, in good agreement with the galaxies in the ARGO simulation, which do not undergo fragmentation. This is true also in the case of run 27, which hosts the most massive galaxy/disc in our sample.
Once again this is consistent with the notion clumps have moderate masses in our simulations, hence they cannot contribute appreciably to the mass budget of the bulge, and therefore the inner surface density slope has to converge to the slope achieved via gas inflows driven by ordinary non-axisymmetric instabilities in absence of clumps.\\
In conclusion we do not have an analog of the subset of classical bulges formed in the cosmological simulations of \citet{Cev15} without a significant contribution of mergers to their mass assembly. This might either reflect the fact that gas inflows in unstable discs can be much stronger in fully cosmological simulations relative to our isolated setups, thus steepening further the bulge profile, or that mergers, despite they do not carry most of the mass to the bulge, are still instrumental in enhancing such gas inflows.
\section{SUMMARY AND DISCUSSION}\label{Discussion}
In this paper we have revisited the susceptibility of gas-rich galactic discs to large-scale fragmentation driven by gravitational instability as well as the typical masses and sizes of clumps eventually produced by fragmentation. We have thus focused only on in-situ clump formation within gas-rich hi-z discs, whereas recent work employing cosmological simulations has shown that such in-situ clumps would co-exist with an ex-situ population produced by minor mergers and galaxy interactions \citep{Mandelker14}.
Previous numerical works studying clumpy galaxies have employed both cosmological simulations and isolated disc simulations. For the systematic study presented in this paper we have decided to use isolated galaxy simulations because 
these allow us to check carefully the effect of every single parameter of galaxy structure (concentration, virial velocity/mass, disc gas fraction), to deeply explore the effect of sub-grid physics, and to reach higher resolution compared to cosmological simulations. 
In passing we note that cosmological simulations, published claiming resolutions similar to those of our standard runs, actually achieve that resolution only for the deepest levels of refinements using  AMR techniques \citep{C10, M14}, while for smaller scale astrophysical discs it has already been shown, using both SPH and AMR codes, that the {\it initial resolution} before the onset of fragmentation is crucial in order to capture fragmentation correctly \citep{D07, MG08}.\\
The bottom line of our investigation is that the formation of giant clumps with masses much above $10^8$ $M_{\odot}$ via disc fragmentation is not a common occurrence when a large sample of galaxies spanning an order of magnitude in mass and a wide range of structural properties and sub-grid physics is considered.
The typical gaseous mass of clumps is in the range $10^7-10^8$ $M_{\odot}$ for gas-rich galaxy discs with masses in the range
$10^10-{10}^{11} M_{\odot}$. which is well understood since it is close to the initial fragmentation mass inferred with the notion of local collapse inside self-gravitating spiral arms explained in Section \ref{Clumps}. The latter fragmentation mass is typically 5-6 times smaller than the Toomre mass.
At late times the stellar and baryonic masses (gas +stars) of clumps are skewed to somewhat larger values, peaking at $\sim 10^8$ $M_{\odot}$, reflecting the role of gas 
accretion and, especially, clump-clump mergers.
Galaxy discs at the high mass tail of the the distribution observed at $z \sim 2$ (e.g. \citealt{Wisnioski15}) are the most favourable hosts for giant clumps. Indeed a typical example is run 27, which at late times has clumps exceeding $10^9$ $M_{\odot}$. 
In this high mass regime our results on the mass spectrum of clumps are consistent with those of previous simulations that indeed focused on very massive
discs (e.g. \citealt{C10, M14}.
However, we argue our galaxy models with $V_{vir} = 100-150$ km/s, which yield $V_{max} \sim 180-250$ km/s, are more representative based on 
the recent observations by \citealt{Livermore12, Livermore15}.\\
The role of feedback is very important since it affects the degree of fragmentation, hence the available mass in the clumpy phase and thus, indirectly, how high the clump mass can become through clump-clump mergers. Feedback thus acts to stabilize the disc against persistent fragmentation but does not affect individual clumps once they form. 
The initial mass of clumps at fragmentation is also independent on feedback. This means that in our case feedback does not suppress systematically lower mass clumps, although we caution that we lack the treatment of other forms of feedback, such as radiation pressure and ionizing radiation from massive stars, that could destroy the clumps from inside.
Without feedback, clump formation is much more vigorous and gives rise to longer-lasting clumps that can grow until they reach very high masses ($> 10^8$ $M_{\odot}$ before sinking in towards the bulge.\\
An important feature of our simulations concerns the way we prepare ``quiet'' initial conditions, by letting the disc relax adiabatically for 1 Gyr and then cool down from a high Toomre parameter configuration, a procedure that is followed in studies of high-z discs, while being quite commonly adopted for reasons of numerical robustness in studies of small scale self-gravitating astrophysical discs \citep{D07, M04, M07}.
Such studies have shown that gravitationally unstable discs, which are artificially started with Toomre parameters close to unity and no relaxation, lead to numerically amplified fragmentation and larger masses of bound condensations. This is likely one important reason why clumps in our discs are smaller and lighter than in previous work. In addition, relaxation gives rise to the formation of a small central density enhancement in the galaxy, with density profile similar to a pseudobulge, which surely helps to stabilise the inner disc.
Cosmological hydrodynamical simulations show that such central enhancement is seen to form already at $z > 4$ in massive disc galaxies as a result of early gas inflows (\citealt{G13, F15}), hence its presence in our models should be regarded as realistic.
Numerical improvements over standard SPH, such as thermal diffusion and Geometric Density SPH, lead to comparable or weaker fragmentation, depending on the specific run, suggesting that our results are conservative. \\
Since in our simulations typical clumps are smaller and lighter than the giant clumps highlighted in previous work on fragmenting discs, being essentially oversized Giant Molecular Clouds (GMCs), we argue that a sizable fraction of the giant clumps identified in observations could have a different origin, such as from minor mergers or other ex-situ contributions (see e.g. \citealt{Mandelker14}), or may result from blending of smaller subunits that are artificially smeared out owing to lack of resolution.
Indeed, all the measurements of the physical properties of clumps suffer from seeing-limited observations. When performed from ground-based telescopes, they can at best reach an angular resolution of 0.5''-0.8" FWHM, unless the adaptive optics is deployed, which allows to improve the angular resolution to an effective $\sim$0.2" FWHM \citep{G11}. This is hence almost comparable to what can be achieved with the Hubble Space Telescope (0.15" FWHM).
However, such an angular resolution still corresponds to a relatively high physical scale of $\sim 1.3$ kpc at $z \sim 2$ \citep{Guo12}. The only way currently to beat this limit is with the help of gravitational lensing, which leads to a stretching of the lensed object such that physical scales as low as $\sim 100$ pc may be reached at $z \sim 2$ \citep{Jones10, Olmstead14, Livermore15}. Note that insufficient resolution in observations, in addition to prevent from disentangling two nearby clumps, could potentially lead to confuse a young bulge in formation for a massive clump since there is often not enough information on the global galaxy morphology (see \citealt{Livermore15}). Note that the bulge in our simulations is more massive than the most massive clumps formed through disc fragmentation but its structural properties and mean stellar age are quite similar to those of massive clumps.
It is then noteworthy that results on the mass function of clumps are in good agreement with the observed mass spectrum of clumps presented  in \citealt{Livermore15}. These authors, owing to gravitational lensing, focus on rather ``common'' galaxies at redshift $1<z<4$ rather than only on the bright end, and find a cut-off mass for the clump mass equal to $10^8$ $M_{\odot}$ for $1<z<1.5$ and $9 \times 10^8$ $M_{\odot}$ for $1.5<z<3$. This cut-off should be thought as the maximum mass a clump can reach given a certain luminosity function.
Moreover the kinematics of the discs in our simulations, which maintains $v/ \sigma > 1$ throughout the fragmentation stage, seems in good agreement with observations (e.g. \citealt{Wisnioski15}), suggesting that the underlying disc dynamics is in a similar regime and that the disc is relatively undisturbed by clump formation, which argues against very large clumps roaming through it.
Our results do not imply an important role of clumps in galaxy evolution, in particular they suggest that clumps are not a major channel of stellar bulge growth. Indeed in our simulations a small bulge is built up before the formation of the clumps and is not modified significantly as a result of inward clump migration.
This conclusion is at odds with that of recent work on the role of clumps in galaxy evolution, that identifies them as a major channel of bulge formation (\citealt{Bo15}).
\\
While our simulations are partially based on input parameters from cosmological simulation, they are still fairly idealised, since discs are evolved in isolation without accreting mass or interacting with other galaxies. \citet{A09} have emphasised how the formation of giant clumps might be triggered in rapidly accreting galaxies at high redshift. 
The importance of gas accretion has been emphasised also in studies of fragmentation of protostellar and protoplanetary discs into sub-stellar companions \citep{B10, Hey11}, which have shown that a disc, that would be stable in isolation, can fragment if gas is damped on a timescale shorter than the typical orbital time (note however the initial clump mass is still well reproduced by our local fragmentation mass scale (see \citealt{B10}).
Strong tidal perturbations by massive companions might also excite the instability. While these effects are built-in in the ARGO cosmological simulation, our study suggests that the absence of fragmentation in the ARGO galaxies might be due to a combination of factors, such as the relatively low masses and virial velocities of galaxies therein (except the very massive central, which has a circular velocity $V_{vir} > 150$ km/s, all the others have $V_{vir}$ close to $80-120$ km/s at $z \sim 3-4$) and the fact most of them have moderate gas fractions, perhaps due to excessive efficient star formation.
We should be warned that the regime of triggered/tidally excited instability might be different from the regime of instability in isolated discs, mainly because the initial perturbation can be stronger and excite larger scale spiral modes, which may have trigger larger collapsing patches. However, once again work done in star and planet formation has taught us that the outcome can go in an unexpected directions. Disc instability can indeed be stifled rather than enhanced in presence of perturbing companions (e.g. \citealt{Hey11}). 
Nevertheless, in light of our new results, we should also revisit the effect of both gas accretion and external perturbers in the specific context of high redshift discs, but once again with controlled experiments that allow to understand the results in depth. 
One first step in this direction is to understand the role of gas accretion alone, in absence of external perturbers.
To this aim in a forthcoming paper we will present new simulations using cooling halo models (e.g. \citealt{K07, H14},) in which the gas and stellar discs are not initialised using a prescribed model, rather they form self-consistently from an accretion flow undergoing radiative cooling inside a dark halo. 
Preliminary results show that the sizes and masses of the clumps are similar to those presented in this paper.
As long as cosmological simulations will not yield naturally a realistic rate of conversion of gas into stars, while simultaneously capturing correctly the internal structure of discs from high to low redshifts, simulations such as those presented in this paper and those now in progress will remain a valuable complement to understand processes occurring in the interstellar medium on galactic and sub-galactic scales.\\
\section{ACKNOWLEDGMENTS}
This work is supported by the STARFORM Sinergia Project funded by the Swiss National Science Foundation. 
We would like to thank Daniel Schaerer, Miroslava Dessauges and Antonio Cava for very useful conversations and clarifications on the current state of observations of high redshift galaxies, and Rok Ro\v{s}kar and Davide Fiacconi for their valuable help during the analysis of the simulations. We also acknowledge useful discussions with Robert Feldmann, Romain Teyssier, Valentin Perret, Piero Madau and Filippo Fraternali.
We would also like to thank the referee, Avishai Dekel, for the many useful remarks and comments that considerably improved the manuscript.
%
%

%
\bsp
\label{lastpage}
\end{document}